\documentclass[a4paper,12pt]{article}
\usepackage{theorem}

\newtheorem{theorem}{Theorem}[section]

\newtheorem{lemma}[theorem]{Lemma}

\newcount\driver
\newcount\bozza

\font\msytw=msbm10 scaled\magstep1                  
                 \font\indbf=cmbx10 scaled\magstep2

{\count255=\time\divide\count255 by 60 \xdef\hourmin{\number\count255}
        \multiply\count255 by-60\advance\count255 by\time
   \xdef\hourmin{\hourmin:\ifnum\count255<10 0\fi\the\count255}}

\let\a=\alpha \let\b=\beta    \let\g=\gamma     \let\d=\delta     \let\e=\varepsilon
\let\z=\zeta  \let\h=\eta      \let\k=\kappa     \let\l=\lambda
\let\m=\mu    \let\n=\nu      \let\x=\xi        \let\p=\pi        \let\r=\rho
\let\s=\sigma \let\t=\tau     \let\f=\varphi       \let\c=\chi
\let\ps=\psi   \let\o=\omega     
\let\G=\Gamma \let\D=\Delta   \let\Th=\Theta    \let\L=\Lambda    
\let\P=\Pi      \let\F=\Phi       
\let\O=\Omega 

\def\PP{{\cal P}}\def\EE{{\cal E}}\def\VV{{\cal V}}
\def\CC{{\cal C}}\def\WW{{\cal W}}
\def\TT{{\cal T}}\def\BB{{\cal B}}
\def\LL{{\cal L}}
\def\DD{{\cal D}}\def\AA{{\cal A}}\def\GG{{\cal G}}
\def\XXX{{\bf X}}
\def\KK{{\cal K}}

\def\pp{{\bf p}}\def\qq{{\bf q}}\def\xx{{\bf x}}
\def\aaa{{\bf a}} \def\cc{{\bf c}} \def\bb{{\bf b}} \def\dd{{\bf d}}
\def\yy{{\bf y}}\def\kk{{\bf k}}\def\nn{{\bf n}}
\def\zz{{\bf z}}\def\uu{{\bf u}}\def\vv{{\bf v}}\def\ww{{\bf w}}
 \def\bP{{\bf P}}

\def\ss{{\underline \sigma}}       \def\oo{{\underline \omega}}
\def\ee{{\underline \varepsilon}}  
\def\un{{\underline \nu}}          
\def\um{{\underline \mu}}          \def\ux{{\underline\xx}}
          
\def\uaa{{\underline \aaa}} \def\ub{{\underline\bb}}
\def\uc{{\underline\cc}} \def\ud{{\underline\dd}}
           
\def\ut{{\underline t}}            \def\uxi{{\underline \xi}}
\def\umu{{\underline \m}}          \def\uv{{\underline\vv}}
           \def\uy{{\underline\yy}}
\def\uz{{\underline \zz}}
\def\uw{{\underline \ww}}          \def\uo{{\underline \o}}
\def\us{{\underline \s}}           \def\xxx{{\underline \xx}}
          \def\uuu{{\underline\uu}}

\def\RRR{\hbox{\msytw R}}

        \def\EE{\hbox{\msytw E}}

\let\dpr=\partial

\let\bs=\backslash
\let\==\equiv

\let\io=\infty
\let\0=\noindent

\def\*{{\hfill\break\null\hfill\break}}

\def\tilde#1{{\widetilde #1}}

\def\lft{\left}
\def\rgt{\right}
\def\der{\hbox{\rm d}}
\def\la{{\langle}}
\def\ra{{\rangle}}

\def\tende#1{\,\vtop{\ialign{##\crcr\rightarrowfill\crcr
             \noalign{\kern-1pt\nointerlineskip}
             \hskip3.pt${\scriptstyle #1}$\hskip3.pt\crcr}}\,}
\def\otto{\,{\kern-1.truept\leftarrow\kern-5.truept\to\kern-1.truept}\,}

\def\lp{{\hskip-1pt:\hskip 0pt}}
\def\rp{{\hskip-1pt :\hskip1pt}}
\def\defi{{\buildrel \;def\; \over =}}

\def\diam{{\rm diam}}

\def\wt#1{\widetilde{#1}}

\def\sqt[#1]#2{\root #1\of {#2}}

\def\ha{{\widehat \a}}\def\hb{{\widehat \b}}

\def\hf{{\widehat \f}}

\def\hp{{\widehat \ps}}

\def\hJ{{\widehat \jmath}}
\def\hJ{{\widehat J}}
\def\hg{{\widehat g}}

\def\PP{{\cal P}}\def\EE{{\cal E}}\def\VV{{\cal V}}
\def\CC{{\cal C}}\def\WW{{\cal W}}
\def\TT{{\cal T}}\def\BB{{\cal B}}
\def\LL{{\cal L}}
\def\DD{{\cal D}}\def\AA{{\cal A}}\def\GG{{\cal G}}
\def\AAA{{\cal A}}

\def\T#1{{#1_{\kern-3pt\lower7pt\hbox{$\widetilde{}$}}\kern3pt}}
\def\VVV#1{{\underline #1}_{\kern-3pt
\lower7pt\hbox{$\widetilde{}$}}\kern3pt\,}
\def\W#1{#1_{\kern-3pt\lower7.5pt\hbox{$\widetilde{}$}}\kern2pt\,}

\def\indica{\leaders \hbox to 0.5cm{\hss.\hss}\hfill}
\def\guida{\leaders\hbox to 1em{\hss.\hss}\hfill}
\mathchardef\oo= "0521

\def\pp{{\bf p}}\def\qq{{\bf q}}\def\xx{{\bf x}}
\def\yy{{\bf y}}\def\kk{{\bf k}}\def\nn{{\bf n}}
\def\dd{{\bf d}}\def\zz{{\bf z}}\def\uu{{\bf u}}\def\vv{{\bf v}}
 \def\bP{{\bf P}}
 
\def\ss{{\underline \sigma}}\def\oo{{\underline \omega}}
\def\xxx{{\underline\xx}}

\def\qed{\raise1pt\hbox{\vrule height5pt width5pt depth0pt}}

\def\indic{\hbox{\raise-2pt \hbox{\indbf 1}}}

\def\RRR{\hbox{\msytw R}}

\def\ins#1#2#3{\vbox to0pt{\kern-#2 \hbox{\kern#1 #3}\vss}\nointerlineskip}

\newdimen\xshift \newdimen\xwidth \newdimen\yshift
\newcount\griglia

\def\insertplot#1#2#3#4#5#6{%
\xwidth=#1pt \xshift=\hsize \advance\xshift by-\xwidth \divide\xshift by 2%
\begin{figure}[ht]
\vspace{#2pt}
\hspace{\xshift}
\begin{minipage}{#1pt}
#3
\ifnum\driver=1 \griglia=#6
\ifnum\griglia=1
\openout13=griglia.ps
\write13{gsave .2 setlinewidth}
\write13{0 10 #1 {dup 0 moveto #2 lineto } for}
\write13{0 10 #2 {dup 0 exch moveto #1 exch lineto } for}
\write13{stroke}
\write13{.5 setlinewidth}
\write13{0 50 #1 {dup 0 moveto #2 lineto } for}
\write13{0 50 #2 {dup 0 exch moveto #1 exch lineto } for}
\write13{stroke grestore}
\closeout13
\includegraphics{griglia.ps}
\fi
\includegraphics{#4.ps}\fi%
\ifnum\driver=2 \fi
\end{minipage}
\caption{#5}
\end{figure}
}

\newdimen\shift \shift=-1.5truecm
\def\lb#1{%
\ifnum\bozza=1
\label{#1}\rlap{\hbox{\hskip\shift$\scriptstyle#1$}}
\else\label{#1} \fi}

\def\be{\begin{equation}}
\def\ee{\end{equation}}
\def\bea{\begin{eqnarray}}\def\eea{\end{eqnarray}}
\def\bean{\begin{eqnarray*}}\def\eean{\end{eqnarray*}}
\def\bfr{\begin{flushright}}\def\efr{\end{flushright}}
\def\bc{\begin{center}}\def\ec{\end{center}}
\def\bal{\begin{align}}\def\eal{\end{align}}
\def\ba#1{\begin{array}{#1}} \def\ea{\end{array}}
\def\bd{\begin{description}}\def\ed{\end{description}}
\def\bv{\begin{verbatim}}\def\ev{\end{verbatim}}
\def\nn{\nonumber}
\def\Halmos{\hfill\vrule height10pt width4pt depth2pt \par\hbox to \hsize{}}
\def\pref#1{(\ref{#1})}
\def\Dim{{\bf Dim. -\ \ }} 
\def\virg{\quad,\quad}

\title{Massless Sine-Gordon
and Massive Thirring Models: proof of the Coleman's equivalence}

\author{G. Benfatto$^1$ \and P. Falco$^2$ \and  V. Mastropietro$^1$\\
\\
{\small $^{1}$ Dipartimento di Matematica, Universit\`a di Roma ``Tor Vergata''}\\ 
{\small via della Ricerca Scientifica, I-00133, Roma}\\
\\
{\small $^{2}$   Mathematics Department, University of British Columbia,}\\
{\small Vancouver, BC Canada, V6T 1Z2}}
\date{}
\begin{document}
\maketitle
\begin{abstract}
We prove the Coleman's conjecture on the
equivalence between the massless Sine-Gordon model with finite volume
interaction and the Thirring model with a finite volume mass term.
\end{abstract}

\section{Introduction}

\subsection{Coleman's Equivalence}

One of the most fascinating aspects of QFT in $d=1+1$ is the
phenomenon of {\it bosonization}; fermionic systems can be mapped
in bosonic ones and viceversa. The simplest example is provided by
the equivalence between free massless Dirac fermions and free
massless bosons with the identifications (see for instance
\cite{ID}):
\be\lb{1a1}
\bar\psi_\xx(1+\s\g_5)\psi_\xx\;\sim\; b_0\, \lp e^{i\s \sqrt{4\p}
\phi_\xx}\rp \virg \bar\psi_\xx\g^\m\psi_\xx\;\sim\; -{1\over
\sqrt{\p}}\, \e^{\m\n}\dpr_\n \phi_\xx
\ee
where  $\s=\pm 1$ and $b_0$ is a suitable constant, depending on
the precise definition of the Wick product. Such equivalence can
be extended to interacting theories; Coleman \cite{C} showed the
equivalence, in the zero charge sector, between the {\it massive
Thirring model}, with Lagrangian (with our conventions)
\be\lb{1b1}
\LL= i Z\bar\psi_\xx \not{\hskip-.05cm\partial}\, \psi_\xx -Z_1 \m
\bar\psi_\xx\psi_\xx - {\l\over 4} Z^2 j_{\m,\xx} j^\m_\xx
\ee
where $Z$ and $Z_1$ are (formal) renormalization constants,
$j_{\m,\xx} = \bar\psi_\xx\g^\m\psi_\xx$ and the massless
Sine-Gordon model, with Lagrangian
\be\lb{1b2}
\LL= {1\over 2} \dpr_\m \f_\xx \dpr^\m \f_\xx + \z\, \lp
\cos(\a\phi_\xx)\, \rp
\ee
with the identifications
\be\lb{1a11}
Z_1\bar\psi_\xx (1+\s\g_5)\psi_\xx\;\sim\; b_0\, \lp
e^{i\a\s\phi_\xx}\rp \virg Z \bar\psi_\xx\g^\m\psi_\xx\;\sim\;
-b_1\, \e^{\m\n}\dpr_\n \phi_\xx
\ee
where $b_0$, $b_1$ are two suitable constants, depending on $\l$
and the details of the ultraviolet regularization. Moreover, this
equivalence is valid if certain relations between the Thirring
parameters $\l$, $\m$ and the Sine-Gordon parameters $\a,\z$ are
assumed. The case $\a^2=4\pi$ is special, as it corresponds to
free fermions ($\l=0$); the choice $\z=0$ (free bosons)
corresponds to massless fermions ($\m=0$).

In order to establish such equivalence, Coleman considered a {\it
fixed} infrared regularizations of the models \pref{1b1} and
\pref{1b2}, replacing $\m$ in \pref{1b1} with $\m\chi_\L(\xx)$ and
$\z$ with $\z\chi_\L(\xx)$, with $\chi_\L(\xx)$ a compact support
function; this means that the mass term in the Thirring model, and
the interaction in the Sine-Gordon is concentrated on a finite
volume $\L$. Such regularization makes possible a perturbative
expansion, respectively in $\m$ for the Thirring model and  $\z$
for the Sine-Gordon model; it turned out that the coefficients of
such series expansions can be explicitly computed (in the case of
the Thirring coefficients this was possible thanks to the explicit
formulas for the correlations of the massless Thirring model given
first in \cite{Ha,K}) and they are {\it order by order} identical
if the identification \pref{1a11} is done and provided that
suitable relations between the parameters are imposed.

The identification of the series expansions coefficients would
give a rigorous proof of the equivalence {\it provided that} the
series are convergent. The issue of convergence, which was mentioned
but not addressed in \cite{C}, is technically quite involved and crucial;
there are several physical examples in which order by order
arguments without convergence lead to uncorrect prediction.

The search for a rigorous proof of Coleman equivalence was the
subject of an intense investigation in the framework of
constructive QFT, leading to a number of impressive results. In
\cite{FS} it was rigorously proven the equivalence between the
{\it massive} Sine-Gordon model (with mass $M$ large enough) at
$\a^2< 4\pi$ and a Thirring model with a large long-range
interaction; similar ideas were also used in \cite{SU}. The
properties of the {\it massive} Sine-Gordon model for $\a^2\ge
4\p$ were later on deeply investigated. In \cite{BGN} and
\cite{NRS} it was proved that the model is stable if one adds a
finite number, increasing with $\a$, of vacuum counterterms, while
the full construction, through a cluster expansion, of the model
was partially realized in \cite{DH}. In \cite{DH} it was also
proved that the correlation functions are analytic in $\z$, for
any $\a^2<8\p$. A proof of analyticity, only based on a multiscale
analysis of the perturbative expansion, was first given in
\cite{B}, for $\a^2< 4\pi$, and then extended in \cite{BK} up to
$\a^2< 16/3\pi$.

Using the results in \cite{DH} for a fixed finite volume, Dimock
\cite{D} was able finally to achieve a proof of Coleman's
equivalence, in the Euclidean version of the models, {\it for the
case $\a^2=4\pi$}; such a value is quite special as it corresponds
to $\l=0$, that is the equivalence is with a {\it free} massive
fermionic system, without current-current interaction. Such
limitation was mainly due to the fact that the constructive
analysis of interacting fermionic systems was much less developed
at that time: indeed a rigorous construction of the massive
Thirring model in a functional integral approach has been achieved
only quite recently \cite{BFM}.

A more physically oriented research on Coleman's equivalence was
focused in recovering bosonization in the framework of the
(formal) path-integral approach, \cite{N,FGS}. The idea is to
introduce a vector field $A_\m$ and to use the identity
\be
\exp \left\{ -{\l\over 4}\int d\xx j_{\m,\xx} j_{\m,\xx} \right\}
= \int DA \exp \left\{ \int d\xx  \left[ - A_{\m,\xx}^2 +
\sqrt{\l} A_{\m,\xx} j_{\m,\xx}) \right] \right\}
\ee
By parameterizing $A_\m$ in terms of scalar fields
$\x_\xx,\phi_\xx$
\be\lb{1.6a}
A_\m=\partial_\m\x_\xx +\e_{\m,\n}\partial_\n\phi_\xx
\ee
it turns out that the massive Thirring model can be expressed in
terms of the boson fields $\x_\xx$ and $\phi_\xx$: the first is a
massless free field, while the second one has an exponential
interaction when $\m\not=0$. In the expectations of the operators
$\bar\psi_\xx(1+\s\g_5)\psi_\xx$ and $j_{\m,\xx}$, the $\x_\xx$
field has no role and it can be integrated out; the resulting
correlations imply the identification \pref{1a11}. Such
computations are however based on formal manipulations of
functional integrals (with no cut-offs, hence formally infinite)
and it is well known that such arguments can lead to incorrect
result (see for instance the discussion in \S 1 in [BFM]).

In this paper we will give the first proof of Coleman's
equivalence between the Euclidean massive Thirring model with a
small interaction and the Euclidean massless Sine-Gordon model
with $\a$ around $4\pi$. We will follow the Coleman strategy, but
an extension of the multiscale techniques developed in \cite{B}
for the Sine-Gordon model and in \cite{BM,BFM} for the Thirring
model allow us to achieve the convergence of the expansion.

\subsection{Main results}
\lb{s1.2}

We start from a suitable regularization of the Sine-Gordon and
Thirring models via the introduction of infrared and ultraviolet
cut-offs, to be removed at the end.

Let us consider first the (Euclidean) {\it Sine-Gordon} model. Let
$\g>1$, $h$ be a large negative integer ($\g^h$ is the {\it
infrared cutoff}) and $N$ be a large positive integer ($\g^N$ is
the {\it ultraviolet cutoff}). Moreover, let $\f_\xx$ a
2-dimensional bosonic field and $P_{h,N}(d\f)$ be the Gaussian
measure with covariance $C_{h,N}(\xx)\defi$ $\sum_{j=h}^N
C_0(\g^j\xx)$, for
\bea \lb{cov0}
C_0(\xx) \defi \frac1{(2\p)^2} \int \frac{d \kk}{\kk^2} \left[
e^{-\kk^2}- e^{-(\g \kk)^2} \right] e^{i\kk\xx} \eea
Given the two real parameters $\z$, the {\it coupling}, and $\a$
(related with the inverse temperature $\b$, in the Coulomb gas
interpretation of the model, by the relation $\b=\a^2$), the
Sine-Gordon model with finite volume interaction and ultraviolet
and infrared cutoffs is defined by the {\it interacting measure}
$P_{h,N}(d\f) \exp \{\z_N V(\f)\}$, with
\be \lb{int}
V(\f) = \int_\L d\xx\; \cos(\a\f_\xx) \virg \z_N= e^{\frac{\a^2}2
C_{0,N}(0)} \z
\ee
where $\L$ is a fixed volume of size $1$. Note that $\z_N V(\f) =
\z \int_\L d\xx\ \lp \cos(\a\f_\xx) \rp$, where
\be \lb{WP}
\lp e^{ia\f_\xx}\rp \; \defi\; e^{ia\f_\xx} e^{\frac{a^2}2
C_{0,N}(0)}
\ee
is the Wick product of $e^{i\a\f_\xx}$, $a\in\RRR$, with respect
to the measure with covariance $C_{0,N}(\xx)$ (for any $h$); hence
$\z_N$ has the role of the {\it bare strength}.

We consider now the {\it Thirring model}. The precise
regularization of the path integral for fermions was already
described in \cite{BFM}, \S1.2, therefore  we only remind the main
features. We introduce in $\L_L\= [-L/2, L/2]\times [-L/2, L/2]$ a
lattice $\L_a$ whose sites represent the space-time points. We
also consider the lattice  $\DD_a$ of space-time momenta
$\kk=(k,k_0)$. We introduce a set of Grassmann spinors
$\psi_\kk,\bar\psi_\kk$, $\kk\in\DD_a$, such that
$\psi_\kk=(\psi^-_{\kk,+},\psi^-_{\kk,-})$, $\bar\psi=\ps^+ \g^0$
and $\psi_\kk^+=(\psi^+_{\kk,+},\psi^+_{\kk,-})$. The $\g$
matrices are explicitly given by
$$\g^0=
 \pmatrix{0&1\cr
          1&0\cr}\;,
 \qquad
 \g^1=
 \pmatrix{0&-i\cr
          i&0\cr}\;,
 \qquad
 \g^5=-i\g^0\g^1
 =
 \pmatrix{1&0\cr
          0&-1\cr}\;.$$

We also define a Grassmann field on the lattice $\L_a$ by Fourier
transform, according to the following convention:
\be
 \ps^{[h,N]\s}_{\xx,\o}\defi
 {1\over L^2} \sum_{\kk\in\DD_a} e^{i\s\kk\xx}
 \hp^{[h,N]\s}_{\kk,\o}\;,
 \qquad \xx\in \L_a\;.
\ee
Sometimes $\ps^{[h,N]\s}_{\xx,\o}$ will be shorten into
$\ps^{\s}_{\xx,\o}$. Moreover, since the limit $a\to 0$ is trivial
[BFM], we shall consider in the following $\ps^{[h,N]\s}_{\xx,\o}$
as defined in the continuous box $\L_L$.

In order to introduce an ultraviolet and an infrared cutoff, we
could use a gaussian cut-off as in \pref{cov}, but for technical
reason,and to use the results of \cite{BFM}, we find more
convenient to use a compact support cut-off. We define the
function $\c_{h,N}(\kk)$ in the following way; let $\c\in
C^\io(\RRR_+)$ be a {\it Gevrey function} of class 2,
non-negative, non-increasing smooth function such that
\be \c(t)=
\lft\{\matrix{ 1\hfill&\hfill{\rm if\ }0\le t\le 1\cr
0\hfill&\hfill{\rm if\ } t\ge \g_0\;,}\rgt.\ee
for a fixed choice of  $\g_0:1<\g_0\le\g$; then we define, for any
$h\le j\le N$,
\be \lb{1.5}
f_j(\kk)= \chi
\lft(\g^{-j}|\kk|\rgt)-\chi\lft(\g^{-(j-1)}|\kk|\rgt)
\ee
and $\c_{h,N}(\kk)=\sum_{j=h}^N f_j(\kk)$; hence $\c_{h,N}(\kk)$
acts as a smooth cutoff for momenta $|\kk|\ge \g^{N+1}$  and
$|\kk|\le \g^{h-1}$.

Given two real parameters, the {\it bare coupling} $\l$ and the
{\it bare mass} $\m$, the Thirring model with finite volume mass
term and ultraviolet and infrared cutoffs is defined by the {\it
interacting measure} $P_{h,N}(d\psi) \exp \{\VV(\psi)\}$, with
\be\lb{VVV}
\VV(\psi)=-{\l\over 4} Z_N^2 \int_{\L_L} \!d\xx\ \lft(\bar \ps_\xx
\g^\m\ps_\xx\rgt)^2 + Z_N^{(1)}\m\int_{\L}\!d\xx\ \bar \ps_\xx
\ps_\xx +E_{h,N}|\L_L|
\ee
and
\bea\lb{1.7}
&& P_{h,N}(d\ps)\defi \der\hp \prod_{\kk\in\DD^{[h,N]}} \left[
L^{-4} Z^2_N|(-|\kk|^2C^2_{h,N}(\kk)
\right]^{-1} \cdot\nn\\
&& \exp\left\{-Z_N {1\over L^2}\sum_{\o=\pm} \sum_{\kk\in
\DD^{[h,N]}} {D_{\o}(\kk)\over \c_{h,N}(\kk)}
\hp^{+}_{\kk,\o}\hp^{-}_{\kk,\o'}\right\}\;,
\eea
where $D_\o(\kk)\defi-ik_0+\o k_1$ and $E_{h,N}$ is constant
chosen so that, if $\m=0$, $\int P_{h,N}(d\psi) \exp \{\VV(\psi)\}
= 1$.
We will prove the following theorem.

\begin{theorem} \lb{t1}
Assume $|\z|,|\l|,|\m|$ small enough, $\a^2< 16\p/3$; then there
exist two constants $\h_-=a\l^2+O(\l^3)$ and $\h_+=b\l+O(\l^2)$,
with $a, b>0$, independent of $\m$ and analytic in $\l$, such
that, if we put
\be\lb{ZZ}
Z_N=\g^{-\h_- N}\virg Z^{(1)}_N= \g^{-\h_+ N}
\ee
then, if $r=0$ and $q\ge 2$ or $r\ge 1$, for any choice of the non
coinciding points $(\xx_1, \ldots, \xx_q, \yy_1, \ldots, \yy_r)$,
and of $\s_i=\pm 1$, $i=1,\ldots,q$, $\n_j=0,1$, $j=1,\ldots,r$,
\bea\lb{asa}
&&\lim_{-h,N\to\io} \la \Bigg [\prod_{i=1}^q \lp e^{i \s_i
\a\f_{\xx_i}}\rp\Bigg] \Bigg[\prod_{j=1}^r (-1) \e^{\n_j \m}\dpr^\m
\phi_{\yy_j}\Bigg] \ra^T_{SG} = \\
&& = \lim_{-h,N\to\io} (b_0Z_N^{(1)})^q  (b_1Z_N)^r \la
\Bigg[\prod_{i=1}^q
\bar\psi_{\xx_i} \left( {1+\s_i\g_5\over 2} \right)
\psi_{\xx_i}\Bigg] \Bigg[\prod_{j=1}^r \bar\psi_{\yy_j}
\g^{\n_j}\psi_{\yy_j}\Bigg] \ra^T_{Th}\nn
\eea
where $\la\cdot\ra^T_{Th}$ and $\la\cdot\ra^T_{SG}$ denote the truncated
expectations in the Thirring (in the limit $L\to\io$) and
Sine-Gordon models, respectively, $b_0$ and $b_1$ are bounded
functions of $\l$ and the following relations between the
parameters of the two models have to be verified:
\be\label{1v}
{\a^2\over 4\pi}=1+ \h_- - \h_+ \virg \z = b_0\m 
\ee
If $q=1$ and $r=0$ both the r.h.s. and the l.h.s. of \pref{asa}
are diverging for $\l\le 0$, while the equality still holds for
$\l> 0$. A divergence also appears, for $\l\le 0$, in the
pressure, but only for the second order term in $\z$ or $\m$;
however, if we add a suitable vacuum counterterm, also the
pressures are equal.
\end{theorem}

This Theorem proves Coleman's equivalence \pref{1a11}. We remark that
the relations between the Sine-Gordon parameters and the Thirring
parameters in \pref{1v} are slightly different with respect to
those in \cite{C}, for $\l\not=0$; this is true in particular for
the first equation, involving only quantities which have a
physical meaning in the removed cutoff limit, if we express them
in terms of $\l$, as Coleman does. This is not surprising, as the
relations between the physical quantities, like the critical
indices $\h_\pm$, and the bare coupling depend on the details of
the regularization, and in our Renormalization Group analysis the
running coupling constants have a bounded but non trivial flow
from the ultraviolet to the infrared scales. Indeed, with a
different regularization of the Thirring model (that is starting
from a non local current-current interaction and performing the
local limit after the limit $N\to\io$), as in \cite{M1,M2}, one
would get a simple relation between $\a$ and $\l$. This new
relation again is not equal to that of \cite{C}, but is in
agreement with the regularization procedure of \cite{J}, see
footnote 7 of \cite{C}.

Another important remark concerns the limit $\L\to\io$. In the
case of Sine-Gordon model, one expects that, in this limit, there
is exponential decrease of correlations (implying the screening
phenomenon in the Coulomb gas interpretation), which is not
compatible with convergence of perturbative expansion (in this
case the correlations would have a power decay as in the free
theory). Up to now, screening has been proved only for $\a^2 <<
4\p$ \cite{Y}, but it is expected to be verified in all range of
validity of the model ($\a^2<8\p$), hence even around $\a^2=4\p$.
However, if the interaction is restricted to a fixed finite
volume, convergence is possible and we could indeed prove it, for
$\a^2<6\p$; in this paper, for simplicity, we give the proof only
for $\a^2<16\p/3$, which is sufficient to state the main result.

The situation for the massive Thirring model is slightly
different, because it has been shown \cite{BFM} that it is well
defined in the limit $\L\to\io$ and that its correlations decay at
least as $\exp(-c\sqrt{|\m|^{1+O(\l)}|\xx|})$. Hence, even if the
power expansion in the mass can be convergent only if we fix the
volume, the proof of Coleman's conjecture strongly supports the
related conjecture that even the Sine-Gordon model is well defined
around $\a^2=4\p$ in the infinite volume limit and has exponential
decrease of correlations.

The proof is organized in the following way. In \S\ref{sec3} we
analyze the massless Sine-Gordon model with finite volume
interaction and $\a^2< 16\p/3$, extending the proof of analyticity
in $\z$ given in \cite{B} for the massive case in the infinite
volume limit and $\a^2<4\p$. With respect to the technique used in
\cite{D}, where only the case $\a^2=4\p$ was analyzed, our method
has the advantage that an explicit expression of the coefficients
can be easily achieved; this is probably possible even with the
other method, but the proof was given only for $\a^2<4\p$ and, as
a consequence, the correlations in the model with $\a^2=4\p$ were
defined as the limit $\a^2\to 4\p$ of those with $\a^2<4\p$.

In \S\ref{sec3a} we use the methods developed in \cite{BFM,M1,M2}
to prove the analyticity in $\m$ of the Thirring model; the
explicit expressions of the coefficients are obtained in
\S\ref{sec4}, by using the explicit expression of the field
correlation functions given in the Appendix (through the solution
of a Schwinger-Dyson equation, based on a rigorous implementation
of Ward Identities) and by a rigorous implementation, in a RG
context, of the point spitting procedure used in theoretical
physics.

\section{The Massless Sine-Gordon Model with a finite volume
interaction}\lb{sec3}

We want to study the measure defined in \S\ref{s1.2} in the limit
of removed cutoff, $-h,N\to\io$. To this purpose, we consider the
{\it Generating functional}, $\KK_{h,N}(J,A,\z)$, defined by the
equation
\bea
&& \KK_{h,N}(J,A,\z) = \log \int\!P_{h,N}(d\f)\; e^{\z_N V(\f)}
\cdot\nn\\
&&\cdot \exp\lft\{ \sum_{\s=\pm 1} \int d\xx J^{\s}_\xx \lp e^{i\a
\s\f_\xx}\rp + \sum_{\n=0,1} \int d\yy A^{\n}_\yy
\lft(\dpr^\n\f_\yy\rgt)\rgt\}
\eea
where  $J^\s_\zz$ and $A^\m_\yy$ are two-dimensional, external
bosonic fields. Then, given two non negative integers $q$ and $r$,
as well as two sets of labels $\us=(\s_1,\ldots,\s_q)$ and
$\un=(\n_1,\ldots,\n_r)$, together with two sets of two by two
distinct points $\uz=(\zz_1, \ldots, \zz_q)$ and $\uy=(\yy_1,
\ldots, \yy_q)$, we consider the Schwinger functions, defined by
the equation
\be\lb{GhN}
K^{(q,r;\z)}_{h,N}(\uz,\uy;\us,\un) \defi {\dpr^{q+r} \KK_{h,N}
\over \dpr J^{\s_1}_{\zz_1}\cdots \dpr J^{\s_q}_{\zz_q} \dpr
A^{\n_1}_{\yy_1}\cdots \dpr A^{\n_r}_{\yy_r}}(0,0,\z)
\ee

\begin{theorem} \lb{t2}
If $|\z|$ is small enough, $\a^2< 16\p/3$ and $q\ge 2$, if $r=0$,
or $q\ge 0$, if $r\ge 1$, the limit
\be
K^{(q,r;\z)}(\uz,\uy;\us,\un) \defi \lim_{-h,N\to+\io}
K^{(q,r;\z)}_{h,N}(\uz,\uy;\us,\un)
\ee
exists and is analytic in $\z$. In the case $q=r=0$ (the
pressure), the limit does exist and is analytic, up to a
divergence in the second order term, present only for $\a^2\ge
4\p$.
\end{theorem}
For clarity's sake, we prefer to give the proof of the above
theorem in the special cases $(q,r)=(k,0)$ and $(q,r)=(0,k)$
separately; the proof in the general case is a consequence of the
very same ideas that will be discussed for the special ones, but
it needs a more involved notation, so we will not report its
details.

\subsection{The free measure}

By the definitions given in \S\ref{s1.2}, the regularized free
measure is the two--dimensional boson Gaussian measure with
covariance
\be \lb{cov}
C_{h,N}(\xx) = \frac1{(2\p)^2} \int \frac{d \kk}{\kk^2} \left[
e^{-(\g^{-N}\kk)^2}- e^{-(\g^{-h+1}\kk)^2} \right] e^{i\kk\xx} =
\sum_{j=h}^N C_0(\g^j\xx)
\ee
The two--dimensional massless boson Gaussian measure is obtained
by taking the limits $h\to -\io$ and $N\to\io$. It is easy to
prove that
\be \lb{dec}
C_0(0) = {\log\g \over 2\p} \virg
\left| \dpr_{x_0}^{q_0} \dpr_{x_1}^{q_1} C_0(\xx) \right| \le
A_{q_0,q_1,\k} e^{-\k |\xx|}
\ee
where $q_0$, $q_1$ are non negative integers and $\k0$ is an
arbitrary positive constant. Let us now consider the function
\be
C_{h,\io}(\xx) = \lim_{N\to\io} C_{h,N}(\xx) = C_{0,\io}(\g^h
\xx)\ee
It is easy to show, by a standard calculation, that there exists a
constant $c$ such that
\be \lb{zero}
| C_{0,\io}(\xx) + {1\over 4\p} \log (c |\xx|^2) | \le C |\xx|^2
\ee
Hence, $C_{h,\io}(\xx)$ diverges for $h\to -\io$ as $-(2\p)^{-1}
\log (\g^h |\xx|)$. However, if we define
\be \lb{defV}
\D^{-1}_{h,\io}(\xx) = C_{h,\io}(\xx) + {1\over 4\p} \log (c
\g^{2h})
\ee
we have, by \pref{zero}:
\be \lb{Vinf}
\D^{-1}(\xx) \defi \lim_{h\to-\io} \D^{-1}_{h,\io}(\xx) = -{1\over
2\p} \log |\xx|
\ee

Then it is natural to define the {\it Coulomb potential with
ultraviolet cutoff} by
\be \lb{defVhN}
\D^{-1}_N(\xx) \defi  \lim_{h\to-\io} \D^{-1}_{h,N}(\xx) \virg
\D^{-1}_{h,N}(\xx) \defi C_{h,N}(\xx) + {1\over 4\p} \log (c
\g^{2h})
\ee

Since $C_{h,N}(\xx) = C_{h,\io}(\xx) - C_{N,\io}(\xx)$, by using
\pref{dec} and \pref{Vinf}, we see that
\be \lb{asinf}
\Big|\D^{-1}_N(\xx) + {1\over 2\p} \log |\xx|\Big| \le C
e^{-\k\g^N |\xx|} \virg \g^N|\xx| \ge 1
\ee
and, by using \pref{zero}, we see that
\be \lb{aszero}
|\D^{-1}_N(\xx) - {1\over 4\p} \log (c \g^{2N})| \le C \g^{2N}
|\xx|^2 \virg \g^N|\xx| \le 1
\ee
We define $\EE_{h,N}$ and $\EE_j$ to be the expectation with
respect to the Gaussian measures with covariance $C_{h,N}(\xx)$
and $C_j(\xx)=C_0(\g^j\xx)$, respectively; a superscript $T$ in
the expectation will indicate a truncated expectation. Recall
that, for a generic probability measure with expectation $\EE$,
and any family of random variables $(f_1, \ldots, f_s)$, $\EE^T$
is defined as
\be\lb{trexp}
\EE^T[f_1;\ldots;f_s] = \sum_{\P} (-1)^{|\P|-1} (|\P|-1)! \prod_{X\in
\P} \EE\lft[\prod_{i\in X} f_i\rgt]
\ee
where $\sum_\P$ denotes the sum over the partitions of the set
$(1,\ldots,s)$. Finally we remind that $\lp e^{i\s\b\f_\xx}\rp$ is
the Wick normal ordering of $e^{i\s\b\f_\xx}$ {\it always} taken
with respect to the measure with covariance $C_{0,N}(\xx)$ (see
definitions in \S\ref{s1.2}).

\begin{lemma} \lb{lm2}
Let $\s_i\in \{-1,+1\}$, $i=1,\ldots,n$, and $\a\in\RRR$. If
$Q\defi\sum_r \s_r$, then
\be \lb{exp}
\lim_{h\to -\io} \EE_{h,N}\lft[ \prod_{r=1}^n \lp
e^{i\a\s_r\f_{\xx_r}}\rp\rgt]= \d_{Q,0} c^{-{\a^2\over 8\p}n}
e^{-\a^2 \sum_{r<s} \s_r\s_s \D^{-1}_N(\xx_r-\xx_s)}
\ee
\end{lemma}

{\bf\0Proof.} We first notice that if the Wick product had been
defined  with respect to the covariance $C_{h,N}$,  then
$\log\EE_{h,N}\lft[\prod_{r=1}^n \lp
e^{i\a\s_r\f_{\xx_r}}\rp\rgt]$ would have been equal to $-\a^2
\sum_{r<s} \s_r\s_s C_{h,N}(\xx_r-\xx_s)$. Hence, by definition
\pref{defVhN}, we get
\bean
&& \log\EE_{h,N}\lft[\prod_{r=1}^n \lp
e^{i\a\s_r\f_{\xx_r}}\rp\rgt] = \frac{\a^2}{4\p} h n\log\g -\a^2
\sum_{r<s} \s_r\s_s
C_{h,N}(\xx_r-\xx_s)= \\
&& = \frac{\a^2}{4\p} h Q^2 \log\g + {\a^2\over 8\p} (Q^2-n) \log
c -\a^2 \sum_{r<s} \s_r\s_s \D^{-1}_{h,N}(\xx_r-\xx_s)
\eean
which immediately implies the lemma.\Halmos

If $\EE$ is the expectation $\EE_{h,N}$ in the limit $-h,N\to\io$,
by taking the limit $N\to\io$ in the r.h.s. of \pref{exp}, we get,
in the case $Q=0$,
\be\lb{expinf}
\EE\lft[\prod_{r=1}^n \lp e^{i\a\s_r\f_{\xx_r}}\rp\rgt]  =
\d_{Q,0} c^{-{\a^2\over 8\p}n} \prod_{r<s} |\xx_r-\xx_s|^{\s_r\s_s
{\a^2\over 2\p}}
\ee
We are now ready to consider the interacting measure.

\subsection{The case $q=r=0$ (the pressure)}\lb{sec2}

To begin with we analyze the pressure:
\be \lb{press}
p(\z) \defi \lim_{-h,N\to\io}  \log Z_{h,N}(\z) \virg Z_{h,N}(\z)
\defi \int\!P_{h,N}(d\f)\; e^{\z_N V(\f)}
\ee
We proceed as in \cite{B}, by studying the multiscale expansion
associated with the following decomposition of the covariance:
\be \lb{decomp}
C_{h,N}(\xx) = \sum_{j=0}^N C_j(\xx) + C_{h,-1}(\xx)\virg C_j(\xx)
\defi C_0(\g^j \xx)
\ee
In comparison with  \cite{B}, where the case $h=0$ - the ``Yukawa
gas'' - was considered, here we are collecting in a single
integration step all scales below $h=0$: as we shall see, this is
effective since the volume size is fixed to be $1$. To simplify
the notation, from now on $\EE_{-1}$ will denote the expectation
w.r.t. $C_{h,-1}(\xx)$, while $\EE_j$ will have the previous
meaning for $j\ge 0$.

Let $\TT^{(N)}_n$, $n\ge 2$, be the family of labelled trees with
the following properties:

\bd
\item{1)} there is a root $r$ and $n$ ordered {\it endpoints}
$e_i$, $i=1,\ldots,n$, which are connected by the tree; the tree
is ordered from the root to the endpoints;

\item{2)} each vertex $v$ carries a {\it frequency label} $h_v$,
which is an integer taking values between $-1$ and $N+1$, with the
condition that $h_{u}<h_v$, if $u$ precedes $v$ in the order of
the tree; moreover, the root has frequency $-1$ and the endpoint
$e_i$ has frequency $h_i+1$, if $h_i$ is the frequency of the
higher vertex preceding it.

\item{3)} The endpoint $e_i$ carries two other labels, the {\it
charge} $\s_i$ and the {\it position} $\xx_i$. \ed
These trees differ from those used in \cite{BFM} for the Thirring
model, because there are no ``trivial vertices'' on the lines of
the tree.

Since $\TT^{(N)}_n \subset \TT^{(N+1)}_n$, then $\TT^{(\io)}_n =
\lim_{N\to\io} \TT^{(N)}_n$ is obtained from  $\TT^{(N)}_n$ by
letting the frequency indices free to vary between $-1$ and $\io$.

We shall also use the following definitions:

\bd \item{a)} Given a tree $\t$, we shall call {\it non trivial}
(n.t. in the following) the tree vertices different from the root
and from the endpoints. If $v\in\t$ is a n.t. vertex, $s_v\ge 2$
will denote the number of lines branching from $v$ in the positive
direction, $v'\in \t$ is the higher non trivial vertex preceding
$v$, if it does exist, or the root, otherwise. Moreover, $X_v$
will be the set of endpoints following $v$ along the tree; $X_v$
will be called the {\it cluster} of $v$ and $n_v$ will denote the
number of its elements. If $v$ is an endpoint, $X_v$ will denote
the endpoint itself. Finally we define $\Phi(X_v)=
\sum_{i\;:\;e_i\in X_v} \s_i \f_{\xx_i}$.

\item{b)} Given a n.t. vertex $v$ and an integer $j\in [-1, N]$,
 we shall denote
\be \lb{Uj}
U_j(v) \defi \sum_{r,m\;:\;e_r,e_m \in X_v} \s_r \s_m
C_j(\xx_r-\xx_m) =\EE_j\lft[\Phi^2(X_v)\rgt]\ge 0\ee
the (double of the) {\it total energy on scale $j$} associated
with its cluster. If $k'+1\le k-1$, we shall also define
\be \lb{Ujj}
U_{k',k}(v) \defi \sum_{j=k'+1}^{k-1} U_j(v) \ee

\item{c)} If $X$ and $Y$ are two disjoint clusters and $j$ is an
integer contained in $[-1, N]$, we denote
\be\lb{wjxy}
 W_j(X,Y) \defi \sum_{r,m\;:\;e_r\in X,\;e_m\in Y}
\s_r \s_m C_j(\xx_r-\xx_m)=\EE_j\lft[\Phi(X)\Phi(Y)\rgt]
\ee
the {\it interaction energy on scale $j$} between $X$ and $Y$.

\item{d)} Given a n.t. vertex $v$, $(v_1, \ldots, v_{s_v})$, will
denote the set of vertices following it along the tree; moreover
we define
\be \lb{Gj}
G_j(v_1, \ldots, v_{s_v}) = \EE^T_j \lft[ e^{i\a \Phi(X_{v_1})};
\ldots; e^{i\a \Phi(X_{v_{s_v}})}\rgt]
\ee

\ed

By proceeding as in \cite{B}, it is easy to see that
\be \lb{ZhN}
Z_{h,N}(\z) = \int\! P_{h,-1}(\f)\; e^{\z V(\f) + \sum_{n=2}^\io
\z^n V^{(N)}_n(\f)}
\ee
where
\be
V^{(N)}_n(\f) = \sum_{\t\in \TT^{(N)}_n} {1\over 2^n}\sum_{\s_1,
\ldots, \s_n} \int_{\L^n} d\xx_1 \cdots d\xx_n\;
e^{i\a\sum_{r=1}^n \s_r \f_{x_r}} V_\t(\ss,\ux)
\ee
\be \lb{Vtau}
V_\t(\ss,\ux) = \lft(\prod_{i=1}^n \g^{{\a^2\over 4\p}
h_{e_i}}\rgt) \prod_{n.t.\ v\in\t} {G_{h_v}(v_1, \ldots,
v_{s_v})\over s_v!} e^{-{\a^2\over 2} U_{h_{v'}, h_v}(v)}
\ee
We note that $V_\t(\ss,\ux)$ is independent of $N$ and $h$.

In order to prove that the pressure, see \pref{press}, is well
defined, the main step is to verify that, uniformly in $h$ and
$N$, $Z_{h,N}(\z) = 1 +O(\z)$. As we will discuss later in this
section, since the only dependence on $h$ in \pref{ZhN} is through
the measure $P_{h,-1}(d\f)$, which has support on smooth functions
for any $h$, the wanted bound for $Z_{h,N}(\z)$ is an easy
consequence of a uniform $C^n$ bound of $V^{(N)}_n(\f)$. Since
\be \lb{btau}
|V^{(N)}_n(\f)| \le \sum_{\t\in \TT^{(N)}_n} b_\t \virg b_\t \defi
 {1\over 2^n}\sum_{\s_1, \ldots, \s_n} \int_{\L^n} d\xx_1 \cdots d\xx_n\; |V_\t
(\ss,\ux)|
\ee
and the number of trees is of order $C^n$, we shall look for a
``good'' bound of $b_\t$. The main ingredients in this task are
the positivity of $U_j(v)$, see \pref{Uj}, and the
Battle-Federbush formula for the truncated expectations (see
\cite{Br}):
\be \lb{Brid}
G_j(v_1, \ldots, v_s) = \sum_{T\in \bar\TT_s} \prod_{\la r,m \ra
\in T} \lft[-\a^2 W_j(X_{v_r}, X_{v_m})\rgt] \int\!dp_T(\ut)\;
 e^{-{\a^2\over 2} U_j(v,\ut)}
\ee
where $s=s_v$, $\bar\TT_s$ is the family of connected tree graphs
on the set of integers $\{1,\ldots,s\}$, ${1\over 2} U_j(v, \ut)$
is obtained by taking a sequence of convex linear combination,
with parameters $\ut$, of the energies of suitable subsets of $X_v
= \cup_i X_{v_i}$ (hence $U_j(v, \ut)\ge 0$) and $dp_T(\ut)$ is a
probability measure.

By using \pref{Vtau}, \pref{Brid} and \pref{wjxy}, we get
\bea \lb{btaub}
&&|V_\t (\ss,\ux)| \le \lft(\prod_{i=1}^n \g^{{\a^2\over 4\p}
h_{e_i}}\rgt) \cdot\nn\\
&&\cdot\prod_{n.t.\ v\in\t} {\a^{2(s_v-1)}\over s_v!} \sum_{T\in
\bar\TT_{s_v}} \prod_{\la r,m\ra\in T} \sum_{e\in X_{v_r}\atop
e'\in X_{v_m}} \lft|C_{h_v}(\xx_e-\xx_{e'})\rgt|
\eea
On the other hand, for any given $\e>0$, we can use the bound
\be
\sum_{T\in \bar\TT_s} \prod_{\la r,m\ra\in T} n_{v_r} n_{v_m} \le
s! \; \e^{-2(s-1)} \prod_{r=1}^s e^{2\e n_{v_i}}
\ee
Moreover, by \pref{decomp} and \pref{dec}, for $h_v\ge 0$,
\be \lb{intC}
\int_{\L}\! d\xx\; |C_{h_v}(\xx)| \le C  \g^{-2h_v}\;.
\ee
The trees, as defined after \pref{decomp}, satisfy the following
identity: $\sum_{w\ge v} (s_w-1) =n_v-1$; as a consequence, if
$v_0$ is the first non trivial vertex of $\t$,
$$\sum_{n.t.\;v\in \t} h_v (s_v-1)
= h_r (n_{v_0}-1) + \sum_{n.t.\;v\in\t} (h_v-h_{v'}) (n_v-1)
$$
where $v'$ is the n.t. vertex immediately preceding $v$ or the
root, if $v=v_0$. This allows us to write:
\bea \lb{btaub1}
&&b_\t \le C_\e^n \lft(\prod_{i=1}^n \g^{{\a^2\over 4\p}
h_{e_i}}\rgt) \prod_{n.t.\ v\in\t} \g^{-2 h_v (s_v-1)} e^{2\e
n_v} \le\\
&& \le C_\e^n \prod_{n.t.\ v\in\t} \g^{-(h_v-h_{v'}) [D(n_v) -2\e
n_v]} \nn
\eea
where the {\it dimension} $D(n)$ is given by
\be \lb{dim}
D(n) = 2(n-1) -{\a^2\over 4\p}n\;.
\ee
Let us consider first the case $\a^2< 4\p$. This condition implies
that $D(n)>0$ for any $n\ge 2$; hence, the bound \pref{btaub1}
implies in the usual way that $|V_\t (\ss,\ux)| \le C_\a^n$, for a
constant $C_\a$, which diverges as $\a^2\to (4\p)^-$; since
$|\L|=1$, this bound is valid also for $b_\t$. By a little further
effort, see below, one can then prove that the pressure $p(\z)$ is
an analytic function of $\z$, for $\z$ small enough.

If $4\p\le \a^2 <16\p/3$, $D(n)>0$ only for $n\ge 3$, so that the
previous bound is divergent for all trees containing at least one
vertex with $n_v=2$. In particular, $V^{(N)}_2(\f)$ diverges as
$N\to\io$; this divergence is related with the fact that the term
of order $\z^2$ and $\s_1=-\s_2$ in the perturbative expansion of
the pressure is really divergent, as one can easily check.  The
only way to cure this specific divergence is to renormalize the
model by subtracting a suitable constant of order $\z^2$ from the
potential, as we shall see below.

However, all other terms, even those associated with a tree
containing at least one vertex with $n_v=2$, are indeed bounded
uniformly in $N$; in order to prove this claim, we need to improve
the bound \pref{btaub1} by the two following lemmas.
\begin{lemma} \lb{lm3}
If $n_v=2$ and  $v_1, v_2$ are the two endpoints following $v$
with positions $\xx_1$, $\xx_2$ respectively and equal charges
$\s_1=\s_2$, then
\be
|G_{h_v}(v_1, v_2) e^{-{\a^2\over 2} U_{h_{v'}, h_v}(v)}| \le C
\g^{-{\a^2\over \p}(h_v -h_{v'})} e^{-\g^{h_v}|\xx_1-\xx_2|}
\ee
\end{lemma}

{\bf\0Proof.} Since $h_{v'}+1\ge 0$, it is easy to check that
\bean
&& G_{h_v}(v_1, v_2) e^{-{\a^2\over 2} U_{h_{v'}, h_v}(v)} =
\lft[e^{-\a^2 C_{h_v}(\xx_1-\xx_2)} -1\rgt] e^{-\a^2 C_0(0)} \;\cdot\\
&& \cdot\; e^{-\a^2 \sum_{k=h_{v'}+1}^{h_v-1}
\big[C_0(0)+C_k(\xx_1-\xx_2)\big]}
\defi F(\xx_1-\xx_2)
\eean
Hence, by using \pref{dec} with $\k>1$ and since $C_0(\xx)\le
C_0(0)$, we get
\bean
&& |F(\zz)| \le C e^{-\k\g^{h_v}|\zz|} e^{-2\a^2 C_0(0) (h_v
-h_{v'})}
e^{-\a^2 \sum_{k=h_{v'}+1}^{h_v-1} \big[C_0(\g^k\zz) -C_0(0)\big]} \le\\
&& \le C \g^{-{\a^2\over \p}(h_v -h_{v'})} e^{2\a^2 R C_0(0)}
e^{-(\k - \k_R)\g^{h_v}|\zz|}
\eean
where $\k_R\defi\a^2 B \sum_{r=R}^\io \g^{-r}$, the constant $B$
is such that $|C_0(\xx) -C_0(0)| \le B|\xx|$ (it exists by
\pref{dec} for $(q_0,q_1)=(1,0),(0,1)$) and $R$ is an arbitrary
positive integer, that we can choose so that $\k -\k_R\ge
1$.\Halmos

\begin{lemma} \lb{lm4}For $j\ge 0$, if the cluster $X$ is made of two endpoints with positions $\xx_1$,
$\xx_2$ and opposite charges, and $Y$ is another arbitrary
cluster, then
\be\lb{Wj1}
|W_j(X,Y)| \le C \g^j |\xx_1-\xx_2| \sum_{\yy\in Y} \int_0^1\!
dt\;
 e^{-\g^j|\xx_2 +t(\xx_1-\xx_2) -\yy|}
\ee
Moreover, if also the cluster $Y$ is made of two endpoints with
opposite charge and positions $\yy_1$, $\yy_2$, then
\be \lb{Wj2}
|W_j(X,Y)| \le C \g^{2j} |\xx_1-\xx_2|\;|\yy_1-\yy_2| \int_0^1\!dt
ds\; e^{-\g^j \big|\xx_2 +t(\xx_1-\xx_2) -\yy_2
-s(\yy_1-\yy_2)\big|}
\ee
\end{lemma}

{\bf\0Proof.} By using the identity
\be
C_0(\xx_1-\yy) - C_0(\xx_2-\yy) = \sum_{a=0,1}(\xx_1 - \xx_2)_a
\int_0^1\! dt\; \big(\dpr_a C_0\big)[\xx_2 + t(\xx_1-\xx_2)-\yy]
\ee
together with \pref{dec}  (with $\k\ge 1$), we get the bound
\be
|C_j(\xx_1-\yy) - C_j(\xx_2-\yy)| \le C \g^j |\xx_1-\xx_2|
\int_0^1\! dt\; e^{-\g^j |\xx_2 +t(\xx_1-\xx_2) -\yy|}
\ee
which immediately implies \pref{Wj1}. The bound \pref{Wj2} is
proved in a similar way, by using the identity
\bea\lb{2.35aa}
&& C_0(\xx_1-\yy_1) - C_0(\xx_2-\yy_1) - C_0(\xx_1-\yy_2) +
C_0(\xx_2-\yy_2)=
\sum_{a,b=0,1} (\xx_1 - \xx_2)_a \cdot\nn\\
&& \cdot(\yy_1-\yy_2)_b \int_0^1\! dtds\; \big(\dpr_a \dpr_b
C_0\big)[\xx_2 +t(\xx_1-\xx_2) -\yy_2 -s(\yy_1-\yy_2)]
\eea
\Halmos

Let us now consider a tree with $n\ge 3$ endpoints. By using Lemma
\ref{lm3}, we can improve the bound \pref{btaub1} by replacing
$D(n_v)$ with $D(n_v)+\a^2/\p$ in all vertices with $n_v=2$ and
$\s_1=\s_2$. Since $D(2)+\a^2/\p = 2+\a^2/(2\p)>0$, this is
sufficient to make the corresponding sum over $h_v-h_{v'}$
convergent. It follows that the sum over all trees with $n\ge 3$
and no vertex with $n_v=2$ and $Q=0$ is finite, uniformly in $h$,
if $\a^2<16\p/3$.

A similar argument can be used to control the vertices with
$n_v=2$ and $Q=0$. In fact, if $v$ is a vertex with $n_v=2$, then
$v'$ is certainly a n.t. vertex, otherwise $n$ would be equal to
$2$ and we are supposing $n\ge 3$. Hence, we can use Lemma
\ref{lm4} in \pref{Brid} for the vertex $v'$, which allows us to
improve the bound of \pref{Brid} for the vertex $v$: since
$\g^{h_{v'}} |\xx_1-\xx_2|\; |G_{h_v}(v_1,v_2)| \le C
\g^{-(h_v-h_{v'})} e^{-\g^{h_v} |\xx_1-\xx_2|/2}$, if $v_1$ and
$v_2$ are the two endpoints following $v$, we can modify the bound
\pref{btaub1} by adding $1$ to the dimension $D(n_v)$ of $v$; this
is sufficient, since $D(2)+1=3-\a^2/(2\p)$ is positive for
$\a^2<6\p$. It follows that $|V_\t (\ss,\ux)| \le C_\a^n$ holds
for all $n\ge 3$, with $C(\a)\to \io$ as $\a^2\to (16\p/3)^-$.
\insertplot{200}{130}
{{\ins{30pt}{83pt}{$v'$}
\ins{113pt}{100pt}{$v$}
\ins{150pt}{109pt}{$v_1$}
\ins{150pt}{78pt}{$v_2$}
\ins{37pt}{35pt}{$h_{v'}$}
\ins{76pt}{30pt}{$\ldots$}
\ins{117pt}{35pt}{$h_v$}
\ins{150pt}{35pt}{$h_v+1$}}%

}%
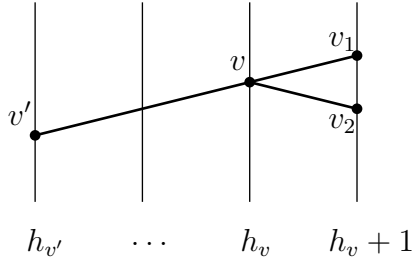
{n1}{\lb{n1} A subtree of $\t$ with $n_v=2$ and $Q=0$. While $v_1$
and $v_2$ are endpoints, therefore their scale has to be $h_v+1$,
$v'$ is the higher non trivial vertex of $\t$ preceding $v$, hence
the only constraint is that $h_{v}-h_{v'}\le N$.}{0}
By a further effort, one could prove that $C(\a)$ can be
substituted with a new constant, which is indeed finite up to
$6\p$, but we do not need here this stronger property.

Let us now come back to the terms of order two. It is easy to see
that
\be \lb{V2}
V_2^{(N)}(\f) = {\z^2\over 2} \sum_{\s=\pm 1} V_{2,\s}^{(N)}(\f)
\ee
\be
V_{2,\s}^{(N)}(\f) = \int_{\L^2}\! d\xx d\yy \; \cos[\a\f_\xx+\s
\a\f_\yy] W_\s^{(N)}(\xx-\yy)
\ee
\be
\hskip-0.58truecm W_\s^{(N)}(\xx-\yy) = \frac12 \sum_{j=0}^N
\g^{\frac{\a^2}{2\p} (j-1)} \left[  e^{-\s\a^2 C_j(\xx-\yy)} -1
\right]  e^{ -\a^2 \sum_{r=0}^{j-1} [C_r(0) +\s C_r(\xx-\yy)]}
\ee
By proceeding as in Lemma \ref{lm3}, it is easy to see that
$V_{2,+}^{(N)}(\f)$ is bounded uniformly in $N$ for any $\a$. This
is not true for $V_{2,-}^{(N)}(\f)$; in fact, if we define
\be
c_N= \EE_{h,-1} (V_{2,-}^{(N)}(\f))
\ee
one can easily check that $c_N$ diverges for $N\to\io$ and that
$\z^2 c_N/2$ is equal to the term of order $\z^2$ and $\s_1=-\s_2$
in the perturbative expansion of the pressure. However, if we
define $\tilde Z_{h,N}(\z) = Z_{h,N}(\z) e^{-{\z^2\over 2} c_N}$,
we can show that the {\it renormalized pressure} (in presence of
the cutoffs) $\tilde p_{h,N}(\z) = \log \tilde Z_{h,N}(\z)$ has a
power expansion uniformly convergent as $-h,N\to\io$, for
$\a^2<16\p/3$ (the result is indeed true for $\a^2< 6\p$).

It is easy to see that
\be \lb{ptilde}
\tilde p_{h,N}(\z) = \z\g^{\a^2 h\over 4\p} + \sum_{n=2}^\io \z^n
\sum_{\t\in \tilde\TT^{(N)}_n} p^{(h)}_\t
\ee
where $\tilde\TT^{(N)}_n$ is a family of trees defined as
$\TT^{(N)}_n$, with the following differences:

\bd

\item{1)} the root has scale $-2$;

\item{2)} there is no tree which has only two endpoint with opposite charge.

\ed

Moreover
\be
p^{(h)}_\t = {1\over 2^n}\sum_{\s_1, \ldots, \s_n} \int_{\L^n}\!
d\xx_1 \cdots d\xx_n\; \tilde V_\t^{(h)}(\ss,\ux)
\ee
\be \lb{Vtauh}
\tilde V_\t^{(h,N)} (\ss,\ux) = \lft(\prod_{i=1}^n \g^{{\a^2\over
4\p} (h_i+1)}\rgt) \prod_{n.t.\ v\in\t} {\tilde
G_{h_v}(v_1,\ldots, v_{s_v})\over s_v!} e^{-{\a^2\over 2}
U_{h_{v'}, h_v}(v)}
\ee
where $\tilde G_{h_v}(v_1, \ldots, v_{s_v})= G_{h_v}(v_1, \ldots,
v_{s_v})$, if $h_v\ge 0$, while, if $h_v=-1$ and $s_v=s$,
\be \lb{G-1}
\tilde G_{-1}(v_1, \ldots, v_{s}) = \EE^T_{h,-1} \left[ F(\f,
X_{v_1}); \ldots; F(\f, X_{v_s}) \right]
\ee
with, given a cluster $X$,
\be
F(\f,X) = \left\{ \ba{lll} \cos[\a \F(X)] -1 &,& \hbox{if\ } |X|=2
\hbox{\ and\ } \s_1=-\s_2\\ \cos[\a \F(X)] &,& \hbox{otherwise}
\ea \right.
\ee
where we subtracted a $-1$ in the terms with $|X|=2$ and
$\s_1=-\s_2$ (without changing the value of the truncated
expectation, since $s\ge 2$), in order to improve the bound in the
corresponding vertex, with an argument similar to that used
before. In fact, in order to bound \pref{G-1}, we shall use the
definition \pref{trexp} and the bound
\bea\lb{expm}
&&\left| \EE_{h,-1}\lft[\prod_{i=1}^m F(\f,X_i)\rgt] \right| \le
\left[
\prod_{i:|X_i|=2} |\xx^{(i)}_1 - \xx^{(i)}_2|^2 \right] \cdot\nn\\
&&\cdot \left( {\a^2\over 2} \right)^{m_2}
\sup_{\yy_1,\ldots,\yy_{m_2} \in\L} \EE_{h,-1}
\lft[|\dpr\f_{\yy_1}|^2 \cdots |\dpr\f_{\yy_{m_2}}|^2\rgt]
\eea
where $m_2\le m$ is the number of clusters with $2$ endpoints and,
for each cluster of this type, $\xx^{(i)}_1$ and $\xx^{(i)}_2$ are
the two endpoint positions; $|\dpr\f_{\yy}|^2\defi
(\dpr_0\f_{\yy})^2+ (\dpr_1\f_{\yy})^2$. On the other hand, it is
easy to see that there is a constant $c_0$, independent of $h$,
such that $\lft|\EE_{h,-1} \lft[\dpr_{a_1}\f_{\xx_1}
\dpr_{a_2}\f_{\xx_2}\rgt]\rgt| \le c_0$. It follows, by using the
Wick Theorem, that $\EE_{h,-1} \lft[|\dpr\f_{\yy_1}|^2 \cdots
|\dpr\f_{\yy_{q}}|^2\rgt]\le 2^qc^q_0 (2q-1)!!\le  C^q q!$, so
that, if we choose $C\ge 1$ (which allows us to substitute $m_2$
with $m$) and use \pref{trexp}, we obtain the following bound
\bea
&&|\tilde G_{-1}(v_1, \ldots, v_{s})|  \le \left[
\prod_{i:|X_{v_i}|=2} |\xx^{(i)}_1 - \xx^{(i)}_2|^2 \right]\cdot\\
&&\cdot \sum_{k=1}^s (k-1)! \frac1{k!} \sum_{m_1, \ldots, m_k \ge
1\atop \sum_{r=1}^k m_r=s} {s!\over m_1! \cdots m_k!}
\prod_{r=1}^k (C^{m_r} m_r!) \nn
\eea
The sum in the second line is equal to $C^s s! \sum_{k=1}^s
\frac1{k} \left( \ba{c} s-1\\k-1 \ea  \right) \le 2^{s-1} C^s s!
$, so that
\be
|\tilde G_{-1}(v_1, \ldots, v_{s})|  \le C^s s! \left[
\prod_{i:|X_{v_i}|=2} |\xx^{(i)}_1 - \xx^{(i)}_2|^2 \right]
\ee
The factors $|\xx^{(i)}_1 - \xx^{(i)}_2|^2$ can be used control
the sum over the scale labels of the vertices with $|X_{v_i}|=2$,
by the same argument used in the discussion following
\pref{2.35aa}. Hence, if we define the function $W_{n,h,N}(\ux)$
so that $\sum_{\t\in \tilde\TT^{(N)}_n} p^{(h)}_\t = \int_{\L^n}\!
d\xx_1 \cdots d\xx_n\; W_{n,h,N}(\ux)$, the previous arguments
imply that there exist positive functions $f_\t(\ux)$, independent
of $h$ and $N$, and a constant C, such that
\be \lb{Wbound}
|W_{n,h,N}(\ux)| \le \sum_{\t\in \tilde\TT^{(N)}_n} f_\t(\ux)
\defi H_{n,N}(\ux) \virg \int_{\L^n}\! d\ux\; H_{n,N}(\ux) \le C^n
\ee
Since $\tilde\TT^{(N)}_n \subset \tilde\TT^{(N+1)}_n$,
$H_{n,N}(\ux)$ is monotone in $N$. Hence, by the Monotone
Convergence Theorem, $H_{n,N}(\ux)$ has a $L^1$ limit $H_n(\ux)$,
as $N\to\io$; by \pref{Wbound}, $|W_{n,h,N}(\ux)| \le H_n(\ux)$.
On the other hand, by definition we have
\be
W_{n,h,N}(\ux) = {1\over n!} {1\over 2^n}\sum_{\s_1,\ldots,\s_n}
\EE^T_{h,N} \lft[ \lp e^{i\a\s_1\f_{\xx_1}} \rp; \ldots;\lp
e^{i\a\s_n\f_{\xx_n}} \rp\rgt]
\ee
and Lemma \ref{lm2}, \pref{asinf} and \pref{trexp} imply that
$W_{n,h,N}(\ux)$ is almost everywhere convergent as $-h,N\to\io$.
Then, by the Dominated Convergence Theorem, \pref{ptilde} and
\pref{Wbound}, $\tilde p(\z) = \lim_{-h,N\to\io} \tilde
p_{h,N}(\z)$ does exist and is an analytic function of $\z$, for
$\z$ small enough; moreover, $\tilde p(\z) =\sum_{n=2}^\io p_n
\z^n$ and, if $n\ge 3$, by \pref{expinf}
\bea
&&p_n = { c^{-{\a^2\over 8\p}n} \over n!} {1\over
2^n}\sum_{\s_1,\ldots,\s_n}^{ Q=0} \int_{\L^n}
d\xx_1 \cdots d\xx_n \cdot\\
&&\left\{ \sum_{\P} (-1)^{|\P|-1} (|\P|-1)! \prod_{Y\in \P}
\prod_{r,s\in Y\atop r<s} |\xx_r-\xx_s|^{\s_r\s_s {\a^2\over 2\p}}
\right\}\nn
\eea
where $\sum_\P$ denotes the sum over the partitions of the set
$\{1,\ldots,n\}$.  If $\a^2<4\p$, the previous expression is well
defined also for $n=2$, and gives the coefficient of order $2$ of
$p(\z)$. We stress that the integral and the sum over the
partitions can not be exchanged.

\subsection{The case $r=0$ (the charge correlation functions)}\lb{3.3}

Let $\x_i=(\zz_i, \s_i)$, $i=1,\ldots,k$, a family of fixed
positions and charges, such that $\zz_i \not= \zz_j$ for $i\not=j$
and $\m_i$, $i=1,\ldots,k$, a set of real numbers. If
\be
Z_{h,N}(\z,\uxi,\umu) =\int P_{[h,N]}(d\f) e^{\z_N V(\f)
+\sum_{r=1}^k \m_r :e^{i\a \s_r \f(\zz_r)}:}
\ee
the {\it charge correlation function of order $k$}, $k\ge 1$,
defined by \pref{GhN}, is given by
\be\lb{GhN0}
K^{(k,\z)}_{h,N}(\uz,\us) = \left. {\dpr^k\over \dpr\m_1 \cdots
\dpr\m_k} \log Z_{h,N}(\z,\uxi,\umu) \right|_{\umu=0}
\ee

By proceeding as in Sect. \ref{sec2}, one can show that
\be \lb{ZhNxi}
Z_{h,N}(\z,\uxi,\umu) = \int P_{[h,-1]}(d\f)
e^{V^{(N)}_{eff}(\z,\f) + B^{(N)}(\z,\f,\uxi,\umu) +
R^{(N)}(\z,\f,\uxi,\umu)}
\ee
where $ V^{(N)}_{eff}(\z,\f) = \z V(\f) + \sum_{n=2}^\io \z^n
V^{(N)}_n(\f)$, $B^{(N)}(\z,\f,\uxi,\umu)$ is the sum over the
terms of order at most $1$ in each of the $\m_r$, and
$R^{(N)}(\z,\f,\uxi,\umu)$ is the rest. \pref{GhN0} implies that
\be\lb{GhN1}
K^{(k,\z)}_{h,N}(\uz,\us) = \left. {\dpr^k\over \dpr\m_1 \cdots
\dpr\m_k} \log \tilde Z_{h,N}(\z,\uxi,\umu) \right|_{\umu=0}
\ee
where
\be \lb{ZhNxi1}
\tilde Z_{h,N}(\z,\uxi,\umu) = \int\! P_{h,-1}(d\f)\;
e^{V^{(N)}_{eff}(\z,\f) -\z^2 c_N/2 + B^{(N)}(\z,\f,\uxi,\umu)}
\ee

In order to describe the functional $B^{(N)}(\z,\f,\uxi,\umu)$, we
need to introduce a new definition. We shall call
$\TT^{(N)}_{n,k}$ the family of labelled trees whose properties
are very similar to those of $\TT^{(N)}_n$, with the only
difference that there are $n+m$ endpoints, $n\ge 0$, $1\le m\le
k$; $n$ endpoints, to be called {\it normal}, are associated as
before to the interaction, while the others, to be called {\it
special}, are associated with $m$ different variables $\xi_i$,
whose set of indices we shall denote $I_\t$, while $\xi_\t$ will
denote the set of variables itself. It is easy to see that
\be
B^{(N)}(\z,\f,\uxi,\umu) = \sum_{n=0}^\io \z^n
B^{(N)}_n(\f,\uxi,\umu)
\ee
\bea\lb{BN}
&&B^{(N)}_n(\f,\uxi,\umu) = \d_{n,0}\sum_{r=1}^k \m_r \lp
e^{i\a\s\f_{\zz_r}}\rp + \sum_{\t\in \TT^{(N)}_{n,k}\atop n+m\ge
2} {1\over 2^n}\sum_{\s'_1, \ldots, \s'_n} \int_{\L^n} d\xx_1
\cdots d\xx_n\; \cdot\nn\\
&&\qquad \cdot \prod_{s\in I_\t} \m_s e^{i\a \left[\sum_{r=1}^n
\s'_r \f_{\xx_r} + \sum_{s\in I_\t} \s_s \f_{\zz_s} \right]}
V_\t(\ss',\ux,\xi_\t)
\eea
where $V_\t(\ss',\ux,\xi_\t)$ is defined exactly as in
\pref{Vtau}, with $(\ss', \us_\t, \ux, \uz_\t)$ in place of
$(\ss,\ux)$.

One can easily check that, if $\a^2\ge 4\p$, the terms with
$n=k=1$ and $\s_1=-\bar\s_1$ in the r.h.s. of \pref{BN} have a
divergent bound as $N\to\io$. This is related to the fact that the
function $K^{(1,\z)}_{h,N}(\zz;\s)$ is indeed divergent at the
first order in $\z$. However, if we regularize these terms by
subtracting their value at $\f=0$, the counterterms give no
contribution to $K^{(k,\z)}_{h,N}(\uz;\us) $, for $k\ge 2$. Hence,
we can proceed as in the bound of the pressure and we get similar
results. There are however a few differences to discuss.

Given a tree $\t$ (with root of scale $-2$) contributing to
$K^{(k,\z)}_{h,N}(\uz;\us)$, we call $\t^*$ the tree which is
obtained from $\t$ by erasing all the vertices which are not
needed to connect the $m\le k$ special endpoints. The endpoints of
$\t^*$ are the $m$ special endpoints of $\t$, which we denote
$e^*_i$, $i=1,\ldots,m$. Given a vertex $v\in\t^*$, we shall call
$\uz_v$ the subset of the positions associated with the endpoints
following $v$ in $\t^*$; moreover, we shall call $s^*_v$ the
number of branches following $v$ in $\t^*$. The positions in
$\uz_v$ are connected in our bound by a {\it spanning tree} of
propagators of scales $j\ge h_v$; hence, if we use the bound
\be
e^{-2\g^h|\xx|} \le e^{-\g^h|\xx|} \cdot e^{-c\sum_{j=0}^h
\g^j|\xx|} \virg c=\sum_{j=0}^\io \g^{-j/2}
\ee
and define $\d = \min_{1\le i<j \le k} |\zz_i- \zz_j|$, it is easy
to see that we can extract, for any $v\in\t^*$, a factor
$e^{-c\g^{h_v}\d}$ from the propagators bound, by leaving a
decaying factor $e^{-\g^j|\xx|}$ for each propagator (of scale
$j$) of the spanning tree. On the other hand, the fact that the
points in $\uz_v$ are not integrated implies that there are
$s^*_v-1$ less integrations to do by using propagators of scale
$h_v$, for each vertex $v\in\t^*$. In conclusion, with respect to
the pressure bound, we have to add, for each tree $\t$, a factor
\be\lb{139}
\prod_{v\in \t^*} \g^{2h_v(s^*_v-1)} e^{-c\g^{h_v}\d} \le
(c\d)^{-2(m-1)} (2m-2)!
\ee
where we used the identity $\sum_{v\in\t*} (s^*_v-1) = m-1$. Since
$m\le k$, the sum over the scale labels can be done exactly as in
the pressure case, up to a $C^k (2k)!$ overall factor.

There is another difference to analyze, related with the fact
that, in the analogue of \pref{G-1}, the function $F(\f,X_v)$
corresponding to a cluster with two endpoints of opposite charge,
one normal and one special, is bounded by $|\xx-\zz|
\sup_\yy|\dpr\f_\yy|$, rather than bounded by $|\xx-\zz|^2
\sup_\yy|\dpr\f_\yy|^2$. The fact that the zero in the positions
is of order one has no consequence, since such a zero is
sufficient to regularize the bound over a cluster with two
endpoints of opposite charges. The fact that $|\dpr\f_\yy|$
appears, instead of its square, is also irrelevant, since the only
consequence is that, in the bound analogue to \pref{expm}, one has
to substitute $\EE_{h,-1} \lft[|\dpr\f_{\yy_1}|^2 \cdots
|\dpr\f_{\yy_{m_2}}|^2\rgt] $ with $\EE_{h,-1}
\lft[|\dpr\f_{\yy_1}| \cdots |\dpr\f_{\yy_{m'_2}}|\rgt] $, with
$m'_2\le 2m_2$. However,  by Schwartz inequality,
$\EE_{h,-1}\big[|\dpr\f_{\yy_1}| \cdots |\dpr\f_{\yy_{m'_2}}|\big]
\le \sqrt{\EE_{h,-1}\big[|\dpr\f_{\yy_1}|^2 \cdots
|\dpr\f_{\yy_{m'_2}}|^2\big]}$ and we can still use the Wick
Theorem to get an even better bound.

The previous arguments allow us to prove that $K^{(k,\z)}(\uz;\us)
= \lim_{-h,N\to+\io}$ $K^{(k,z)}_{h,N}(\uz,\us)$ does exist and is
an analytic function of $\z$ around $\z=0$, with a radius of
convergence independent of $\d$ (the minimum distance between two
points in  $\uz$). On the other hand, it is easy to check the well
known identity
\bea
K^{(k,\z)}_{h,N}(\uz;\us) = \sum_{n=0}^\io {\z^n\over n!} {1\over
2^n}\sum_{\us'} \int\! d\xx_1\cdots d\xx_n
\nn\\
\EE^T_{h,N} \lft[ \lp e^{i\a\s_1\f_{\zz_1}}\rp;\ldots; \lp
e^{i\a\s_k\f_{\zz_k}}\rp; \lp e^{i\a\s'_1\f_{\xx_1}}\rp;\ldots; \lp
e^{i\a\s'_n\f_{\xx_n}}\rp\rgt]
\eea
An argument similar to that used at the end of \S\ref{sec2} allows
us to prove that the power expansion of $K^{(k,\z)}(\uz;\us)$ is
obtained by the previous equation, by substituting in the r.h.s
$\EE^T_{h,N}$ with $\EE^T$. Hence, by using \pref{expinf}, we get
that $K^{(k,\z)}(\uz;\us)= \sum_{n=0}^\io \z^n g_{k,n}(\uz;\us)$,
with
\bea\lb{pn1}
&& g_{k,n}(\uz;\us) = { c^{-{\a^2\over 8\p}(n+k)} \over n!}{1\over
2^n} \sum_{\s'_1,\ldots,\s'_n\atop \sum_{i=1}^n \s'_i
+\sum_{r=1}^k \s_r=0} \int_{\L^n}\;
d\xx_1 \cdots d\xx_n \cdot\\
&&\left\{ \sum_{\P} (-1)^{|\P|-1} (|\P|-1)! \prod_{Y\in \P}
\prod_{r,s\in Y\atop r<s} |\yy_r-\yy_s|^{\bar\s_r\bar\s_s
{\a^2\over 2\p}} \right\}\nn
\eea
where $\uy=(\ux,\uz)$, $\bar\us=(\us',\us)$ and $\sum_\P$ denotes
the sum over the partitions of the set $(1,\ldots,n+k)$.

\subsection{The case $r>0$ (the $\dpr\f$ correlation functions)}

Let $\uy=(\yy_1, \ldots, \yy_k)$ a set of $k\ge 1$ distinct fixed
positions, $\un=(\n_1, \ldots, \n_k)$ a set of derivative indices
and $\um=(\m_1,\ldots, \m_k)$ a set of real numbers. If
\be
Z_{h,N}(\z,\uy,\un,\umu) =\int P_{[h,N]}(d\f) e^{\z_N V(\f)
+\sum_{r=1}^k \m_r \dpr^{\n_r}\f(\yy_r)}
\ee
the {\it $\dpr\f$ correlation function of order $k$}, $k\ge 1$, is
given by
\be\lb{GhNf}
K^{(k,\z)}_{h,N}(\uy;\un)  = \left. {\dpr^k\over \dpr\m_1 \cdots
\dpr\m_k} \log Z_{h,N}(\z,\uy,\un,\umu) \right|_{\umu=0}
\ee

We can proceed as in \S\ref{3.3} and we can represent
$K^{(k,\z)}_{h,N}(\uy;\un)$ as in \pref{GhN1}, that is we can
substitute in \pref{GhNf} $Z_{h,N}(\z,\uy,\un,\umu)$ with
\be \lb{ZhNxi2}
\tilde Z_{h,N}(\z,\uy,\un,\umu) = \int P_{[h,-1]}(d\f)
e^{V^{(N)}_{eff}(\z,\f) -2\z^2 c_N + B^{(N)}(\z,\f,\uy,\un,\umu)}
\ee
It is not hard to see that $B^{(N)}(\z,\f,\uy,\un,\umu) =
\sum_{n=0}^\io \z^n$ $B^{(N)}_n(\f,\uy,\un,\umu)$, with
\bea\lb{BN1}
&&B^{(N)}_n(\f,\uy,\un,\umu) = \d_{n,0} \sum_{r=1}^k \m_r
\dpr^{\n_r} \f(\yy_r) + \sum_{\t\in \TT^{(N)}_{n,k}\atop n+m\ge 2}
{1\over 2^n} \sum_{\s_1, \ldots, \s_n} \int_{\L^n} d\xx_1 \cdots
d\xx_n\;\cdot\nn\\
&&\qquad \cdot e^{i\a \sum_{r=1}^n \s_r \f(\xx_r)} \tilde
V_\t(\ss,\un,\ux,\uy_\t) \; \prod_{r\in I_\t}\m_r
\eea
where $\TT^{(N)}_{n,k}$ is defined exactly as in \pref{BN}, except
for the fact that the $m$ special endpoints ($1\le m \le k$) are
associated with the $\dpr\f$ terms; moreover $\tilde
V_\t(\ss,\ux,\uy_\t)$ is defined in a way similar to
$V_\t(\ss,\ux,\uxi_\t)$, but, before giving its expression, we
need a few new definitions.

If $v$ is a non trivial vertex, we shall call $I_v\subset I_\t$
the set of special endpoints immediately following $v$ (that is
the set of $\dpr\f$ endpoints which are contracted in $v$), $\bar
s_v$ the number of vertices immediately following $v$, which are
not special endpoints, and $s^*_v=|I_v|$ (hence $s_v=\bar s_v+
s^*_v$). Moreover, we shall use $X_v$ to denote the set of normal
endpoints (instead of all endpoints) following $v$. Then we can
write
\be \lb{Vtau1}
\tilde V_\t(\ss,\un,\ux,\uy_\t) = \lft(\prod_{i=1}^n
\g^{{\a^2\over 4\p} h_{e_i}}\rgt) \prod_{n.t.\ v\in\t} {\tilde
G_{h_v}(v_1, \ldots, v_{s_v})\over s_v!}  e^{-{\a^2\over 2}
U_{h_{v'}, h_v}(v)}
\ee
with $\tilde G_j(v_1, \ldots, v_{s}) = \EE^T_j \lft[F_{v_1}(\f);
\ldots; F_{v_s}(\f)\rgt]$, where $F_v(\f)=\dpr^\n\f_\yy$, if the
vertex $v$ is a special endpoint with position $\yy$ and label
$\n$,
 otherwise $F_v(\f)= \exp(i\a\F(X_v))$. We can always rearrange the
order of the arguments so that the first possibility happens for
$i=1,\ldots,m$. If $m=0$, we can use the identity \pref{Brid},
otherwise we can write
\be \lb{Gj2}
\tilde G_j(v_1, \ldots, v_s) = {\dpr^m\over \dpr\l_1 \cdots
\dpr_{\l_m}} H_j(\l_1,\ldots,\l_m)\Bigg|_{\underline\l=0}
\ee
where
\be
H_j(\underline\l) = \EE^T_j (e^{\l_1\dpr^{\n_1}\f_{\yy_1}};
\ldots; e^{\l_m\dpr^{\n_m}\f_{\yy_m}}; e^{i\a\F(X_{v_{m+1}})};
\ldots; e^{i\a\F(X_{v_s})})
\ee
is a quantity which satisfies an identity similar to \pref{Brid},
that is
\be \lb{Brid1}
H_j(\underline\l) = \sum_{T\in \bar\TT_s} \prod_{\la a,b\ra\in T}
c_{a,b} \int\! dp_T(\ut)\; e^{-\frac12 \tilde U_j(v, \ut,
\underline\l)}
\ee
where %
\be c_{a,b}\defi
\lft\{\matrix{ \tilde c_{a,b}\defi -\a^2 W_j(X_{v_a}, X_{v_b})
\hfill&\hfill {\rm if\ }a,b>m\cr \l_a \tilde c_{a,b}\defi
i\a\l_a\sum_{r\;:\;e_r\in X_{v_b}} \s_r
\lft(\dpr^{\n_a}C_j\rgt)(\yy_a-\xx_r) \hfill&\hfill {\rm if\ }a\le
m <b\cr \l_a\l_b \tilde c_{a,b} \defi - \l_a \l_b
\lft(\dpr^{\n_a}\dpr^{\n_b}C_j\rgt)(\yy_a-\yy_b) \hfill&\hfill{\rm
if\ } a,b\le m\;.}\rgt.\ee
It follows that
\be\lb{Gj3}
\tilde G_j(v_1, \ldots, v_s) = \sum_{T\in \bar\TT'_s} \prod_{\la
a,b\ra \in T} \tilde c_{a,b} \int dp_T(\ut) e^{-\frac12 \tilde
U_j(v, \ut, 0)}
\ee
where $\bar\TT'_s$ is the set of $T\in \bar\TT_s$, such that all
special endpoints are leaves of $T$. Note that $\tilde U_j(v, \ut,
0)$ is a positive quantity, since it is a convex combination of
``interaction energies'' which do not involve the special
vertices; hence we can safely bound the r.h.s. of \pref{Gj3}, as
in the previous sections. Let us define
\be
\tilde b_\t(\uy) = {1\over 2^n}\sum_{\un,\us} \int_{\L^n}\! d\ux\;
|\tilde V_\t(\ss,\un,\ux,\uy_\t)|
\ee
The bound of $\tilde b_\t(\uy)$ differs from the r.h.s. of
\pref{btaub1} for the following reasons:

\bd

\item{1)} there is a $\g^{k_i}$ factor more, coming from the
field derivative, for the $i$-th special endpoint, if $k_i$ is the
scale label of the higher n.t. vertex preceding it (the vertex
where it is contracted);

\item{2)} there is a factor $\g^{h_v(s^*_v-1)}$ more, which takes
into account the fact that the special endpoints positions are not
integrate, for each n.t. vertex $v$ such that $s^*_v>0$;

\item{3)} if $\d = \min_{1\le i<j \le k} |\yy_i-\yy_j|$, there is a
factor $\exp (-c\g_{h_v}\d)$ for each n.t. vertex $v$ such that
$s^*_v>0$, coming from the same argument used in the case of the
charge correlation functions.

\ed

Hence, if $m_\t\ge 1$ is the number of special endpoints in $\t$,
we get
\bea
\tilde b_\t(\uy) &\le& C_\e^{n+m_\t} \lft(\prod_{i=1}^n
\g^{{\a^2\over 4\p} h_{e_i}} \rgt) \lft(\prod_{n.t.\ v\in\t}
\g^{-2 h_v (s_v-1)} e^{2\e n_v}
\prod_{i=1}^{m_\t} \g^{k_i}\rgt) \cdot\nn\\
&\cdot& \prod_{n.t. v: s^*_v>0} \g^{2h_v(s^*_v-1)}
e^{-c\g^{h_v}\d}
\eea
The last product can be bounded as in \pref{139}, so that, by
``distributing along the tree'' the other factors, we get
\be
\tilde b_\t(\uy) \le C_\e^{n+k} (\d)^{-2(m_\t-1)} (2m_\t-2)!
\prod_{n.t.\ v\in\t} \g^{-(h_v-h_{v'}) (\tilde D(n_v,m_v)-2\e
n_v)}
\ee
where $n_v$ and $m_v$ denote the number of normal and special
endpoints following $v$, respectively, and
\be
\tilde D(n,m) = 2(n-1) -{\a^2\over 4\p}n + m
\ee

Let us consider first the case $\a^2<4\p$. Since $n_v+m_v\ge 2$,
$\tilde D(n_v,m_v)$ is always positive, except if $n_v=0$ and
$m_v=2$. However, no tree may have a non trivial vertex of this
type, except the trees with only two special endpoints and no
normal endpoint, that is the trees belonging to $\TT^{(N)}_{0,2}$,
and it is very easy to see that
\be
\sum_{\t\in \TT^{(N)}_{0,2}} \tilde V_\t(\n_1,\n_2, \yy_1,\yy_2) =
-\sum_{j=0}^N \g^{2j} \lft(\dpr^{\n_1}\dpr^{\n_2}
C_{0}\rgt)\lft(\g^j(\yy_1-\yy_2)\rgt)
\ee
By \pref{dec}, this quantity has a finite limit as $N\to\io$, if
$\yy_1\not= \yy_2$, as we are supposing. Hence there is no
ultraviolet divergence in the expansion \pref{BN1} of
$B^{(N)}_n(\f,\uy,\un,\umu)$ and we have only to check that there
is no infrared problem related with the integration over the $\f$
field in \pref{ZhNxi2}. This follows as in \S\ref{sec2}, by using
the identity \pref{trexp}; it is sufficient to observe that
\be
\left| \EE_{h,-1}\lft[\lft(\prod_{i=1}^m
\lft(\dpr^{\n_i}\f_{\yy_i}\rgt)\rgt) \prod_{j=1}^s
e^{i\a\F(X_{v_j})}\rgt] \right| \le \sqrt{\EE_{h,-1}
\lft[\prod_{i=1}^m \lft|\dpr\f_{\yy_i}\rgt|^2\rgt]}
\ee
and then apply the arguments used in \S\ref{sec2} to bound the sum
over the partitions.

Let us now suppose that $4\p \le \a^2 < 16\p/3$. In this case
$\tilde D(n_v,m_v)$ can be non positive only if either $m_v=0$ and
$n_v=2$ or $m_v=n_v=1$. The vertices satisfying the first
condition can be regularized as before, for the others we can use
the factor $e^{-{\a^2\over 2} U_{h_{v'}, h_v}(v)} =
\g^{-{\a^2\over 4\p}(h_v-h_{v'}-1)}$ to make their dimension
positive; in fact $\tilde D(1,1) +\a^2/(4\p) =1$. The integration
over the $\f$ field in \pref{ZhNxi2} can now be done by an obvious
modification of the argument used for the charge correlation
functions.

It is now easy to prove, as in the previous sections, that
$K^{(k,\z)}_{h,N}(\uy;\un)$ has a finite limit, as $-h,N\to\io$,
if $\d>0$, and that this limit is an analytic function of $\z$
around $\z=0$, with a radius of convergence independent of $\d$.
On the other hand,
\bea\lb{kkz}
K^{(k,\z)}_{h,N}(\uy;\un)= \sum_{n=0}^\io {\z^n\over n!} {1\over
2^n}\sum_{\us}
\int\! d\xx_1\cdots d\xx_n\;\nn\\
\EE^T_{h,N}
\lft[\dpr^{\n_1}\f_{\yy_1};\ldots;\dpr^{\n_k}\f_{\yy_k}; \lp
e^{i\a\s_1\f_{\xx_1}}\rp;\ldots; \lp e^{i\a\s_n\f_{\xx_n}}\rp \rgt]
\eea
An argument similar to that used at the end of \S\ref{sec2} allows
us to prove that the power expansion of $K^{(k,\z)}(\uy;\un)$ is
obtained by the previous equation, by substituting in the r.h.s
$\EE^T_{h,N}$ with $\EE^T$. Moreover, it is not hard to check
that, if $n>0$, $Q \=\sum_{i=1}^n \s_i$ and $
h_{k,n}(\ux,\uy;\us,\un)$ is the limiting value of the truncated
expectation in \pref{kkz}, we have
\bea\lb{2.87}
&& h_{k,n}(\ux,\uy;\us,\un)= \d_{Q,0} \; c^{-{\a^2\over 8\p}n}
\lft(\prod_{r=1}^k W(\yy_r,\ux;\n_r,\us)\rgt)
\cdot\\
&&\cdot  \left\{ \sum_{\P} (-1)^{|\P|-1} (|\P|-1)! \prod_{Y\in \P}
\prod_{r,s\in Y\atop r<s} |\xx_r-\xx_s|^{\s_r\s_s {\a^2\over 2\p}}
\right\} \nn
\eea
where $\sum_\P$ denotes the sum over the partitions of the set
$(1,\ldots,n)$ and
\be\lb{2.88}
W(\yy,\ux;\us,\n) = i\a\sum_{i=1}^n \s_i \lft(\dpr^\n
\D^{-1}\rgt)(\yy-\xx_i) = {i\a\over 2\p} \sum_{i=1}^n \s_i
{(\xx_i-\yy)^\n\over |\yy-\xx_i|^2}
\ee
while, if $n=0$,
\be\lb{2.94}
h_{k,0}(\uy,\un) = \d_{k,2} \EE^T
\Big[\lft(\dpr^{\n_1}\f_{\yy_1}\rgt);\lft(\dpr^{\n_2}\f_{\yy_2}\rgt)\Big]
= \d_{k,2}h^{\n_1,\n_2}(\yy_1-\yy_2)
\ee
with
\be\lb{hnn}
h^{\n_1,\n_2}(\yy)= {1\over 2\p|\yy|^2} \left[ \d^{\n_1,\n_2} - 2
{\yy^{\n_1} \yy^{\n_2}\over |\yy|^2} \right]
\ee
Hence, we get that $K^{(k,\z)}(\uy;\un) = \sum_{n=0}^\io \z^n \bar
h_{k,n}(\uy;\un)$, with
\be\lb{gkn}
\bar h_{k,n}(\uy;\un) = {1 \over n!}{1\over 2^n}
\sum_{\s_1,\ldots,\s_n\atop \sum_{i=1}^n \s_i=0} \int_{\L^n}\!
d\xx_1 \cdots d\xx_n\; h_{k,n}(\ux,\uy;\us,\un)
\ee
Note that $\bar h_{1,n}(\yy,\n)=0$ for any $n$, since
$W(y,\ux,\us,\n)$ is odd in $\us$ and the sum in \pref{gkn} is
restricted to the $\us$ such that $Q=0$; hence
$K^{(1,\z)}(\yy,\n)=0$.

\section{The Thirring model with a finite volume mass term}
\lb{sec3a}

The {\it Generating Functional}, $\WW_{h,N}(J,A,\m)$, of the
Thirring model with cutoff and  with a mass term in finite volume
is defined by the equation
\bea\lb{1.6}
&&\WW_{h,N}(J,A,\m) \defi \log \int\!\!P_{h,N}(d\ps)\; \exp
\Big\{-\l Z_N^2 V_L(\ps)+\m Z^{(1)}_N\int_\L \!d\xx\ \bar
\ps_\xx \ps_\xx +\nn\\
&&+ Z_N^{(1)}\sum_{\s=\pm 1} \int\!d\xx\; J^\s_{\xx}\lft(\bar
\ps_\xx\G^\s\ps_\xx\rgt) +Z_N\sum_{\n=0,1} \int\!d\xx\;
A^\n_{\xx}\lft(\bar \ps_\xx\g^\n\ps_\xx\rgt)\Big\}
\eea
where the free measure $P_{h,N}(d\ps)$ is defined by \pref{1.7},
$Z_N$ and $Z^{(1)}_N$ are defined in \pref{ZZ}, $J^\s_\zz$ and
$A^\m_\yy$ are two-dimensional, external bosonic fields and
\be\lb{1}
Z_N^2 V_L(\ps)\defi{1\over 4} \int_{\L_L} \!d\xx\ \lft(Z_N \bar
\ps_\xx \g^\m\ps_\xx\rgt)^2 + E_{h,N}|\L_L| \virg \G^\s \defi {I
+\s \g^5\over 2}
\ee
$E_{h,N}$ being the {\it vacuum counterterm} introduced in
\pref{VVV}; it is chosen so that $\WW_{h,N}(0,0,0)=0$.

Given the set of non coinciding points ${\ux}= (\xx_1, \ldots,
\xx_q)$ and the set $\us=(\s_1, \ldots, \s_q)$, $\s_i=\pm 1$, we
want to study the Schwinger functions
\be\lb{1.10}
G_{h,N}^{(q,r;\m)}(\ux,\uy;\us,\un)\defi \lim_{a^{-1},L\to\io}
{\partial^q\WW_{h,N} \over\partial J^{\s_1}_{\xx_1} \cdots
\partial J^{\s_q}_{\xx_q} \dpr A^{\n_1}_{\yy_1} \cdots \partial
A^{\n_r}_{\yy_r}}(0,0,\m)\;.
\ee

\begin{theorem} \lb{t2a}
If $\m$  and $\l$ are small enough and $q\ge 2$, if $r=0$, or
$q\ge 0$, if $r\ge 1$, the limit
\be
G^{(q,r;\z)}(\uz,\uw;\us,\un) \defi \lim_{-h,N\to+\io}
G^{(q,r;\z)}_{h,N}(\uz,\uw;\us,\un)
\ee
exists and is analytic in $\m$. In the case $q=r=0$ (the
pressure), the limit does exist and is analytic, up to a
divergence in the second order term, present only for $\l\le 0$.
\end{theorem}
As in \S\ref{sec3}, we shall give the proof of the above theorem
only in the special cases $(q,r)=(k,0)$ and $(q,r)=(0,k)$
separately.

In order to prove Theorem \ref{t2a}, we note first that definition
\pref{1.10} and the identity $\bar\psi_\xx \psi_\xx= \sum_\s \bar
\ps_\xx\G^\s\ps_\xx$ imply that
\be \lb{Gqm0}
G^{(q,r;\m)}_{h,N}(\uz, \uy;\us,\un) = \sum_{p=2n-q\atop n\ge 0}
\sum_{\us'} {\m^p\over p!} \int\!d\ux\; \overline\c_\L(\ux)
S^{(2n,r)}_{h,N}(\uz\hskip1pt\ux,\uy;\us\hskip1pt\us',\un)
\ee
where $\uz\hskip1pt\ux=(\zz_1,\ldots,\zz_q,\xx_1,\ldots, \xx_p)$,
$\us\hskip1pt\us'=(\s_1,\ldots,\s_q,\s'_1,\ldots, \s'_p)$, we
defined
\be
\overline\c_\L(\ux)\defi \c_\L(\xx_1)\cdots \c_\L(\xx_p)\;, \qquad
S^{(m,r)}_{h,N}(\ux,\uy;\us,\un)\defi
G^{(m,r;0)}_{h,N}(\ux,\uy;\us,\un)\;.
\ee
and we used the fact that $G^{(m,r;0)}_{h,N}(\ux,\uy;\us,\un)$ can
be different from $0$ only if $\sum_{i=1}^m \s_i=0$, implying in
particular that $m$ is even.

In the following we shall give a bound for the functions
$S^{(m,r)}_{h,N}(\ux,\uy;\us,\un)$, uniform in the cutoffs and
implying (by an argument similar to that used for the Sine-Gordon
model, that we shall skip here) that the limit exists, is
integrable and is exchangeable with the integral in \pref{Gqm0}.
It follows that
\be \lb{Gqm}
G^{(q,r;\m)}(\uz, \uy;\us,\un) = \sum_{p=2n-q\atop n\ge 0}
\sum_{\us'} {\m^p\over p!} \int\!d\ux\; \overline\c_\L(\ux)
S^{(2n,r)}(\uz\hskip1pt\ux,\uy;\us\hskip1pt\us',\un)
\ee
We remark that $\overline\c_\L(\ux)$ {\it is not a regular test
function} since it is not vanishing for coinciding points, and
hence we could encounter divergences caused by the ultraviolet
problem. Indeed, as we shall see, the integration of $G^{(2,0;0)}$
will be finite only for $\l>0$ (and small in absolute value), so
that the pressure $G^{(0,0;\m)}$ and the, if $\xx\in\L$,
``density'' $G^{(1,0;\m)}(\xx,\s)$ are really divergent for $\l\le
0$, since this is true for the terms with $2n=2$ and $r=0$ in the
r.h.s. of \pref{Gqm}.

As announced in the introduction, we  first consider the case
$q,r=0$ then we discuss the case $q>0$ and $r=0$; and finally the
case  $q=0$ and $r>0$.

\subsection{Case $q=r=0$ (the pressure)}\lb{3.1}

Our definitions imply that $S^{(0,0)}=0$. If $m\ge 2$ and even
(otherwise it is $0$ by symmetry), the $m$-points Schwinger
function $S^{(m,0)}(\ux,\us)$ is obtained as the $m$-th order
functional derivative of the generating function
$\WW_{h,N}(J,0,0)$ with respect to $J^{\s_1}_{\xx_1}, \ldots,
J^{\s_m}_{\xx_m}$ at $J=0$. We can proceed as in \cite{BFM} and we
get an expansion similar to eq. (2.28) of that paper, which we
refer to for the notation. The only difference is that the special
endpoints of type $J$ are associated with the terms $Z^{(1)}_j
\sum_\s \bar \ps_\xx\G^\s\ps_\xx = Z^{(1)}_j \sum_\s
\psi^+_{\xx,-\s} \psi^-_{\xx,\s}$ instead of $Z_j \sum_\s
\psi^+_{\xx,\s} \psi^-_{\xx,\s}$, but this does not change the
structure of the expansion; we only have to add, for each special
endpoint of scale $h_i$, a factor $Z^{(1)}_{h_i}/Z_{h_i}$, which
can be controlled by studying the flow of $Z^{(1)}_j$. It turns
out that there are two constants $\h_+(\l) = b\l + O(\l^2)$,
$b>0$, and $c_+(\l)=1+ O(\l)$, such that, in the limit $N\to\io$,
$Z^{(1)}_j= c_+(\l) \g^{-\h_+ j}$; this result is obtained by an
argument similar to that used in \cite{BFM} to prove that there
are two constants $\h_-(\l) = a\l^2 + O(\l^3)$, $a>0$, and
$c_-(\l)=1+ O(\l)$, such that, in the limit $N\to\io$, $Z_j=
c_-(\l) \g^{-\h_- j}$ (in \cite{BFM} $c_-(\l)=1$, since the
definition of $Z_N$ differs by a constant chosen so to get this
result). In analogy to eq. (2.40) of \cite{BFM}, we can write
\be\lb{2.26a}
S^{(m,0)}(\ux,\us)= m!\lim_{|h|,N\to \io}
 \sum_{n=0}^\io \sum_{j_0=-\io}^{N-1}
\sum_{\t\in\TT_{j_0,n}^{0,m}} \sum_{\bP\in \PP}
S_{0,m,\t,\us}(\xxx)\;, \ee
Given a tree $\t$ contributing to the r.h.s. of \pref{2.26a}, we
call $\t^*$ the tree which is obtained from $\t$ by erasing all
the vertices which are not needed to connect the $m$ special
endpoints (all of type $J$). The endpoints of $\t^*$ are the $m$
special endpoints of $\t$, which we denote $v^*_i$,
$i=1,\ldots,m$; with each of them a space-time point $\xx_i$ and a
label $\s_i$ are associated. Given a vertex $v\in\t^*$, we shall
call $\uuu_v$ the set of the space-time points associated with the
normal endpoints of $\t$ that follow $v$ in $\t$ (in \cite{BFM}
they were called {\it internal points}); $\ux_v$ will denote the
subset of $\xxx$ made of all points associated with the endpoints
of $\t^*$ following $v$.

Furthermore, we shall call $s^*_v$ the number of branches of
$\t^*$ following $v\in \t^*$, $s^{*,1}_v$ the number of branches
containing only one endpoint and $s^{*,2}_v = s^*_v- s^{*,1}_v$.
For each n.t. vertex or endpoint $v\in\t^*$, shortening the
notation of $s^*_v$ into $s$, we choose one point in $\ux_v$, let
it be called $\ww_v$, with the only constraint that, if
$v_1,\ldots, v_{s}$ are the n.t. vertices or endpoints following
$v$, then $\ww_v$ is one among $\uw_v \=
\{\ww_{v_1},\ldots,\ww_{v_s}\}$.

The bound of $S_{0,m,\t,\us}(\xxx)$ will be done as in \cite{BFM},
by comparing it with the bound of its integral over $\xxx$, given
by eq. (2.36) of that paper. However, we shall slightly modify the
procedure, to get an estimate more convenient for our actual
needs.

Given the space-time points $\uv=(\vv_1,\ldots,\vv_p)$ connected
by the tree $T$, we shall define, if $\vv_{l,i}$ and $\vv_{l,f}$
denote the endpoints of the line $l\in T$,
\be
D_T(\uv)\defi\|T\|\defi\sum_{l\in T} \sqrt{|\vv_{l,i}-\vv_{l,f}|}
\ee
Now we want to show that, from the bounds of the propagators
associated with the lines $l$ of the spanning tree $T_\t=\bigcup_v
T_v$, we can extract a factor $e^{-c' \sqrt{\g^{h_v}}
D_{C_v}(\uw_v)}$ for each n.t. vertex $v\in\t^*$, where $C_v$ is a
chain of segments that only depends on $\t$ and $T_\t$, and
connects the space-time points $\uw_v$.

Indeed, given a n.t. $v\in\t^*$, there is a subtree $T_v^*$ of
$T_\t$ connecting the points $\uw_v$ together with a subset of
$\ux_v\cup \uuu_v$. Since $T_v^*$ is made of lines of scale $j\ge
h_v$, the decaying factors in the bounds of the propagator in
$T_v^*$ can be written as
\be\lb{2.30}
e^{-c\sqrt{\g^h|\xx|}} = e^{-\frac{c}{2} \sqrt{\g^h|\xx|}} \cdot
e^{-2 c' \sum_{j=-\io}^h \sqrt{\g^j|\xx|}} \virg
\ee
for $c'=c/ \lft[4 \sum_{j=0}^\io \g^{-j/2}\rgt]$. Hence,
collecting  the latter factor for each of the lines $T_v^*$ we
obtain $e^{-2 c' \sqrt{\g^{h_v}}\|T^*_v\|}$.

We finally would like to replace, in the previous bound,
$\|T^*_v\|$ with $\|C_v\|$, up to a constant, {\it for a $C_v$
which does not depend on the position of the internal points of
$T^*_v$}. This is possible as a consequence of the following
lemma.

\begin{lemma}\lb{lm3.1}
Let $T$ be a tree graph connecting the points $\{\ww_j\}_{j=1}^l$
together with other ``internal points'', $\{\uu_j\}_{j=1}^q$. Then
there exists a chain $C$ connecting all and only the points
$\{\ww_j\}_{j=1}^l$ such that $2 \|T\|\ge \|C\|$ and $C$ only
depends on $T$.
\end{lemma}
{\bf\0Proof.} Suppose that the points $\{\uu_j\}_{j=1}^q$ are
fixed in an arbitrary way and let us consider the oriented closed
path $\bar\CC$ obtained by ``circumnavigating'' $T$, for example
in the clockwise direction; this path contains twice each branch
of $T$, with both possible orientations. We shall call $\CC$ the
oriented closed path obtained by continuous deformation of
$\bar\CC$, as the points $\{\uu_j\}_{j=1}^q$ vary in $\RRR^2$. The
path $\CC$ allows us to reorder the points $\ww_1,\ldots,\ww_l$
into $\ww_{t(1)},\ldots,\ww_{t(l)}$, by putting $t(1)=1$ and by
choosing $t(i+1)$, $1\le i\le l-1$, so that $\ww_{t(i+1)}$ is the
point following $\ww_{t(i)}$ on $\CC$. The chain $C$ is obtained
by joining with a segment $\ww_{t(i)}$ and $\ww_{t(i+1)}$, for
$i=1,\ldots,l-1$; the condition $2 \|T\|\ge \|C\|$ then easily
follows from the triangle inequality for the function
$x\to|x|^{1/2}$.\hfill\qed\hskip1em\null
\\

As a consequence of the above lemma and \pref{2.30}, we can
extract from the propagator bounds, for each choice of $T_v$, a
factor $\prod_{v\in\t^*}e^{-c'\sqrt{\g^{h_v}} D_{C_v}(\uw_v)}$,
which does not depend on the internal points positions, by leaving
a factor $e^{-(c/2)\sqrt{\g^j|\xx-\yy|}}$ for each propagator of
$T_\t$, to be used for bounding the integral over the internal
points.

The final bound of $S_{0,m,\t,\us}(\xxx)$ will be obtained by
``undoing'', in the r.h.s of eq. (2.36) of \cite{BFM}, the sum
over $T_v$ for any $v\in \t^*$ (note that $C_v$ depends on $T_v^*$
and hence on $T_v$), then adding the factors coming from the
previous considerations, together with a factor taking into
account that there are $1+ \sum_{v\in \t^*} (s^*_v-1)= m$
integrations less to do. By suitably choosing them, the lacking
integrations produce in the bound an extra factor $\prod_{v\in
\t^*} \g^{2 h_v (s_v^*-1)}L^{-2}$, so that we get
\bea\lb{2.31}
&&|S_{0,m,\t,\us}(\xxx)| \le  C^m(C\bar\l_{j_0})^n \g^{-j_0(-2+
m)}\left[ \prod_{\rm v\ not\ e.p.} \lft({Z_{h_v}\over
Z_{h_v-1}}\rgt)^{|P_v|/2} \g^{-d_v} \right]\cdot\nn\\
&&\lft(\prod_{i=1}^m {Z^{(1)}_{h_i} \over Z_{h_i}} \rgt)
\lft(\prod_{n.t. v\in\t^*}{1\over s_v!} \sum_{T_v}\g^{2 h_v
(s^*_v-1)} e^{-c'\sqrt{\g^{h_v}} D_{C_v}(\uw_v)}\rgt)
\eea
where $h_i$ is the scale of the $i$-th endpoint of type $J$ and
\be
d_v\defi -2+m_v+|P_v|/2 + z_v\;, \ee
with $m_v=|\XXX_v|$ and $z_v$ equal to the parameter $\tilde
z(P_v)$ defined by eq. (2.38) of \cite{BFM}.

We have now to bound the integral of $S_{0,m,\t,\us}(\xxx)
\overline\c_\L(\ux)$, let us call it $I_{m,\t,\us}$. In order to
exploit the improvement related with the restriction of the
integration variables to a fixed volume of size $1$, we shall
proceed in a way different with respect to that followed in
\cite{BFM}, that is we bound the integral {\it before} the sums
over the trees $T_v$. We use the bound:
\be\lb{3.9}
\int\!d\xx\ \c_\L(\xx)e^{-c'\sqrt{\g^h |\xx-\yy|}} \le C \lft\{
\begin{array}{lc}
\g^{-2h} & {\rm if\ }h>0\cr 1 & {\rm if\ }h\le 0
\end{array}
\rgt.
\ee
The sum over the tree graphs is done in the usual way and we get
\bea
&&I_{m,\t,\us} \le C^m(C\bar\l_{j_0})^n \g^{-j_0(-2+m)}
\lft(\prod_{n.t. v\in\t^*} \g^{2 h_v (s^*_v-1)}
 \rgt) \lft(\prod_{i=1}^m {Z^{(1)}_{h_i} \over
Z_{h_i}}\rgt) \cdot\nn\\
\lb{2.31a}&&\cdot  \lft[ \prod_{\rm v\ not\ e.p.}
\lft({Z_{h_v}\over Z_{h_v-1}}\rgt)^{|P_v|/2} \g^{-d_v}\rgt] \left(
\prod_{n.t. v\in\t^*}^{h_v> 0} \g^{-2 h_v (s^*_v-1)} \right)
\eea

Let us now call $E_i$ the family of trivial vertices belonging to
the branch of $\t^*$ which connects $v^*_i$ with the higher non
trivial vertex of $\t^*$ preceding it and note that, by the remark
preceding \pref{2.26a}, $Z^{(1)}_{h_i}/ Z_{h_i} \le C \g^{-h_i\bar
\h}$, with $\bar \h = c_0\l + O(\bar\l_{j_0}^2)$, $c_0>0$. Hence,
the definition of $s^{*,1}_v$ implies that, if $E=\cup_i E_i$,
\be\lb{2.32}
\prod_{i=1}^m {Z^{(1)}_{h_i} \over Z_{h_i}} \le C^m
\lft(\prod_{v\in E} \g^{-\bar\h} \rgt)\prod_{n.t. v\in \t^*}
\g^{-h_v\bar\h s^{*,1}_v} \;.\ee

Let $v^*_0$ be the first vertex with $s^*_v\ge 2$ following $v_0$
(recall that $v_0$ is the vertex immediately following the root of
$\t$, of scale $j_0+1$); since $m\ge 2$, this vertex is certainly
present. Then, since $m_v=m$ for $v_0 \le v \le v^*_0$, we have
the identity
\be \lb{2.33}
\g^{-j_0(-2+ m)} \prod_{v_0 \le v <v^*_0} \g^{-d_v} =
\g^{-h_{v^*_0} (-2+ m_{v^*_0})} \prod_{v_0 \le v <v^*_0}
\g^{-\tilde d_v}\;,\ee
where we used the definition $\tilde d_v = d_v - \left(-2+
m_v\right) = \frac{|P_v|}{2} + z_v$; note that $\tilde d_v \ge
1/2$, for any $v\in \t^*$, $v>v_0$.

By inserting \pref{2.32} and \pref{2.33} in the r.h.s. of
\pref{2.31a}, we get
\bea
&&I_{m,\t,\us} \le  C^m(C\bar\l_{j_0})^n \left[ \prod_{v\notin
\t^*\atop v \ not\ e.p.} \g^{-d_v} \right]
\left[ \prod_{v\in E} \g^{-d_v-\bar\h} \right] \left[ \prod_{v_0
\le v <v^*_0} \g^{-\tilde d_v}\right]\cdot\nn\\
&&\lb{2.35} \cdot\left[ \prod_{v\ not\ e.p.} \lft({Z_{h_v}\over
Z_{h_v-1}}\rgt)^{|P_v|/2} \right]\;
\left( \prod_{n.t. v\in\t^*}^{h_v> 0} \g^{-2 h_v (s^*_v-1)}
\right)\; F_\t \;,
\eea
where
\be\lb{2.36}
F_\t = \g^{-h_{v^*_0} (-2+ m_{v^*_0})}
\left[\prod_{n.t. v\in\t^*}  \g^{h_v \lft[2(s^*_v-1) -\bar \h
s^{*,1}_v\rgt]}\right] \left( \prod_{v\in(\t^*\bs E)}^{v\ge v^*_0
} \g^{-d_v} \right) \;.
\ee
Given a n.t. vertex $v\in\t^*$, let $s=s^*_v$, $s_1=s^{*,1}_v$,
$s_2=s-s_1$ and $v_1, \ldots, v_{s_2}$ the n.t. vertices
immediately following $v$ in $\t^*$. Since $m_v=s_1 +
\sum_{i=1}^{s_2} m_{v_i}$, we can write
\bea\lb{2.37}
 -(-2+m_v) + [2(s-1) -\bar \h s_1]
= s_1(1-\bar \h) - \sum_{i=1}^{s_2} (-2 +m_{v_i})\;.
\eea
This identity, applied to the vertex $v^*_0$, implies that, if
$v_1, \ldots, v_{s_2}$, $s_2=s^*_{v_0^*} - s^{*,1}_{v_0^*}$, are
the n.t. vertices immediately following $v^*_0$ in $\t^*$, then
\bea
&& \g^{-h_{v^*_0} (-2+ m_{v^*_0})} \g^{h_{v^*_0}
\lft[2(s^*_{v^*_0}-1) -\bar\h s^{*,1}_{v^*_0}\rgt]}  =\nn\\
&&\lb{2.38} = \g^{{\a'}_{v^*_0} h_{v^*_0}} \left[
\prod_{i=1}^{s_2} \g^{-h_{v_i} (-2+m_{v_i})} \cdot \prod_{v\in
\CC_i} \g^{-2+m_v} \right] \;,
\eea
where $\CC_i$ is the path connecting $v^*_0$ with $v_i$ in $\t^*$
(not including $v_i$) and we used the definition
\be\lb{2.39a} {\a'}_v= s^{*,1}_v(1-\bar\h)\;. \ee

The presence of the factor $\g^{-h_{v_i} (-2+ m_{v_i})}$ for each
vertex $v_i$ in the r.h.s. of  \pref{2.38} implies that an
identity similar to \pref{2.38} can be used for each n.t. vertex
$v\in\t^*$. It is then easy to show that
\be\lb{2.40}
F_\t =  \lft[ \prod_{n.t. v\in\t^*} \g^{{\a'}_v h_v} \right]
\left[ \prod_{v\in(\t^*\bs E)}^{v\ge v^*_0 } \g^{-\tilde d_v}
\right] \;.
\ee
By inserting this equation in \pref{2.35}, we get
\bea
I_{m,\t,\us} &\le& C^m (C\bar\l_{j_0})^n \lft( \prod_{n.t.
v\in\t^*} \g^{{\a'}_v h_v}\rgt) \left( \prod_{n.t. v\in\t^*}^{h_v>
0} \g^{-2 h_v (s^*_v-1)} \right) \cdot\nn\\
\lb{2.35a} &\cdot& \left[ \prod_{v\ not\ e.p.} \lft({Z_{h_v}\over
Z_{h_v-1}}\rgt)^{|P_v|/2} \g^{-{d'}_v} \right]\;,
\eea
where
\be\lb{2.41a}
{d'}_v = \cases{
\tilde d_v & if $v\in\t^*$, $v_0\le v\notin E$\cr
d_v+\bar\h & if $v\in E$\cr
d_v & otherwise\cr }
\ee
and it is always strictly greater than zero, if $\bar\h$ (hence
$\l$) is small enough. This allows us to control the sum over the
$\t$ scale labels in the usual way, by keeping fixed $h_{v_0^*}$.
However, before doing that, it is necessary to extract from the
r.h.s. of \pref{2.35a} some factors needed to control the sum over
$h_{v_0^*}$ too. First of all, in order to control the sum over
the negative values of $h_{v_0^*}$, we try to replace the
non-negative quantity ${\a'}_v$ with another non-negative one,
$\a_v$, s.t. $\a_{v_0^*}$ is {\it strictly positive}, paying a
price in the dimension of the vertices; this can be easily
achieved by fixing $\e>0$ and using the inequality
\be
1 \le \g^{\e h_{v_0^*}} \lft(\prod_{n.t. v\in\t^*}
\g^{\e(s^{*,2}_v-1)h_v}\rgt) \prod_{v\in\t^*, v\notin E}\g^\e
\ee
which allows us to replace \pref{2.35a} with
\bea
I_{m,\t,\us} &\le& C^m (C\bar\l_{j_0})^n \g^{\e h_{v_0^*}} \lft(
\prod_{n.t. v\in\t^*} \g^{\a_v h_v}\rgt) \left( \prod_{n.t.
v\in\t^*}^{h_v>
0} \g^{-2 h_v (s^*_v-1)} \right) \cdot\nn\\
\lb{2.35c} &\cdot& \left[ \prod_{v\ not\ e.p.} \lft({Z_{h_v}\over
Z_{h_v-1}}\rgt)^{|P_v|/2} \g^{-{\bar d}_v} \right]\;,
\eea
with $\a_v={\a'}_v+\e(s^{*,2}_v-1)$ and $\bar d_v=d'_v-\e$, if
$v_0^*\le v \not\in E$; and $\bar d_v=d'_v$ otherwise.

Let us now define $\c_v=1$ if $h_v>0$ and $\c_v=0$ for $h_v\le 0$.
If we put $w=v_0^*$, we can write
\bea
\lb{33}&& \g^{\e h_{v_0^*}}\lft[ \prod_{n.t. v\in\t^*} \g^{\a_v
h_v}\rgt] \left( \prod_{n.t. v\in\t^*}^{h_v> 0} \g^{-2 h_v
(s^*_v-1)}
\right) =\\
&& = \g^{\big[\a_{w}+\e-2\c_w(s^*_w-1)\big] h_{w}}  \prod_{n.t.
v\in\t^*\atop v\neq w} \g^{\big[\a_v-2\c_v(s^*_v-1)\big] h_v}\nn
\eea
and, if $|\l|<<|\e|<1/2$, we use the two straightforward
inequalities
\bea\lb{3.25}
\a_v &=& \e(s^*_v-1)+s^{*,1}_v(1-\bar\h-\e)\ge\e
\nn\\
\a_v-2(s^*_v-1) &=& (2-\e)-(1+\bar \h)s^*_v
-s^{*,2}_v(1-\e-\bar\h)\nn\\
&\le& (2-\e)-(1+\bar \h)s^*_v<0
\eea
Hence, the two  terms in square brackets in the r.h.s. of
\pref{33} can be bounded by $c^m$, {\it irrespective of the sign
of $\bar \h$}, that is the sign of $\l$.

Thanks to these arguments, we can replace \pref{2.35c} with
\be\lb{Im}
\hskip-0.5truecm I_{m,\t,\us} \le C^m (C\bar\l_{j_0})^n
\g^{\big[\a_{w}+\e-2\c_w(s^*_w-1)\big] h_{w}} \left[ \prod_{v\
not\ e.p.} \lft({Z_{h_v}\over Z_{h_v-1}}\rgt)^{|P_v|/2} \g^{-{\bar
d}_v} \right]
\ee
Now the sum over the tree scale labels, as well as the sum over
the trees with a fixed value of $h_w$, can be performed in the
usual way, using the fact that $\bar d_v$ is always strictly
positive and bounded below by a quantity proportional to $\bar
d_v$ itself; this gives a $C^m$ bound. We finally have to bound
the sum over $h_{w}$; note that
\be
\sum_{h_w=-\io}^{+\io} \g^{\big[\a_{w}+\e-2\c_w(s^*_w-1)\big] h_w}
=\sum_{h>0} \g^{\big[\a_w+\e-2(s^*_w-1)\big] h} + \sum_{h\le 0}
\g^{\big[\a_w+\e\big] h}
\ee
The second sum is always finite since $\a_w+\e \ge \e s^*_w\ge
2\e$. Regarding the first sum, we note that
\bea
\a_w+\e-2(s^*_w-1) &=& 2-(1+\bar \h)s^*_w -s^{*,2}_w(1-\e-\bar\h)
\nn\\
&\le& 2-(1+\bar \h)s^*_w
\eea
Hence, the sum is always bounded, except  in the case $\bar\h\le0$
(that is $\l\le 0$) with $s^{*}_w=s^{*,1}_w=2$. It follows, by
\pref{2.26a}, that
\be\lb{3.29}
\int\!d\ux\; \overline\c_\L(\uz) |S^{(m)}(\uz,\us)| \le m! C^m
\virg m\ge 3
\ee
so that, by \pref{Gqm}, the pressure can be defined only by
subtracting  from $G^{(0,\m)}$ the term with $m=2$. The {\it
renormalized pressure} is analytic in $\m$, for $\m$ small enough.

\subsection{Case $q\ge 2$, $r=0$}\lb{3.2}

We have for $S^{(m)}(\uz\hskip1pt\ux; \us,\us')$ an expansion
analogous to \pref{2.26a}, but now the special endpoints are
associated with two different types of space-time points, those
which have to be integrated as before ($\ux$) and those which are
fixed ($\uz$). We denote by $\ux_v$ and $\uz_v$ the points
following $v$ of the two types and we slightly modify the
definition of the point $\ww_v$ to be one point in $\uz_v$, if
$\uz_v\neq\emptyset$, or one point in $\ux_v$, otherwise; we still
require that $\ww_v\in \uw_v$.

We want to mimic the strategy used for the Sine-Gordon
correlations functions. Therefore we introduce a new tree $\t^o$,
that is obtained from $\t^*$ by erasing all the vertices which are
not needed to connect the $q$ special endpoints carrying a
space-time point of type $\zz$.  (We remark that the roles of the
trees $\t$ and $\t^*$ of the bosonic theory here are  played by
$\t^*$ and $\t^o$ respectively). Correspondingly, we define
$s^o_v$ the number of the branches of $\t^o$ following $v\in
\t^o$; mote that the space-time points associated with the
endpoints of $\t^o$ following $v\in \t^o$ are those in $\uz_v$,
hence $\sum_{w\ge v}(s^o_w-1)=|\uz_v|-1$.


A bound similar to \pref{2.31} holds. In this case, anyway, we
prefer to have a separate decaying factor in the distance of the
points $\uz$: for each nontrivial vertex $v$ of $\t^o$
\bea\lb{3.30}
D_{C_v}(\uw_v) \ge {1 \over 2} D_{\tilde C_v}(\uz_v\cap \uw_v)+
{1\over 2} D_{C_v}(\uw_v)\;.
\eea
where $\tilde C_v$ denotes the ordered path connecting the points
in $(\uz_v\cap \uw_v)$, made of lines which connect a point with
that following it in the ordered path $C_v$, see Lemma
\ref{lm3.1}.

Therefore, in place of \pref{2.31},  we have:
\bea\lb{2.31p}
&&|S_{0,m,\t,\us}(\uz,\ux)| \le  C^m(C\bar\l_{j_0})^n \g^{-j_0(-2+
m)} \left[ \prod_{\rm v\ not\ e.p.} \lft({Z_{h_v}\over
Z_{h_v-1}}\rgt)^{|P_v|/2} \g^{-d_v} \right] \cdot\nn\\
&&\cdot \lft(\prod_{i=1}^{m} {Z^{(1)}_{h_i} \over Z_{h_i}}\rgt)
\prod_{n.t. v\in\t^*}{1\over s_v!} \sum_{T_v}\g^{2 h_v (s^*_v-1)}
\exp\lft\{-{c'\over 2}\g^{h_v\over
2}D_{C_v}(\uw_v)\rgt\}\cdot \nn\\
&&\cdot \prod_{n.t. v\in\t^o}\exp\lft\{-{c'\over 2}\g^{h_v\over 2}
D_{\tilde C_v}(\uz_v\cap \uw_v)\rgt\}
\eea
We can repeat, with no essential modification, the steps that from
\pref{2.31} have led to \pref{2.31a}. Hence, if we call
$I_{m,\t,\us}(\uz)$ the integral over $\ux$ of
$S_{0,m,\t,\us}(\uz\ux) \overline\c_\L(\ux)$, we get the bound:
\bea\lb{p2.31a}
&&I_{m,\t,\us}(\uz) \le C^m(C\bar\l_{j_0})^n \g^{-j_0(-2+m)}
\lft(\prod_{n.t. v\in\t^*} \g^{2 h_v (s^*_v-1)}
 \rgt) \cdot\nn\\
&&\cdot \lft(\prod_{i=1}^{m} {Z^{(1)}_{h_i} \over Z_{h_i}}\rgt)
\lft[ \prod_{\rm v\ not\ e.p.} \lft({Z_{h_v}\over
Z_{h_v-1}}\rgt)^{|P_v|/2} \g^{-d_v}\rgt] \left(\prod_{n.t.
v\in\t^*}^{h_v> 0} \g^{-2 h_v (s^*_v-1)} \right)
 \cdot\nn\\
&&\cdot \left(\prod_{n.t. v\in\t^o}^{h_v> 0} \g^{2 h_v (s^o_v-1)}
\right) \prod_{n.t. v\in\t^o}\exp\lft\{-{c'\over 2}\g^{h_v\over 2}
D_{\tilde C_v}(\uz_v\cap \uw_v)\rgt\}
\eea
Indeed, we observe that the chain $C_v$ is a spanning tree of
propagators with root in one of the $\zz_v$ points (if any, see
the definition of $\ww_v$). Hence, integrating down the position
of the vertices $\ux_v$ from the endpoints of such a tree to the
root, in the case at hand there are, with respect to the procedure
for $q=0$, $s^o_v-1$ missing integration for each nontrivial
vertex $v$ of the tree $\t^o$. By \pref{3.9}, this means a factor
$\g^{-2h_v(s^o_v-1)}$ less, if $h_v>0$, and a constant factor
less, if $h_v\le 0$; this explains the last line of \pref{p2.31a}.
Going on in parallel with \S \ref{3.1}, we obtain the analogous of
\pref{Im}; recalling that  $w$ is the lowest n.t. vertex of the
tree $\t^*$,
\bea\lb{Imp}
\hskip-0.36cm &&I_{m,\t,\us}(\uz) \le C^m (C\bar\l_{j_0})^n
\g^{\big[\a_{w}+\e-2\c_w(s^*_w-1)\big] h_{w}} \left[ \prod_{v\
not\ e.p.} \lft({Z_{h_v}\over Z_{h_v-1}}\rgt)^{|P_v|/2} \g^{-{\bar
d}_v} \right] \cdot\nn\\
&&\cdot \left(\prod_{n.t. v\in\t^o}^{h_v> 0} \g^{2 h_v (s^o_v-1)}
\right) \prod_{n.t. v\in\t^o}\exp\lft\{-{c'\over 2}\g^{h_v\over 2}
D_{\tilde C_v}(\uz_v\cap \uw_v)\rgt\}
\eea
At this point, in contrast with the pressure bound, we want to
take advantage of the exponential fall off in the diameter of
$\uz_v\cap \uw_v$ to prove the convergence of the correlations
(with $q\ge 2$) {\it for any sign of $\bar\h$}.

Note that our definitions imply that $\cup_{n.t. v\in \t^0}
\uz_v\cap \uw_v =\uz$ and that $\cup_{n.t. v\in \t^0} \tilde C_v$
is a tree connecting all the points in $\uz$. This remark,
together with the trivial bound $h_v\ge h_{v^*_0}$, implies that
\be
\prod_{n.t. v\in\t^o}\exp\lft\{-{c'\over 4}\g^{h_v\over 2}
D_{\tilde C_v}(\uz_v\cap \uw_v)\rgt\} \le \exp\lft\{-{c'\over
4}\sqrt{\g^{h_{v_0^*}}\diam(\uz)}\rgt\}
\ee
On the other hand, since $|\uz_v\cap \uw_v|\ge 2$ for any $n.t.\
v\in \t^0$, if we define $\d\defi\min_{i,j}|\zz_i-\zz_j|$, we have
\be
\g^{2 h_v (s^o_v-1)} \exp\lft\{-{c'\over 4}\g^{h_v\over 2}
D_{\tilde C_v}(\uz_v\cap \uw_v)\rgt\} \le \left( {C\over \d}
\right)^{2(s_v^0-1)} (s_v^0-1)^{4(s_v^0-1)}
\ee
so that, by using also the identity $\sum_{v\in\t^0} (s_v^0-1) =
q-1$,
\bea
\left(\prod_{n.t. v\in\t^o}^{h_v> 0} \g^{2 h_v (s^o_v-1)} \right)
\prod_{n.t. v\in\t^o} \exp\lft\{-{c'\over
  2}\sqrt{\g^{h_v}\diam(\uz_v\cap \uw_v)}\rgt\}
\nn\\
\le [(q-1)!]^4 \lft({C\over \d}\rgt)^{2(q-1)} \exp\lft\{-{c'\over
4}\sqrt{\g^{h_{v_0^*}}\diam(\uz)}\rgt\}
\eea
Let us now remark that the quantity
\be\lb{3.35p}
\lft[ 1 + \diam(\uz)^{(\a_w+\e)}
\rgt]\sum_{h=-\io}^{+\io}\g^{[\a_{w}+\e-2\c_w(s^*_w-1)] h}
\exp\lft\{-c_0\sqrt{\g^{h}\diam(\uz)}\rgt\}
\ee
is bounded by a constant. In fact, the series is convergent also
without the exponential, as shown before, and this is sufficient,
if $\diam(\uz)\le 1$; if $\diam(\uz)=\g^{-h_0}$, $h_0 \le 0$, we
can bound the series by $2\sum_{h=-\io}^{+\io} \g^{(\a_w+\e)h}
\exp[-c_0\sqrt{\g^h}]$, which is convergent, sice $\a_w+\e\ge
2\e$. Hence we get, by using $2\e \le a_w+\e \le  q(1+\e-\bar\h)$,
that there is a constant $C_q$, such that
\be\lb{3.29p}
\int\!d\ux\; \overline\c_\L(\ux) |S^{(m)}(\uz,\ux,\us)| \le
m!\lft(1+\d^{-2(q-1)}\rgt) {C_q\over 1+\diam(\uz)^{2\e}}
\ee

\subsection{Case $r\ge 1$, $q=0$}
This case is very similar to the previous one; therefore we limit
ourself to the discussion of the differences.

Formula \pref{2.31p} still holds, with  $\uy_v$ in place of
$\uz_v$ (to be consistent with notation in \pref{Gqm}) and with
the replacement $\prod_{i=1}^m (Z^{(1)}_{h_i}/ Z_{h_i})
\longrightarrow \prod_{i=1}^p (Z^{(1)}_{h_i}/ Z_{h_i})$, following
from the fact that the strength renormalization of the field
$\bar\ps\g^\m\ps$ is equal to $Z_h$. It is easy to go along the
developments of \S \ref{3.2} again, up to a couple of differences.
The minor one is that in formulas \pref{2.32} and \pref{2.41a} the
set $E$ has to be replaced with the set $E\backslash Y$, where $Y$
is the family of trivial vertices of $\t^*$ belonging to the
branches ending up with an endpoint of type $\yy$; but this is not
a problem, since the dimensions of all the vertices remain
strictly positive. The major difference is that in \pref{2.32}, in
the case at hand, there is $h_v\bar\h(s^{*,1}_v-t^{*,1}_v)$ in
place of $h_v\bar\h s^{*,1}_v$, if $t^{*,1}_v$ is the number of
branches departing from $v$ and ending up with one endpoint of
type $\yy$ (hence $0\le t^{*,1}_v\le s^{*,1}_v$). At the end of
the developments, the latter fact generates a new $\a_v$, that we
have to prove to be  positive in order to control the bound in the
vertices $v\neq w$ such that $h_v\ge 0$ (as done in \pref{3.25}
for the old one). With simple computations we find:
\bea\lb{3.25p}
\a_v =\e(s^*_v-1)+(s^{*,1}_v-t^{*,1}_v) (1-\bar\h-\e)+t^{*,1}_v
(1-\e)\ge\e
\eea
Also, we need to prove that $\a_v-2(s^*_v-1)$ is negative, in
order to to control the bound in the vertices $v\neq w$ such that
$h_v>0$; and indeed:
\bea\lb{3.25pp}
\a_v-2(s^*_v-1)&=& (2-\e)-s^*_v-(s^{*,1}_v-t^{*,1}_v) \bar\h
-s^{*,2}_v(1-\e)\nn\\
&\le& (2-\e)-(1-|\bar \h|)s^*_v<0
\eea
Finally, the summation on the scale of $w$ is controlled by the
the exponential fall off in the diameter of $\uy$, as in
\pref{3.35p}.

\section{Explicit expression of the coefficients in the mass
expansion and proof of Theorem \ref{t1}} \lb{sec4}

\subsection{The case $r=0$}\lb{sec2.3}

As explained in the remark preceding \pref{Gqm}, in order to get
an explicit expression for the coefficients of the expansion
\pref{Gqm}, it is sufficient to calculate the correlations
$S^{(m,0)}(\ux,\s)$. We now show how to get this result by
computing the correlations of the $\psi$ field at non coinciding
points. We consider the following generating function
\be\lb{1l}
\hskip-0.5truecm \WW_{N,\e}(J)=\lim_{h\to-\io} \log
\int\!P_{h,N}(d\psi) e^{-\l Z_N^2 V(\ps) + \bar Z_N^{(1)}
\sum_\s\int d\xx d\yy J^\s_\xx \d_\e(\xx-\yy) \bar\psi_\xx \G^\s
\psi_\yy}
\ee
where $\d_\e(\xx)$ is a smooth approximation of the delta
function, rotational invariant, whose support does not contain the
point $\xx=0$; for definiteness we will choose $\d_\e(\xx) =
\e^{-2} v(\e^{-1}|\xx|)$, $v(\r)$ being a function on $\RRR^1$
with support in $[1,2]$, such that $\int d\r \r v(\r)=(2\p)^{-1}$
(so that $\int d\xx \d_\e(\xx)=1$). We define
\be
\bar S^{(m)}_{N,\e}(\ux,\us)= {\partial^m\over \partial
J_{\xx_1}^{\s_1}...\partial J_{\xx_m}^{\s_m}} \WW_{N,\e}(J)|_{J=0}
\ee
while $S^{(m)}_{N,0}(\ux,\us)$ will denote the analogous quantity
with $\d_\e(\xx-\yy)\to \d(\xx-\yy)$. Note that
$S^{(m,0)}(\ux,\us) = \lim_{N\to\io} S^{(m)}_{N,0}(\ux,\us)$.

\begin{lemma}\lb{lm2.2}
If $\l$ is small enough, there exists a constant $c_1= 1+O(\l)$,
such that, if we put $\bar Z_N^{(1)}=c_1 \e^{\h_+}$, then, for any
set $\ux$ of $m$ distinct points,
\be
\lim_{\e \to 0} \lim_{N\to\io} \bar S^{(m)}_{N,\e}(\ux,\us) =
\lim_{N\to\io} S^{(m)}_{N,0}(\ux,\us)
\ee
\end{lemma}

\0\Dim The proof of the Lemma is based on a multiscale analysis of
the functional $\WW_{N,\e}(J)$, performed by using the techniques
explained in sect. 2 of \cite{BFM}. We shall not give here the
detailed proof, but we shall stress only the relevant differences
with respect to the case studied there.

First of all, the external field $\f$ is zero and the free measure
has mass zero. Moreover the terms linear in $J$ and quadratic in
$\psi$ contains the monomial $\psi_{\xx,-\s}^+\psi_{\yy,\s}^- =
\bar\psi_\xx \G^\s \psi_\yy$, instead of
$\psi_{\xx,\s}^+\psi_{\yy,\s}^-$. This difference is unimportant
from the point of view of the dimensional analysis, so that, in
the case $\e=0$, we can essentially repeat the analysis of
\cite{BFM} with obvious minor changes. The situation is different
for $\e>0$, since in this case these terms (which are marginal)
are not local on the scale $N$, so that they need a more accurate
discussion.

Let us call $B_J^{(j)}(\psi)$ the contribution to the effective
potential on scale $j$, which is linear in $J$ and has as external
fields $\psi_{\xx,\o}^{[h,j]+}$ and $\psi_{\yy,-\o}^{[h,j]-}$ and
let $h_\e$ be the largest integer such that $\g^{-h_\e}\ge \e$ and
let $N>h_\e$. We want to show that, if $N\ge j\ge h_\e$, this
term, which is dimensionally marginal, is indeed irrelevant, so
there is no need to localize it. This follows from the observation
that $B_J^{(j)}(\psi)$ is of the form
\bea\lb{BJ}
&& B_J^{(j)}(\psi) = \bar Z_N^{(1)} \sum_\o\int d\xx d\yy J_\zz^\o
\d_\e(\xx-\yy) \psi_{\xx,-\o}^{[h,j]+}
\psi_{\yy,\o}^{[h,j]-} +\\
&&+ \sum_\o\int d\zz J_\zz^\o \int  d\bar\zz d\xx d\yy
\d_\e(\zz-\bar\zz) W_j(\zz,\bar\zz,\xx,\yy)
\psi_{\xx,-\o}^{[h,j]+} \psi_{\yy,\o}^{[h,j]-}\nn
\eea
where $W_j(\zz,\bar\zz,\xx,\yy)$ is the kernel of the sum over all
graphs containing at least one $\l$ vertex. It is easy to see that
it is of the form
\be
W_j(\zz,\bar\zz,\xx,\yy) = \tilde W_j(\zz,\xx) \tilde
W_j(\bar\zz,\yy)+ \bar W_j(\zz,\bar\zz,\xx,\yy)
\ee
where the second term is given by the sum over the graphs which
stay connected after cutting the line $\d_\e$, while the first
term is associated with the other graphs. The first term do not
need a localization, even for $j<h_\e$, because $\tilde
W_j(\zz,\xx)$ and $\tilde W_j(\bar\zz,\yy)$ are sum over graphs
with two external lines, one (the one contracted with the $J$
vertex) of scale $h_1 > j$, the other one of scale $h_2\le j$. The
momentum conservation and the compact support properties of the
single scale propagators imply that $h_1=j+1$, so that there is no
diverging sum associated with $h_1$, as one could expect since the
first term has a bound $C|\l|$. On the other hand, it easy to see
that the second term satisfies the bound
\be\lb{Wb}
\int d\bar\zz d\xx d\yy \d_\e(\zz-\bar\zz) |\bar
W_j(\zz,\bar\zz,\xx,\yy)| \le C |\l| \g^{-2 (j-h_\e)}
\ee
This immediately follows by comparing this bound with the
analogous one for $\e=0$, which is $C|\l|$ for dimensional
reasons. With respect to the case $\e=0$, we have a new vertex
$\bar\zz$, which is linked to the graph by the line $\d_\e$ and a
propagator of scale $j'>j$. The bound \pref{Wb} is obtained by
using the decaying properties of this propagator to integrate over
$\bar\zz$ and by bounding $\d_\e$ by $C\e^{-2}$.

Note that this procedure is convenient only because $j\ge h_\e$,
otherwise it would be convenient to integrate over $\bar\zz$ by
using $\d_\e$ and we should get the dimensional bound $C|\l|$ of
the case $\e=0$. It follows that, starting from $j=h_\e$, we have
to apply to $B_J^{(j)}(\psi)$ the localization procedure; then we
define, if $j\ge h_\e$,
\be
\LL B_J^{(j)}(\psi) = \sum_\o \bar Z^{(1)}_j \int d\zz J_\zz^\o
\psi_{\zz,-\o}^{[h,j]+} \psi_{\zz,\o}^{[h,j]-}
\ee
and we perform the limit $N\to\io$. In this limit, $\bar
Z^{(1)}_j$ can be represented as an expansion in terms of trees,
which have one special vertex (the $J$ vertex) and an arbitrary
number of normal vertices, the normal vertices being associated
with the limiting value $\l_{-\io}$ of the running coupling (whose
flow is independent of the $\bar Z^{(1)}_j$ flow). It follows that
$\bar Z^{(1)}_{h_e} = c_1\g^{-h_\e \h_+} [1+O(\l)]$ and that, if
$j<h_\e$,
\be\lb{Zbar}
\bar Z^{(1)}_{j-1} = \bar Z^{(1)}_j \g^{\h_+} + O(|\l|
\g^{-h_\e\h_+} \g^{-(h_\e-j)/2})
\ee
where the first term comes from the trees with the special vertex
of scale $\le h_\e$; it is exactly equal to the term one would get
in the theory with $\e=0$, in the limit $N\to\io$. The second term
is the contribution of the trees with the special vertex of scale
$>h_\e$ (these trees must have at least one normal vertex); it is
of course proportional to $\e^{\h_+}$ and takes into account the
``short memory property'' (exponential decrease of the irrelevant
terms influence). The flow \pref{Zbar} immediately implies that,
for any fixed $j$ and $|\h_+|<1/2$, $\lim_{\e\to 0} (\bar
Z^{(1)}_{j-1}/ \bar Z^{(1)}_j) = \g^{\h_+} = (Z^{(1)}_{j-1}/
Z^{(1)}_j)$ and that $\bar Z^{(1)}_j = c_1 [1+O(\l)] Z^{(1)}_j$.
Hence, by suitably choosing $c_1$, we can get $\lim_{\e\to 0} \bar
Z_j^{(1)} = Z^{(1)}_j$. \Halmos

\vskip.5cm

Note that $S^{(m)}(\ux,\oo)$ is different from $0$ only if $m$ is
even and $\sum_i \o_i=0$; moreover the truncated correlations can
be written as sums over the non truncated ones. Hence, in order to
get an explicit formula for $S^{(m)}(\ux,\oo)$, it is sufficient
to calculate the correlation
\be\lb{Kn}
K^{(n)}(\ux,\uuu) = \lim_{-h,N\to\io} (Z_N^{(1)})^{2n} \la
\prod_{j=1}^n \lft(\bar\ps_{\xx_j}\G^+\ps_{\xx_j}\rgt)
\lft(\bar\ps_{\uu_j}\G^-\ps_{\uu_j}\rgt)\ra
\ee
where $\la \cdot \ra$ denotes the expectation with respect to the
zero mass Thirring measure. By using Lemma \ref{lm2.2}, we have
\bea
&& K^{(n)}(\ux,\uuu) = c_1^{2n} \lim_{\e\to 0} \e^{2 n\h_+} \cdot\\
&&\cdot \int d\uy d\uv
[\prod_{i=1}^n\d_\e(\xx_i-\yy_i)\d_\e(\uu_i-\vv_i)] \tilde
K^{(2n)}(\ux,\uy,\uuu,\uv)\nn
\eea
where
\be
\tilde K^{(2n)}(\ux,\uy,\uuu,\uv) = \la  \prod_{j=1}^n
\lft(\bar\ps_{\yy_j}\G^+\ps_{\xx_j}\rgt)
\lft(\bar\ps_{\vv_j}\G^-\ps_{\uu_j}\rgt)\ra
\ee
On the other hand, by using the results of \cite{BFM}, see Theorem
\ref{t10} below, one can prove that, if $\la\cdot \ra_0$ is the
mean value for $\l=0$ and $\ps_i^-
\defi\ps^-_{\xx_i,\o_i}$, $\ps_i^+
\defi \ps^+_{\yy_i,\o'_i}$,
\bea
\lb{SOL} &&\la \ps^-_n \cdots \ps^-_1 \ps^+_1 \cdots \ps^+_n\ra=
c_0^{\l A(a-\bar a)n} \la \ps^-_n \cdots \ps^-_1 \ps^+_1 \cdots
\ps^+_n\ra_{0}\cdot\nn\\
&&\cdot {\prod_{s,t\in X}^{s<t} |\xx_s-\xx_t|^{ {\l A\over
4\p}(a-\bar a\o_s\o_t)} \cdot \prod_{s,t\in X}^{s<t}
|\yy_s-\yy_t|^{ {\l A\over 4\p}(a-\bar a\o'_s\o'_t)} \over
\prod_{s,t\in X}|\xx_s-\yy_t|^{ {\l A\over 4\p}(a- \bar
a\o_s\o'_t)} }
\eea
where $c_0$ is an arbitrary constant, to be determined by fixing,
for example, the value of the 2-points function at some value of
$\xx_1-\yy_1$, while $a$ and $\bar a$ are the parameters (function
of $\l$) defined in eq. (1.6) of \cite{BFM} and $A$ is equal to
the expression $[1-\l \sum_{\e=\pm 1} A_\e (\a_\e+\r_\e)]^{-1}$,
appearing in eq. (1.36) of \cite{BFM}. Hence, since $\bar\psi_\xx
\G^\o \psi_\yy = \psi^+_{\xx,-\o} \psi^-_{\yy,\o}$, we get
\bea
&&\hskip-1cm \tilde K^{(2n)}(\ux,\uy,\uuu,\uv) = c_0^{2 \l
A(a-\bar a)n} \la \prod_{j=1}^n
\lft(\bar\ps_{\yy_j}\G^+\ps_{\xx_j}\rgt)
\lft(\bar\ps_{\vv_j}\G^-\ps_{\uu_j}\rgt)\ra_{0} \cdot\nn\\
&&\hskip-1cm \cdot {\prod_{s,t\in X}^{s<t} |\xx_s-\xx_t|^{ {\l
A\over 4\p}(a-\bar a)} |\uu_s-\uu_t|^{ {\l A\over 4\p}(a-\bar a)}
\cdot \prod_{s,t\in X} |\xx_s-\uu_t|^{ {\l A\over 4\p}(a+\bar a)}
\over \prod_{s,t\in X}|\xx_s-\yy_t|^{ {\l A\over 4\p}(a+\bar a)}
|\uu_s-\yy_t|^{ {\l A\over 4\p}(a-\bar a)}} \cdot\\
&&\hskip-1cm \cdot {\prod_{s,t\in X}^{s<t} |\yy_s-\yy_t|^{ {\l
A\over 4\p}(a-\bar a)} |\vv_s-\vv_t|^{ {\l A\over 4\p}(a-\bar a)}
\cdot \prod_{s,t\in X} |\yy_s-\vv_t|^{ {\l A\over 4\p}(a+\bar a)}
\over \prod_{s,t\in X} |\uu_s-\vv_t|^{ {\l A\over 4\p}(a+\bar a)}
|\xx_s-\vv_t|^{ {\l A\over 4\p}(a-\bar a)}}\nn
\eea
A well known identity for the free fermions correlations,
equivalent to the so called {\it Cauchy Lemma} \cite{H} (since
$g_\o^{-1}(\xx)= 2\p(x_0+i\o x_1)$), is
\bea\lb{Cauchy}
&&\la  \prod_{j=1}^n \ps^-_{\xx_j,\o} \ps^+_{\vv_j,\o}\ra_{0}
=\sum_{\p\in P(1,\ldots,n)}(-1)^\p \prod_{j=1}^n
g_\o(\xx_j-\vv_{\p(j)}) =\nn\\
&& = (-1)^{n(n-1)\over 2} {\prod_{i,j\in X}^{i<j}
g_\o^{-1}(\xx_i-\xx_j) g_\o^{-1}(\vv_i-\vv_j) \over \prod_{i,j\in
X} g_\o^{-1}(\xx_i-\vv_j)}
\eea
Then
\bea
&&\hskip-1cm \la  \prod_{j=1}^n
\lft(\bar\ps_{\yy_j}\G^+\ps_{\xx_j}\rgt)
\lft(\bar\ps_{\vv_j}\G^-\ps_{\uu_j}\rgt)\ra_{0} = (-1)^n \la
\prod_{j=1}^n \ps^-_{\xx_j,+} \ps^+_{\vv_j,+}\ra_{0} \la
\prod_{j=1}^n \ps^-_{\uu_j,-} \ps^+_{\yy_j,-}\ra_{0} \nn\\
&&\hskip-1cm  = {\prod_{i,j\in X}^{i<j} g_+^{-1}(\xx_i-\xx_j)
g_+^{-1}(\vv_i-\vv_j) \over \prod_{i,j\in X}
g_+^{-1}(\vv_i-\xx_j)} \cdot {\prod_{i,j\in X}^{i<j}
g_-^{-1}(\uu_i-\uu_j) g_-^{-1}(\yy_i-\yy_j) \over \prod_{i,j\in X}
g_-^{-1}(\uu_i-\yy_j)}
\eea
and since $g_+^{-1}(\xx-\vv)g_-^{-1}(\xx-\vv)= (2\p)^2
|\xx-\vv|^2$, we get, by some straightforward calculations,
\be\lb{Ktilde}
\tilde K^{(2n)}(\ux,\uy,\uuu,\uv) = \left( {c_0^{\l A(a-\bar a)}
\over 2\p} \right)^{2n} {F(\ux,\uy,\uuu, \uv) \over \prod_s
(|\xx_s-\yy_s| |\uu_s-\vv_s|)^{ {\l A\over 4\p}(a+\bar a)}}
\ee
where $F(\ux,\uy,\uuu, \uv)$ is a continuous function such that
\be
\lim_{\yy_j\to\xx_j\atop \vv_j\to\uu_j} F(\ux,\uy,\uuu, \uv) =
{\prod_{s<t} |\xx_s-\xx_t|^{ 2(1-{\l A\over 2\p}\bar a)}
|\uu_s-\uu_t|^{ 2(1-{\l A\over 2\p}\bar a)} \over \prod_{s,t}
|\uu_s-\xx_t|^{ 2(1-{\l A\over 2\p}\bar a)} }
\ee
By using the previous identities, together with \pref{Kn}, we get
\bea
K^{(n)}(\ux,\uuu) &=& \left( {c_1 c_0^{\l A(a-\bar a)} \over 2\p}
\right)^{2n} \lim_{\e\to 0} \int d\uy d\uv
[\prod_{i=1}^n\d_\e(\xx_i-\yy_i)
\d_\e(\uu_i-\vv_i)] \cdot\nn\\
&\cdot& {\e^{2 n\h_+} F(\ux,\uy,\uuu, \uv) \over \prod_s
(|\xx_s-\yy_s| |\uu_s-\vv_s|)^{ {\l A\over 4\p}(a+\bar a)}}
\eea
By using the tree expansion, one can see that the limit $\e\to 0$
is bounded and different from zero, at least for $n=1$. It
follows, by taking into account the support properties of
$\d_\e(\xx)$, that $\h_+={\l A\over 4\p}(a+\bar a)$; note that in
\cite{BFM} it is stated that $\h_-={\l A\over 4\p}(a-\bar a)$, so
that we have
\be\lb{etapm}
\h_\s = {\l A\over 4\p} (a +\s \bar a)
\ee
Hence it is easy to see that, if we put $c_3(\h)=\int d\xx
\d_0(\xx) |\xx|^{-\h}$,
\be\lb{Kn1}
K^{(n)}(\ux,\uuu) = \left( {c_3(\h_+) c_1 c_0^{\l A(a-\bar a)}
\over 2\p} \right)^{2n} {\prod_{s<t} |\xx_s-\xx_t|^{ 2(1-{\l
A\over 2\p}\bar a)} |\uu_s-\uu_t|^{ 2(1-{\l A\over 2\p}\bar a)}
\over \prod_{s,t} |\uu_s-\xx_t|^{ 2(1-{\l A\over 2\p}\bar a)} }
\ee

If we compare \pref{Kn1} with \pref{pn1} and use the remark at the
beginning of this section, we get the formal equivalence
\be
\lim_{-h,N\to\io} Z_N^{(1)} \bar\psi_\xx \G^\s\psi_\xx \sim b_0\,
\lp e^{i\a\s\f_\xx}\rp
\ee
with
\be
b_0= c^{\a^2\over 8\p} {c_3(\h_+) c_1 c_0^{\l A(a-\bar a)} \over
2\p}
\ee
if the following relation between $\a$ and $\l$ is satisfied:
\be\lb{alfa}
{\a^2\over 4\p} = 1-{\l A\over 2\p}\bar a = 1+ \h_- - \h_+
\ee
where we also used \pref{etapm}.

This completes the proof of Theorem \ref{t1} for $r=0$.

\subsection{The case $r>0$}

Let us define
\be
j_\xx^\m= \bar\psi_\xx \g^\m \psi_\xx \virg \r_{\xx,\o} =
\psi^+_{\xx,\o} \psi^-_{\xx,\o} \virg \P_{\xx,\o} =
\psi^+_{\xx,-\o} \psi^-_{\xx,\o}
\ee
We have
\be\lb{4.22}
j_\xx^0 = \sum_\o \psi^+_{\xx,\o} \psi^-_{\xx,\o} \virg j_\xx^1 =
i\sum_\o \o\psi^+_{\xx,\o}\psi^-_{\xx,\o}
\ee
Hence, in order to calculate
$S^{(2n,r)}(\uz\hskip1pt\ux,\uy;\uo\hskip1pt\uo',\un)$, it is
sufficient to calculate the correlation function
\bea
&&D_{k_+,k_-,n}(\uaa,\uc,\uy,\uv) \= \lim_{-h,N\to\io}
(Z_N)^{k_+ +k_-} (Z_N^{(1)})^{2n}\cdot\nn\\
&&\la(\prod_{i=1}^{k_+} \r_{\aaa_i,+}) (\prod_{i=1}^{k_-}
\r_{\cc_i,-}) (\prod_{i=1}^n \P_{\yy_i,+}) (\prod_{i=1}^n
\P_{\vv_i,-})\ra^{(h,N)}
\eea
where $<\cdot>^{(h,N)}$ denotes the expectation w.r.t. the
massless Thirring measure.

By an obvious extension of Lemma \ref{lm2.2}, we know that there
are two constants $c_1$ and $c_2$ (smooth functions of $\l$, equal
to $1$ for $\l=0$), such that
\bea\lb{Dkk}
&& D_{k_+,k_-,n}(\uaa,\uc,\uy,\uv) = \lim_{\e_1,\e_2\to 0}
(c_2\e_2^{\h_-})^{(k_+ +k_-)} (c_1\e_1^{\h_+})^{2n} \int d\ub\,
d\ud\, d\ux\, d\uu\, \cdot\nn\\
&& \cdot \d_{\e_2}(\ub-\uaa) \d_{\e_2}(\ud-\ub) \d_{\e_1}(\ux-\uy)
\d_{\e_1}(\uu-\uv) \O(\uaa,\ub,\uc,\ud,\uy,\ux,\uv,\uu)
\eea
where, if $\ux=(\xx_1, \ldots,\xx_k)$, $\d_\e(\ux) = \prod_{i=1}^k
\d_\e(\xx_i)$, with $\d_\e(\xx)$ defined as in \S\ref{sec2.3};
moreover,
\bea
\O(\uaa,\ub,\uc,\ud,\uy,\ux,\uv,\uu) &=& \la(\prod_{i=1}^{k_+}
\psi^+_{\aaa_i,+} \psi^-_{\bb_i,+}) (\prod_{i=1}^{k_-}
\psi^+_{\cc_i,-} \psi^-_{\dd_i,-}) \cdot\nn\\
&\cdot& (\prod_{i=1}^n \psi^+_{\yy_i,-} \psi^-_{\xx_i,+})
(\prod_{i=1}^n \psi^+_{\vv_i,+} \psi^-_{\uu_i,-})\ra
\eea
By using the identities \pref{SOL} and \pref{Cauchy} and by doing
some simple algebra, one can see that, if $\uz=(\uaa,\ub,\uc,\ud)$
and $\uw=(\uy,\ux,\uv,\uu)$,
\bea
\O(\uz,\uw) &=& c_0^{\l A(a-\bar a)(k_+ +k_-)} {F_1(\uz,\uw) \over
\prod_{i=1}^{k_+} g_+^{-1}(\aaa_i-\bb_i)
\prod_{i=1}^{k_-} g_-^{-1}(\cc_i-\dd_i) } \cdot\nn\\
&& \hskip1cm\cdot {F_2(\uz,\uw)\over \prod_{i=1}^{k_+}
|\bb_i-\aaa_i|^{\h_-} \prod_{i=1}^{k_-} |\dd_i-\cc_i|^{\h_-}} \;
\tilde K^{2n}(\uw)
\eea
where $\tilde K^{2n}(\uw)$ is defined as in \pref{Ktilde}, and
\bea\lb{F1}
F_1(\uz,\uw) &=& \prod_{s<t} \tilde
h_{s,t}^{(+)}(\aaa_s,\bb_s,\aaa_t,\bb_t)
\prod_{s<t} \tilde h_{s,t}^{(-)}(\cc_s,\dd_s,\cc_t,\dd_t)\cdot\nn\\
&\cdot& \prod_{s,t} \tilde h_{s,t}^{(+)}(\aaa_s,\bb_s,\vv_t,\xx_t)
\prod_{s,t} \tilde h_{s,t}^{(-)}(\cc_s,\dd_s,\yy_t,\uu_t)
\eea
\bea\lb{F2}
F_2(\uz,\uw) &=&
\prod_{s<t} h_{s,t}^{(-)}(\aaa_s,\bb_s,\aaa_t,\bb_t)
\prod_{s<t} h_{s,t}^{(-)}(\cc_s,\dd_s,\cc_t,\dd_t)\cdot\nn\\
&\cdot& \prod_{s,t} h_{s,t}^{(-)}(\aaa_s,\bb_s,\vv_t,\xx_t)
\prod_{s,t} h_{s,t}^{(-)}(\cc_s,\dd_s,\yy_t,\uu_t)\cdot\\
&\cdot& \prod_{s,t} h_{s,t}^{(+)}(\aaa_s,\bb_s,\cc_t,\dd_t)
\prod_{s,t} h_{s,t}^{(+)}(\aaa_s,\bb_s,\yy_t,\uu_t)
\prod_{s,t} h_{s,t}^{(+)}(\cc_s,\dd_s,\vv_t,\xx_t)\nn
\eea
We defined
\be
\tilde h_{s,t}^{(\o)}(\aaa_s,\bb_s,\vv_t,\xx_t) =
{g_\o^{-1}(\aaa_s-\vv_t) g_\o^{-1}(\bb_s-\xx_t) \over
g_\o^{-1}(\aaa_s-\xx_t) g_\o^{-1}(\bb_s-\vv_t)}
\ee
\be
h_{s,t}^{(\s)}(\aaa_s,\bb_s,\vv_t,\xx_t) = \left( {|\aaa_s-\vv_t|
|\bb_s-\xx_t| \over |\aaa_s-\xx_t| |\bb_s-\vv_t|} \right)^{\h_\s}
\ee

Let us first evaluate, in the r.h.s. of \pref{Dkk}, the limit
$\e_1\to 0$. This can be done exactly as in \S\ref{sec2.3} and we
get
\bea\lb{Dkk1}
&& D_{k_+,k_-,n}(\uaa,\uc,\uy,\uv) = \lim_{\e_2\to 0}
(c_2\e_2^{\h_-})^{(k_+ +k_-)} \int d\ub\,
d\ud\, d\ux\, d\uu\, \cdot\nn\\
&& \cdot \d_{\e_2}(\ub-\uaa) \d_{\e_2}(\ud-\ub)
\O_0(\uaa,\ub,\uc,\ud,\uy,\uv)
\eea
where, if we put $\uw_0=(\uy,\uv)$,
\bea
\O_0(\uz,\uw_0) &=& c_0^{\l A(a-\bar a)(k_+ +k_-)}
{F_1(\uz,\uy,\uy,\uv,\uv) \over \prod_{i=1}^{k_+}
g_+^{-1}(\aaa_i-\bb_i)
\prod_{i=1}^{k_-} g_-^{-1}(\cc_i-\dd_i) } \cdot\\
&&\hskip1cm \cdot {F_2(\uz,\uy,\uy,\uv,\uv)\over \prod_{i=1}^{k_+}
|\bb_i-\aaa_i|^{\h_-} \prod_{i=1}^{k_-} |\dd_i-\cc_i|^{\h_-}} \;
K^{(n)}(\uw_0)\nn
\eea
$K^{(n)}(\uw_0)$ being defined as in \pref{Kn}; its explicit
expression is given by \pref{Kn1}.

Let us now perform the limit $\e_2\to 0$. We use the identity,
following from obvious symmetry arguments,
\bea
&& \lim_{\e_2 \to 0} \int d\ub\, d\ud\, \d_{\e_2}(\ub-\uaa)
\d_{\e_2}(\ud-\ub) {\e^{\h_-(k_+ + k_-)} \over \prod_{i=1}^{k_+}
|\bb_i-\aaa_i|^{\h_-} \prod_{i=1}^{k_-} |\dd_i-\cc_i|^{\h_-}}
\cdot\nn\\
&&\hskip1cm \cdot {F(\uaa,\ub,\uc,\ud) \over \prod_{i=1}^{k_+}
g_+^{-1}(\bb_i-\aaa_i) \prod_{i=1}^{k_-} g_-^{-1}(\dd_i-\cc_i)}
=\nn\\
&& = \lim_{\e_2 \to 0} \int d\ub\, d\ud\, \d_{\e_2}(\ub-\uaa)
\d_{\e_2}(\ud-\ub) \prod_{i=1}^{k_+} \left({\e_2 \over
|\bb_i-\aaa_i|}\right)^{\h_-} \prod_{i=1}^{k_-} \left({\e_2 \over
|\dd_i-\cc_i|}\right)^{\h_-}
\cdot\nn\\
&&\hskip1cm \cdot {\prod_{i=1}^{k_+} [(\bb_i-\aaa_i)\cdot
\dpr_{\bb_i}] \prod_{i=1}^{k_-} [(\dd_i-\cc_i) \cdot \dpr_{\dd_i}]
F(\uaa,\uaa,\uc,\uc)\over \prod_{i=1}^{k_+} g_+^{-1}(\bb_i-\aaa_i)
\prod_{i=1}^{k_-} g_-^{-1}(\dd_i-\cc_i)} =\cdot\nn\\
&&= \left({c_3(\h_-)\over 4\p}\right)^{k_+ + k_-}
\prod_{i=1}^{k_+} D^{-}_{\bb_i} \prod_{i=1}^{k_+} D^{+}_{\dd_i}
F(\uaa,\uaa,\uc,\uc)
\eea
where $c_3(\h)$ is defined as in \S\ref{sec2.3} and $D^{\o}_\xx \=
{\dpr\over \dpr x_0} +i\o {\dpr\over \dpr x_1}$.

To complete the calculation, note that, up to terms which are of
the second order in at least one of the differences $\bb_s-\aaa_s$
or $\dd_s-\cc_s$ (these terms give no contribution in the limit
$\e_2\to 0$),
\be
h_{s,t}^{(\s)}(\aaa_s,\bb_s,\vv_t,\yy_t) \simeq 1 + \h_\s
(\bb_s-\aaa_s)\cdot \left[{\aaa_s-\yy_t\over |\aaa_s-\yy_t|^2} -
{\aaa_s-\vv_t\over |\aaa_s-\vv_t|^2} \right]
\ee
\bea
&& h_{s,t}^{(\s)}(\aaa_s,\bb_s,\aaa_t,\bb_t) \simeq 1 + \h_\s
\Bigg[ - {(\bb_s-\aaa_s)\cdot (\bb_t-\aaa_t) \over
|\aaa_s-\aaa_t|^2}
+\nn\\
&&\hskip1cm + 2{[(\bb_s-\aaa_s)\cdot (\aaa_s-\aaa_t)]
[(\bb_t-\aaa_t)\cdot (\aaa_s-\aaa_t)]\over |\aaa_s-\aaa_t|^4}
\Bigg]
\eea
\bea
&& h_{s,t}^{(\s)}(\aaa_s,\bb_s,\cc_t,\dd_t) \simeq 1 + \h_\s
\Bigg[ - {(\bb_s-\aaa_s)\cdot (\dd_t-\cc_t) \over
|\aaa_s-\cc_t|^2}
+\nn\\
&&\hskip1cm + 2{[(\bb_s-\aaa_s)\cdot (\aaa_s-\cc_t)]
[(\dd_t-\cc_t)\cdot (\aaa_s-\cc_t)]\over |\aaa_s-\cc_t|^4} \Bigg]
\eea
\be
\tilde h_{s,t}^{(\s)}(\aaa_s,\bb_s,\vv_t,\yy_t) \simeq 1 +
g^{-1}_\s(\bb_s-\aaa_s) \left[{1\over g_\s^{-1}(\aaa_s-\yy_t)} -
{1\over g_\s^{-1}(\aaa_s-\vv_t)} \right]
\ee
\be
\tilde h_{s,t}^{(\s)}(\aaa_s,\bb_s,\aaa_t,\bb_t) \simeq 1 +
g^{-1}_\s(\bb_s-\aaa_s) g^{-1}_\s(\bb_t-\aaa_t) {1\over
[g_\s^{-1}(\aaa_s-\aaa_t)]^2}
\ee
In order to get the final result, we have to substitute these
expression in the r.h.s of \pref{F1} and \pref{F2}, expand their
product, keep the terms which are of the first order in all the
differences $\bb_i-\aaa_i$ and $\dd_i-\cc_i$ and, finally, apply
to them the differential operator $\prod_{i=1}^{k_+} D^{-}_{\bb_i}
\prod_{i=1}^{k_+} D^{+}_{\dd_i}$, whose effect can be easily
obtained by using the trivial identities
\be
D^{-\o}_{\bb_s} g^{-1}_\o(\bb_s-\aaa_s) = 4\p
\ee
\be
D^{\o}_{\bb_s} {(\bb_s-\aaa_s)\cdot \zz\over |\zz|^2} = 2\p
g_{-\o}(\zz)
\ee
\be
D^{\o}_{\bb_s} D^{\o}_{\bb_t} (\bb_s-\aaa_s)\cdot(\bb_t-\aaa_t) =0
\ee
\be
D^{\o}_{\bb_s} D^{-\o}_{\bb_t} (\bb_s-\aaa_s)\cdot(\bb_t-\aaa_t) =2
\ee

Let us consider, for example, the case $n=0$. Then we see
immediately that $D_{k_+,k_-,0}(\uaa,\uc)$ is different from $0$
only if $k_+ +k_-=2m$ and that, if we put $\uz=(\uaa,\uc)$ and
$\o_i=+1$ if $\zz_i\in\uaa$, $\o_i=-1$ if $\zz_i\in\uc$,
$D_{k_+,k_-,0}(\uaa,\uc)$ satisfies the Wick Theorem with
covariance $C_{\o_1,\o_2}(\zz_1-\zz_2) = \lim_{-h,N\to\io}
(Z_N^{(2)})^2 <\r_{\zz_i,\o_i} \r_{\zz_j,\o_j}>$, that is
$Z_N^{(2)} \r_{\xx,\o}$, in the removed cutoffs limit, is a
Gaussian field in the massless Thirring model. It is easy to check
that
\be
C_{\o_1,\o_2}(\zz_1-\zz_2) = \d_{\o_1,\o_2} {
\left[{c_0^{\l A(a-\bar a)} c_2 c_3\over 2\p} \right]^2
(1+\h_-/2)\over [(z_{1,0}-z_{2,0}) +i\o (z_{1,1}-z_{2,1})]^2}
\ee
It follows that, if we call $D^T_{k_+,k_-,n}(\uaa,\uc,\uy,\uv)$
the truncated expectation corresponding to
$D_{k_+,k_-,n}(\uaa,\uc,\uy,\uv)$, we have
\be
D^T_{k_+,k_-,0}(\uaa,\uc) = \d_{k,2} C_{\o_1,\o_2}(\zz_1-\zz_2)
\ee
Hence, by using \pref{4.22} and the definition \pref{hnn}, we get
\be\lb{S0k}
S^{(0,k)}(\zz_1,\zz_2;\n_1,\n_2) = -\d_{k,2} b^2
h^{\n_1,\n_2}(\zz_1-\zz_2)
\ee
with
\be\lb{bq}
b^2 = {1\over \p} [c_0^{\l A(a-\bar a)} c_2 c_3]^2 (1+{\h_-\over
2})
\ee

Let us now consider the case $n>0$; in this case we shall give the
explicit expression of $D^T_{k_+,k_-,n}(\uaa,\uc,\uy,\uv)$. This
quantity can be obtained, by expanding $K^{(n)}(\uw_0)$ in terms
of products of connected expectations in the usual way and then
trying to get a connected quantity by using the terms which
survive to the limit $\e_2\to 0$, see discussion above. It is
obvious that a connected contribution can be obtained only by
keeping the products of different first order zeros coming from
the functions $h_{s,t}^{(\s)}(\aaa_s,\bb_s,\vv_t,\yy_t)$, $\tilde
h_{s,t}^{(\s)}(\aaa_s,\bb_s,\vv_t,\yy_t)$ and the analogous with
$(\cc_s,\dd_s)$ in place of $(\aaa_s,\bb_s)$, together with the
truncated expectation in the expansion of $K^{(n)}(\uw_0)$, that
we shall call $K_T^{(n)}(\uw_0)$. It follows that
\be\lb{4.46}
D^T_{k_+,k_-,n}(\uaa,\uc,\uy,\uv) = K_T^{(n)}(\uw_0) \;
\prod_{r=1}^{k_+ + k_-} W(\zz_r,\o_r,\uw_0,\us)
\ee
where we defined $\us$ so that $\s_i=+1$ if $\ww_i\in \uy$,
$\s_i=-1$ if $\ww_i\in \uv$ and
\be\lb{4.47}
W(\zz,\o,\uw_0,\us) = -\o\, c_0^{\l A(a-\bar a)} c_2 c_3
\left(1+{\h_- -\h_+\over 2}\right) \sum_{j=i}^n \s_j
g_{\o}(\zz-\ww_j)
\ee
By using \pref{4.22}, we easily get the final result
\be\lb{4.48}
S^{(2n,k)}(\uw,\uy;\us,\un) = S^{(2n,0)}(\uw;\us) \; \prod_{r=1}^k
\tilde W^{\n_r}(\yy_r,\uw,\us)
\ee
with
\be\lb{4.49}
\tilde W^\n(\yy,\uw_0,\us) = {i\over\p} c_0^{\l A(a-\bar a)} c_2
c_3 \left(1+{\h_- -\h_+\over 2}\right) \sum_{j=i}^n \s_j
{\e_{\n,\m} (\yy-\ww_j)^\m\over |\yy-\ww_j|^2}
\ee

It follows that \pref{4.48} has the same structure than
\pref{2.87}, so that, by comparing \pref{4.49} with \pref{2.88},
as well as \pref{S0k} with \pref{2.94}, we get, in agreement with
the considerations after \pref{1.6a}, the equivalence
\be
\lim_{-h,N\to\io} Z_N^{(2)} j^\n_\xx \sim - \e_{\n,\m} (b_1
\dpr^\m\f_\xx + b_2 \dpr^\m\xi_\xx)
\ee
where $\xi$ is a free boson field of zero mass, independent of
$\phi$, and
\be\lb{buno}
 b_1= {2\over \a} c_0^{\l A(a-\bar a)} c_2 c_3
\left(1+{\h_- -\h_+\over 2}\right)
\ee
One can check, by using the relations $a^{-1} =
1-\l/(4\p)+O(\l^2)$, $\bar a^{-1} = 1+\l/(4\p)+O(\l^2)$ and
$A=1+O(\l^2)$, that, if $b^2$ is the constant defined in
\pref{bq},
\be
b_2^2 = b^2-b_1^2 =O(|\l|^3)
\ee
However, one can prove that $b_2=0$. This follows from the remark
that
\be\lb{b12}
{b_1^2\over b^2} = {\left(1+{\h_- - \h_+\over 2} \right)^2 \over
(1+\h_- - \h_+) \left(1+{\h_-\over 2} \right)}
\ee
where we used \pref{buno}, \pref{bq} and \pref{alfa}. Hence, in
order to prove that $b_2=0$, it is sufficient to prove that the
r.h.s. of \pref{b12} is equal to $1$; by a simple calculation one
can check that this condition is equivalent, since $\h_->0$, to
the condition
\be\lb{ideta}
(1+\h_-)^2 = 1+\h_+^2
\ee
Our solution of the Thirring model allows us to represent $\h_-$
and $\h_+$ as well defined power series in the physical value
$\l_{-\io}$ of the running coupling (see eq. (2.35) of \cite{BFM}
for $\h_-$). This representation is not convenient to verify the
identity \pref{ideta}; however, \pref{ideta} is independent of the
details of the ultraviolet regularization of the model, hence it
can be checked also by using the explicit (rigorous)
representations of $\h_-$ and $\h_+$ in terms of the bare
coupling, which were found in \cite{M1} and \cite{M2} with a
different ultraviolet regularization (by the way, they are also in
agreement with the heuristic procedure proposed in \cite{J} and
\cite{K}). In this approach, if we call $\tilde\l$ the bare
constant and put $x=\tilde\l/(4\p)$, one gets
\be
\h_- = {2x^2\over 1-x^2} \virg \h_+ = {2x\over 1-x^2}
\ee
and one can check that \pref{ideta} is indeed satisfied.

This completes the proof of Theorem \ref{t1} for $r>0$.

\appendix

\section{The explicit formula for the field correlation functions}

\subsection{The Schwinger-Dyson equation}
In this appendix we will derive the explicit expression of the
$n$-point Schwinger functions \pref{SOL}, by extending the
arguments used in [BFM], to which we refer for details, to analyze
the 2-point function. Let us define $\WW(J,\f) = \log \int
P_{h,N}(d\psi)$ $\exp\{\VV^{(N)} (\sqrt{Z_N}\psi,J,\f)\}$, where
the free measure $P_{h,N}(d\psi)$ is defined by \pref{1.7},
$\VV^{(N)} = -\l V\lft(\sqrt{Z_N}\psi\rgt) + \sum_\o \int d\xx [
J_{\xx,\o} Z_N \psi^+_{\xx,\o} \psi^-_{\xx,\o} +
\f^+_{\xx,\o}\ps^-_{\xx,\o} + \ps^+_{\xx,\o}\f^-_{\xx,\o}]$ and
the fields $\f^\pm$ are anticommuting between themselves and with
$\psi^\pm$. We shall introduce the Fourier transform of various
fields. In doing that, we shall consider the fields
$\r=\psi^+\psi^-$, $\ps^+$ e $\f^+$ as incoming fields, while
$\a$, $J$, $\ps^-$ e $\f^-$ will be outcoming fields.

First of all, we note that the Schwinger-Dyson equations are
generated by the identity
\be\lb{SDE}
D_\o(\kk){\partial e^{\WW} \over\partial \hf^+_{\kk,\o}} =
\c_{h,N}(\kk) \left[ {\hf^-_{\kk,\o}e^\WW \over Z_N}-
\l\int\!{d\pp\over(2\p)^2}\ {\partial^2 e^{\WW}\over
\partial\hJ_{\pp,-\o} \partial\hf^+_{\kk+\pp,\o}} \right]
\ee
Indeed,  given any $F(\psi)$ which is a power series in the field,
we have, by the Wick Theorem,
\be
\la \hp^-_{\kk,\o}F(\ps)\ra_0 ={\hg^{[h,N]}_{\o}(\kk)\over Z_N} \la
{\dpr F(\ps)\over \dpr \hp^+_{\kk,\o}}\ra_0.
\ee
where $\la \cdot\ra_0$ is the mean value with respect to $P_{h,N}$
and $\hg^{[h,N]}_{\o}(\kk) = \c_{h,N}(\kk)/ D_\o(\kk)$. Then,
\pref{SDE} is a consequence of the identity
\bea
{\partial e^{\WW} \over\partial \hf^+_{\kk,\o}} = \la
\hp^-_{\kk,\o}e^{\VV^{(N)}\lft(\sqrt{Z_N}\psi,\f\rgt)}\ra_0
={\hg^{[h,N]}_{\o}(\kk)\over Z_N} \la {\dpr\over \dpr
\hp^+_{\kk,\o}}e^{\VV^{(N)}\lft(\sqrt{Z_N}\psi,\f\rgt)}\ra_0
\eea
and the remark that $V(\psi) = \int d\xx
\psi^+_{\xx,+}\psi^-_{\xx,+} \psi^+_{\xx,-}\psi^-_{\xx,-}$.

\subsection{The approximate Ward--Takahashi identities}
We consider a new generating functional, $\WW_\AAA (\a,J,\f) =
\log \int P_{h,N}(d\ps)$ \break $ \cdot\exp \{ \VV^{(N)}(\sqrt{Z_N}\psi,J,\f) +
[Z_N\AAA_0 + Z_N\sum_{\s=\pm} \n^{(\s)}_N\AAA_{\s}](\a,\ps) \}$
where $\n^{(\pm)}_N$ are two suitable constants, to be chosen
below as functions of $\l$, and
\bea
&&\AAA_0 (\a,\ps)
\defi
\sum_{\o=\pm}\int\! {d\qq\;d\pp\over (2\p)^4}\
 C_\o(\qq,\pp)\ha_{\qq-\pp,\o}\hp^+_{\qq,\o}\hp^-_{\pp,\o}\;,\\
&&\AAA_\s(\a,\ps)
\defi
\sum_{\o=\pm}\int\! {d\qq\;d\pp\over (2\p)^4}\
D_{\s\o}(\pp-\qq)\ha_{\qq-\pp,\o}\hp^+_{\qq,\s\o} \hp^-_{\pp,\s\o}
\eea
having defined, as in \cite{BFM}, $D_\o(\kk)=-ik_0 +\o k_1$ and
$C_\o(\qq,\pp) = [\c_{h,N}^{-1}(\pp)-1] D_\o(\pp)
-[\c_{h,N}^{-1}(\qq)-1] D_\o(\qq)$.

By doing the transformation $\psi^{\pm}_{\xx,\o} \to e^{i\a_{\xx,\o}}
\psi^{\pm}_{\xx,\o}$, see \cite{BM} for a rigorous definition, we get
\bea\lb{12}
D_\m(\pp){\partial \WW \over \partial \hJ_{\pp,\m}}(J,\f) &=&
\int\!{d\kk\over (2\p)^2} \left[\hf^+_{\kk+\pp,\m} {\partial
\WW\over \partial \hf^+_{\kk,\m}}- {\partial \WW\over
\partial \hf^-_{\kk+\pp,\m}} \hf^-_{\kk,\m}\right]-\nn\\
&-& {\partial \WW_{\AAA_0}\over \partial \ha_{\pp,\m}}(0,J,\f);
\eea
where $\WW_{\AAA_0}$ is the same as  $\WW_{\AAA}$ but neglecting
the interactions $\AAA_\s(\a,\ps)$. The last term in the r.h.s. of
\pref{12} is not negligible in the removed cutoffs limit, but we
can extract its leading contribution by introducing suitable
counterterms \cite{BFM}, so that the rest will vanish, by putting
\bea\lb{grez}
&& (1-\n^{(+)}_N) D_\m(\pp){\partial \WW\over \partial
\hJ_{\pp,\m}}(J,\f) -\n^{(-)}_N D_{-\m}(\pp){\partial \WW\over
\partial \hJ_{\pp,-\m}}(J,\f)=\\
&&= \int\!{d\kk\over (2\p)^2}
 \left[\hf^+_{\kk+\pp,\m}
 {\partial \WW\over \partial \hf^+_{\kk,\m}}(J,\f)-
 {\partial \WW\over \partial \hf^-_{\kk+\pp,\m}}(J,\f)
 \hf^-_{\kk,\m}\right]
 -{\partial \WW^{(h)}_\AAA\over \partial \ha_{\pp,\m}}(0,J,\f)\nn
\eea
If we define
\be
a_N = [1- \n^{(-)}_N - \n^{(+)}_N]^{-1} \virg \bar{a}_N = [1+
\n^{(-)}_N -\n^{(+)}_N]^{-1}
\ee
by some simple algebra we obtain the identity
\bea\lb{WT1}
&& {\partial e^{\WW}\over \partial
\hJ_{\pp,\s}}(0,\f) +\sum_{\s'}{A_{N,\s\s'}\over D_{\s}(\pp)}
 {\partial e^{\WW_\AAA}\over \partial
 \ha_{\pp,\s'}}(0,0,\f)
=\\
&&=\sum_{\s'} {A_{N,\s\s'}\over D_{\s}(\pp)} \int\! {d \kk\over
(2\p)^2}\ \left[\hf^+_{\kk+\pp,\s'} {\partial e^{\WW}\over
\partial \hf^+_{\kk,\s'}}(0,\f)- {\partial e^{\WW}\over
\partial \hf^-_{\kk+\pp,\s'}}(0,\f) \hf^-_{\kk,\s'}\right]\nn
\eea
where $A_{N,\s} = (a_N + \s\bar{a}_N)/2$. With the argument explained
in \cite{BFM}, it would be easy to prove that the term in
$\WW_\AA$ is vanishing in the limit of removed cutoff; anyway this is
not our current objective.

\subsection{A Closed Equation for the field correlation functions}
\lb{secA3}

By doing an arbitrary number of functional derivatives with
respect to the $\f$ external field in \pref{SDE} and then putting
$\f=0$, one can obtain an infinite number of relations between the
field correlation functions and other correlations involving
several fields and one current, integrated over the current
momentum. We want now to show that, by using the identity
\pref{WT1}, it is possible to get a closed equation for the field
correlation functions, in the limit of removed cutoffs. Let us
define $\dpr^\o_\xx= \dpr_0 +i\o\dpr_1$.

\begin{lemma}\lb{l13}
For $|\l|$ small enough, there exists a constant $A=1+O(\l)$, such
that the equations of motion for the truncated Schwinger functions
-except the two point Schwinger function- in the limit of removed
cutoffs are generated, at non coinciding points, by the identity:
\bea\label{SD10}
&&\hskip-1truecm \dpr_{\xx_1}^{\o_1} {\dpr \WW\over \dpr
\f_{\xx_1,\o_1}^+} =\l A \sum_{\m} A_{-\o_1\m} \int\!d\zz\
g_{-\o_1}(\xx_1-\zz) \lft[\f^+_{\zz,\m}{\dpr^2 \WW\over \dpr
\f^+_{\zz,\m}\dpr \f^+_{\xx_1,\o_1}} -{\dpr^2 \WW\over \dpr
\f^+_{\xx_1,\o_1}\dpr \f^-_{\zz,\m}}
\f^-_{\zz,\m}\rgt]\nn\\
&&\hskip-.2truecm +\l A \sum_{\m}A_{-\o_1\m} \int\!d\zz\
g_{-\o_1}(\xx_1-\zz) \lft[\f^+_{\zz,\m}{\dpr \WW\over \dpr
\f^+_{\zz,\m}}{\dpr \WW\over \dpr \f^+_{\xx_1,\o_1}} -{\dpr
\WW\over \dpr \f^+_{\xx_1,\o_1}}{\dpr \WW\over \dpr
\f^-_{\zz,\m}}\f^-_{\zz,\m}\rgt]
\eea
\end{lemma}

{\bf\0Proof.} If we make a derivative with respect to
$\hf^+_{\kk+\pp,\o}$ in both sides of \pref{WT1}, with $\o=\s$,
and then integrate over $\pp$ (which is meaningful, since the
correlation functions can not have a singularity at $\pp=0$ and
have compact support in $\pp$ for $h$ and $N$ finite), we get
\bea\lb{WT11}
&&\hskip-0.2cm \int\!{d\pp\over (2\p)^2}\ {\dpr^2 e^{\WW}\over
\dpr \hJ_{\pp,-\o}\partial \hf^+_{\kk+\pp,\o}}= -\sum_\m
 \int\!{d\pp\over (2\p)^2}\
 {A_{N,-\o\m}\over D_{-\o}(\pp)}
 {\dpr^2e^{\WW_\AAA}\over\dpr\ha_{\pp,\m}\dpr\hf^+_{\kk+\pp,\o}}+\\
&&+\sum_\m
 \int\!{d\pp\;d\qq\over (2\p)^4}
 {A_{N,-\o\m}\over D_{-\o}(\pp)}
 \left[\hf^+_{\qq+\pp,\m}
 {\dpr^2 e^{\WW}\over \dpr \dpr\hf^+_{\qq,\m} \hf^+_{\kk+\pp,\o}}-
 {\dpr^2 e^{\WW}\over \dpr\hf^+_{\kk+\pp,\o}\dpr\hf^-_{\qq+\pp,\m}}
 \hf^-_{\qq,\m}\right]\nn
\eea
where both sides are calculated at $J=\a=0$ and we used the fact
that $D_\o(\pp)$ is odd in $\pp$ to cancel one term in the r.h.s.
of \pref{WT11}.

We introduce the generating functionals $\WW_{\TT,\m}(\b,\f)$, for
$\m=\pm$, defined as
\bea\lb{gfter}
&&e^{\WW_{\TT,\m} (\b,\f)} \defi\int\!P_{h,N}(d\ps)\
e^{-\l_{N}Z^2_N  V(\ps)} \exp\left\{ \int\!
\f^+_{\xx,\o}\ps^-_{\xx,\o}
+ \int\! \ps^+_{\xx,\o}\f^-_{\xx,\o}\rgt\}\cdot\nn\\
&&\cdot\exp\left\{ \left[\TT^{(\m)}_0+\sum_{\s=\pm}\n_N^{(\s)}
\TT^{(\m)}_{\s}\rgt]
\left(\sqrt{Z_N}\ps,\sqrt{Z_N}\b\right)\right\}\cdot\\
&&\cdot\exp\left\{  \sum_{\o=\pm} \lft[-\a^{(\m\o)}\l\BB^{(3)}_\o
 -\r^{(\m\o)}\BB^{(1)}_\o\right]
 \left(\sqrt{Z_N}\ps,\sqrt{Z_N}\b\right)\right\}\nn
\eea
with $\{\a^{(\m)}\}_{\m=\pm}$, $\{\r^{(\m)}\}_{\m=\pm}$, four real
parameters to be fixed later and
\bea
&&\TT^{(\m)}_0(\ps,\b)
\defi
 \sum_{\o=\pm}
 \int\! {d\kk\;d\pp\;d\qq \over (2\p)^6}\;
 {C_\m(\qq,\qq-\pp)\over D_{-\o}(\pp)}
 \hb_{\kk,\o}\hp^-_{\kk+\pp,\o}\hp^+_{\qq,\m}\hp^-_{-\pp+\qq,\m}\;,
\cr\cr &&\TT^{(\m)}_{\s}(\ps,\b)
\defi \sum_{\o=\pm}
 \int\!{d\kk\;d\pp\;d\qq \over (2\p)^6}\;
 { D_{\s\m}(-\pp)\over D_{-\o}(\pp)}\
 \hb_{\kk,\o}\hp^-_{\kk+\pp,\o}\hp^+_{\qq,\s\m}\hp^-_{-\pp+\qq,\s\m}\;,
\cr\cr &&\BB^{(3)}_\o(\ps,\b)
 \defi
\int\!{d\kk\;d\pp\;d\qq \over (2\p)^6}\;
 \hb_{\kk,\o}\hp^-_{\kk+\pp,\o}\hp^+_{\qq,-\o}\hp^-_{-\pp+\kk,-\o}\;,
\cr\cr && \BB^{(1)}_\o(\b,\ps)
 \defi
\int\!{d\kk\over (2\p)^2}\;
 \hb_{\kk,\o} D_\o(\kk)\hp^-_{\kk,\o}
\eea
We remark that $\WW_{\TT,\m} (\b,\f)$ differs from the analogous
generating functional introduced in \cite{BFM} because of the
presence of the {\it interactions} $\BB^{(1)}_\o(\b,\ps)$ and
$\BB^{(3)}_\o(\b,\ps)$, that in the cited paper  were - in a sense
- reconstructed a posteriori; here we describe a faster way to
implement the same procedure of \cite{BFM}. We have the following
identity:
\bea\lb{ce1}
&&\int\!{d\pp\over (2\p)^2}\ {1\over D_{-\o}(\pp)}
 {\partial^2e^{\WW_\AAA}\over \partial \ha_{\pp,\m}
 \partial \hf^+_{\kk+\pp,\o}}(0,0,\f) ={1\over Z_N}
 {\partial e^{\WW_{\TT,\m}}\over \partial \hb_{\kk,\o}}(0,\f)+
 \\
&&+\a^{(\m\o)} \l\int\! {d\pp\over (2\p)^2}\
 {\partial^2 e^{\WW}\over
\partial \hJ_{\pp,-\o} \partial \hf^+_{\kk+\pp,\o}}(0,\f)
+ \r^{(\m\o)} D_{\o}(\kk) {\partial e^{\WW}\over
\partial \hf^+_{\kk,\o}}(0,\f) \nn
\eea
which, plugged into \pref{WT11}, gives
\bea\lb{WT111}
&&\lft(1+\l\sum_\m A_{N,-\m}\ \a^{(\m)}\rgt) \int\!{d\pp\over
(2\p)^2} {\dpr^2 e^{\WW}\over \dpr \hJ_{\pp,-\o}\partial
\hf^+_{\kk+\pp,\o}}=
\nn\\
&&=-\sum_\m
 {A_{N,-\o\m}\over Z_N}
 {\partial e^{\WW_{\TT,\m}}\over \partial \hb_{\kk,\o}}
-\lft(\sum_\m A_{N,-\m}
 \r^{(\m)}\rgt) D_{\o}(\kk)
 {\partial e^{\WW}\over
 \partial \hf^+_{\kk,\o}}+\\
&&+\sum_\m
 \int\!{d\pp\;d\qq\over (2\p)^4}\;
 {A_{N,-\o\m}\over D_{-\o}(\pp)}
 \left[\hf^+_{\qq+\pp,\m}
 {\dpr^2 e^{\WW}\over \dpr\hf^+_{\qq,\m}\dpr\hf^+_{\kk+\pp,\o}}-
 {\dpr^2 e^{\WW}\over \dpr\hf^+_{\kk+\pp,\o}\dpr\hf^-_{\qq+\pp,\m}}
 \hf^-_{\qq,\m}\right]\;.\nn
\eea
This equation, together with \pref{SDE}, the identity
$\d_{\m,\o}=(1+\m\o)/2$ and the remark that $\WW_{\TT,\m}(0,\f)
=\WW(0,\f)$, implies that
\bea\lb{DSE3}
&&\hskip-.5truecm D_\o(\kk){\partial \WW \over\partial
\hf^+_{\kk,\o}} = \c_{h,N}(\kk) {B_N\over Z_N}\hf^-_{\kk,\o}+
\c_{h,N}(\kk) {\l A_N\over Z_N}\sum_\m A_{N,-\o\m} {\partial
\WW_{\TT,\m}\over \partial \hb_{\kk,\o}}(0,\f)+\\
&&\hskip-1truecm  + \l A_N\sum_\m  \c_{h,N}(\kk) \int\!
{d\qq\;d\pp\over (2\p)^4}\; {A_{N,-\o\m}\over  D_{-\o}(\pp)}
e^{-\WW} \left[\hf^+_{\qq,\m}
 {\partial^2 e^{\WW}\over \dpr \hf^+_{\qq+\pp,\m} \dpr\hf^+_{\kk-\pp,\o}}
- {\partial^2 e^{\WW}\over \partial \hf^+_{\kk-\pp,\o}\partial
\hf^-_{\qq,\m}} \hf^-_{\qq+\pp,\m}\right]\nn
\eea
where $A_N = [1+\l\sum_\m A_{N,-\m} (\a^{(\m)}-\r^{(\m)})]^{-1}$
and $B_N = [1+\l \sum_\m A_{N,-\m}\a^{(\m)}] A_N$.

Before doing the limit $-h,N\to\io$, we can rewrite the previous
identity in the space coordinates, by doing the Fourier transform
in both sides. Since we want to get an identity involving only the
correlations with at least four points, the first term in the
r.h.s. gives no contribution. Hence, it is easy to see that we get
the identity \pref{SD10}, with $A=\lim_{N\to\io} A_N$, if the
correlations obtained from derivatives of the last term are proved
to be vanishing in the limit of removed cutoffs and if, in this
limit, we can safely substitute $\c_{h,N}(\kk) $ with $1$. Let us
first consider this problem, without giving the technical details.
If we make a certain number of derivatives with respect to the
field $\f$ at non coinciding points and put $\f=0$, we are faced
with the problem of calculating the limit of expressions of the
type
\be\lb{intd}
\int d\zz\; \d_{h,N}(\zz) g_\o(\xx_1-\xx_2-\zz) G_{h,N}(\xx_1-\zz,
\yy_2, \ldots, \yy_n)
\ee
where $\d_{h,N}(\zz)$ is the Fourier transform of $\c_{h,N}(\kk)$
and the points $\xx_1, \xx_2, \yy_2, \ldots, \yy_n$ are all
different, except $\xx_2$ and $\yy_2$, which can be equal. The
function $G_{h,N}(\uy)$ is a (truncated) Schwinger function, which
was proved in \cite{BFM} (eq. 2.58) to decay as $|\yy_i|^{-\e'}$,
$0<\e'<1$, if $|\yy_i|\to\io$, while the other points are fixed. By
using the bound 2.52 of \cite{BFM} it is also possible to prove
that it diverges as $|\yy_i-\yy_j|^{-\e''}$, $0<\e''<1$, if
$|\yy_i-\yy_j|\to 0$, while the other points stay constant. On the
other hand, it is easy to prove that $|\d_{h,N}(\zz)| \le
C(\g^{-2N} +|\zz|^2)^{-1}$. These properties and the good
convergence properties of $G_{h,N}(\uy)$ as $-h,N\to\io$ (uniform
if the points vary in non intersecting neighborhoods of the
arguments) imply that one can make without any problem in
\pref{intd} the limit $h\to -\io$ and substitute $G_{h,N}(\uy)$
with $G(\uy)=\lim_{-h,N\to\io} G_{h,N}(\uy)$. The previous remarks imply also
that the function $g_\o(\xx_1-\xx_2-\zz)\, G(\xx_1-\zz,
\yy_2, \ldots, \yy_n)$ is a $L^1$ function of $\zz$ with a finite
number of singularities; moreover, $\d_{-\io,N}(\zz) \to \d(\zz)$
as $N\to\io$. It follows that the limit of \pref{intd} does exist
and is given by $g_\o(\xx_1-\xx_2)\, G(\xx_1, \yy_2, \ldots,
\yy_n)$; one can also see that the limit is uniform, if the
$\xx_i$ vary in small non intersecting neighborhoods, so implying
that one can exchange the derivative with the removed cutoff limit
in the l.h.s. of \pref{DSE3}, written in the space coordinates.

We still have to discuss the main point, that is the fact that the
correlations obtained from derivatives of the last term of
\pref{DSE3} vanish in the removed cutoff limit.
Since we assume some familiarity with [BM], we shall do that
very briefly, by
studying the flow of the marginal terms proportional to the field
$\b$ in the effective potential related with the generating
functional \pref{gfter}.

After the integration of the fields $\ps^{(j')}, j'>j$, we obtain
an expression of the type:
\bea
\lb{FWY} &&e^{\WW_{\TT,\m}^{(h)}(\b,\f)} =\int\!P_{[h,j]}(d\ps)\
 \exp\left\{
 \VV^{(j)}\left(\f,\sqrt{Z_j}\ps\right)
+\WW^{(j)}_{\TT,{\rm irr}}\left(\b,\f,\sqrt{Z_j}\ps\right)\right\}
\cdot\nn\\
&&\cdot \exp\left\{
 \Big[\left({Z_N\over Z_j}\right)^2\TT^{(\m)}_0
 +{Z_N\over Z_j}\sum_{\s=\pm}\n_j^{(\s)}
 \TT^{(\m)}_{\s}\Big]
 \left(\sqrt{Z_j}\ps,\sqrt{Z_j}\b\right)\right\}
\cdot\nn\\
&&\cdot \exp\left\{
 \Bigg[
 \wt\z^{(3,\m\o)}_j\BB^{(3)}
 +{Z_N\over Z_j}\sum_{k=j}^N\wt\z^{(1,\m\o)}_k\BB^{(1)}\Bigg]
 \left(\sqrt{Z_j}\ps,\sqrt{Z_j}\b\right)\right\}
\eea
where $ \VV^{(j)}$ is the effective potential for $\b=0$,
$\WW^{(j)}_{\TT,{\rm irr}}$ is the irrelevant part of the terms of
order at least $1$ in $\b$, while the rest represents the
corresponding marginal terms, written in terms of two running
coupling constants:
\bea
\wt\z^{(3,\m)}_j
\defi\lft\{
\matrix{-\a^{(\m)}_N\l\hfill&\hfill \quad{\rm for\ } j=N\cr\cr
        \wt\l^{(\m)}_j-\a^{(\m)}_N\l_j
        \hfill&\hfill\quad{\rm for\ } j\le N-1\;;}\rgt.
\cr\cr\cr\cr \wt\z^{(1,\m)}_j
\defi\lft\{
\matrix{-\r_N^{(\m)}\hfill&\hfill \quad{\rm for\ } j=N\cr\cr
        \lft(\wt z^{(\m)}_j-\a^{(\m)}_N z_j\rgt)
        {Z_j\over Z_N} \hfill&\hfill\quad{\rm for\ } j\le N-1\;;}\rgt.
\eea
where $\{\tilde \l^{\m}_j\}$ and  $ \{\tilde z^\m_j\}$ are {\it
exactly} the coupling studied in \cite{BM}; while $\{\l_j\}$ and
$\{z_j\}$ are {\it exactly} the effective coupling and the field
renormalization of the original generating functional, $\WW$. In
\cite{BM} (equation (144)) it was proved that there exist
$\a^{\m}_N$ such that the following two bounds are both satisfied
\be\lb{tljz}
|\wt\l^{(\m)}_j-\a^{(\m)}_N\l_j|\le C\g^{-(N-j)/2} \virg |\wt
z^{(\m)}_j-\a^{(\m)}_N z_i|\le C\g^{-(N-j)/2}
\ee
Thereby, if we put $\r^{(\m)}_N=\sum_{j\le N-1}\wt\z^{(1,\m)}_j $,
the factor in front of $\BB^{(1)}$ in \pref{FWY} is
\bea
{Z_N\over Z_j}\sum_{k=j}^N\wt\z^{(1,\m)}_k ={Z_N\over
Z_j}\lft(\sum_{k=j}^{N-1}\wt\z^{(1,\m)}_k-\r_N^{(\m)}\rgt)
\cr\cr\cr =-{Z_N\over Z_j}\sum_{k\le
j-1}\wt\z^{(1,\m)}_k=\sum_{k\le j-1}(\wt z^{(\m)}_k-\a^{(\m)}_N z_k)
{Z_k\over Z_j}
\eea
and the last term, by the second \pref{tljz}, can be bounded by
$C\g^{-(N-j)/4}$. This remark, together with the first of
\pref{tljz}, allows us to prove that contribution to the
correlation functions of the last term in \pref{FWY} vanishes in
the limit of removed cutoffs, as a consequence of the short memory
property, see [BM] for details.

\subsection{Solution of the closed equations}

Let us define $a=\lim_{N\to\io} a_N$, $\bar a=\lim_{N\to\io} \bar
a_N$, $\D^{-1}(\xx|\yy)={1\over 2\p} \ln\lft({|\yy|\over
|\xx|}\rgt)$ and
\be
G^{(2n)}_{\uo,\uo'}(\ux,\uy) = e^{-\cal W(0,0)} {\dpr^{2n} e^{\cal
W}\over \dpr\f^+_{\xx_n,\o_n} \cdots \dpr\f^+_{\xx_1,\o_1}
\dpr\f^-_{\yy_1,\o'_1} \cdots \dpr\f^-_{\yy_n,\o'_n} }(0,0)
\ee

\begin{theorem}\lb{t10}
For $|\l|$ small enough and $\xx\not=\yy$,
\be\lb{S2}
S^{(2)}_{\o,\o'}(\xx,\yy) \= G^{(2)}_{\o,\o'}(\xx,\yy) =
\d_{\o,\o'} g_\o(\xx-\yy) e^{{\l A (a-\bar a)\over
2}\D^{-1}(\xx-\yy|\zz)}
\ee
where $\zz$ is a fixed, non-zero position, whose arbitrariness
reflects the arbitrariness of a factor in front of
$S^{(2)}_{\o,\o'}(\xx,\yy)$. Furthermore, if $n>1$ and $(\xx_1,
\ldots, \xx_n,\yy_1, \ldots,\yy_n)$ is a family of two by two
distinct points,
\be\lb{sol}
G^{(2n)}_{\uo,\uo'}(\ux,\uy)=\sum_{\p\in P_X}(-1)^\p
\GG^{(2n)}_{\uo,\p(\uo')}(\ux,\p(\uy))
\ee
where $X\defi\{1,\ldots,n\}$, $P_X$ is the set of the permutations
of the elements of $X$ and
\bea\lb{sol2}
&&\GG^{(2n)}_{\uo,\uo'}(\ux,\uy)= \lft(\prod_{j=1}^n
S^{(2)}_{\o_j,\o'_j}(\xx_j,\yy_j)\rgt)\cdot\nn\\
&&\cdot \prod_{s,t\in X}^{s<t} e^{\l A {a-\bar a\o_s\o'_t\over 2}
\lft[\D^{-1}(\xx_s-\yy_t|\xx_s-\xx_t)-\D^{-1}(\yy_s-\yy_t|\yy_s-\xx_t)\rgt]}
\eea
\end{theorem}

{\bf\0Proof.} The equation \pref{S2} has been proved in
\cite{BFM}. The proof of \pref{sol} will be done by checking that
the {\it truncated} correlation functions corresponding to the
functions \pref{sol}, assumed to be the right {\it not truncated}
correlation functions, solve the identity \pref{SD10} for $n>1$.
The reason for this procedure is that, as we have discussed
\S\ref{secA3}, we were able to get a closed equation, in the limit
of removed cutoffs, only for the truncated correlation functions
with $n>1$. However, it is worthwhile to give first the heuristic
argument which allows us to conjecture that \pref{sol} is the
right expression for the not truncated correlation functions.

In the limit $N\to\io$, if we put $Z=\lim_{N\to\io}$ and we ignore
the fact that $Z=0$, the identity \pref{DSE3} can be written in
terms of the space coordinates as
\bea\lb{2.33a}
&&\dpr_{\o_1}^{\xx_1}{\dpr e^{\WW}\over \dpr \f_{\xx_1,\o_1}^+}
={B\over Z} e^\WW \f_{\xx_1,\o_1}^- + \l A \sum_{\e}A_{-\o_1\e}\cdot\nn\\
&&\cdot \int\!d\zz\ g_{-\o_1}(\xx_1-\zz) \lft[\f^+_{\zz,\e}{\dpr
e^W\over \dpr \f^+_{\zz,\e}\dpr \f^+_{\xx_1,\o_1}} -{\dpr e^W\over
\dpr \f^+_{\xx_1,\o_1}\dpr \f^-_{\zz,\e}}\f^-_{\zz,\e}\rgt]
\eea
This implies, if $\h=\l AA_-$ and $Z=\lim_{N\to\io} Z_N$ (we will
ignore then fact that $Z=0$), that $S^{(2)}_{\o,\o'}(\xx,\yy) =
\d_{\o,\o'} S_\o(\xx-\yy)$, with
\be
\dpr_\o S_\o(\xx) = {B\over Z} \d(\xx) - \h g_{-\o}(\xx) S_\o(\xx)
\ee
Hence, since, for any value of $\zz$, $g_{-\o}(\xx) = -\dpr_\o
\D^{-1}(\xx|\zz)$,  we get
\be\lb{solf}
S_\o(\xx) = {B\over Z} e^{\h \big[\D^{-1}(\xx|\zz) -
\D^{-1}(0|\zz)\big]} g_\o(\xx)
\ee
where $\D^{-1}(0|\zz)=+\io$, which should balance, in this formal
calculation, the fact that $Z=0$. In fact, this equation implies
the correct value \pref{S2} of $S_\o(\xx)$, if we choose $\zz$ so
that
\be\lb{2.39}
{B\over Z} = e^{\h \D^{-1}(0|\zz)}
\ee
Hence, we are encouraged to pursue this formal procedure. If we
take $2n-1$ suitable functional derivatives in both sides of
\pref{2.33a} and we call $\ux_j$ the vector $\ux$ without the
element $\xx_j$, we find the following equation:
\bea
&&\dpr_{\o_1}^{\xx_1}G^{(2n)}_{\uo,\uo'}(\ux,\uy) ={B\over
Z}\sum_{k=1}^n (-1)^{k-1}\d_{\o_1,\o'_k}\d(\xx_1-\yy_k)
G^{(2n-2)}_{\uo_1,\uo'_k}(\ux_1,\uy_{k})\\
&&+\l A \lft[\sum_{k=2}^n A_{-\o_1\o_k} g_{-\o_1}(\xx_1-\xx_k)
-\sum_{k=1}^n A_{-\o_1\o'_k}
g_{-\o_1}(\xx_1-\yy_k)\rgt]G^{(2n)}_{\uo,\uo'}(\ux,\uy)\nn
\eea
By using \pref{2.39}, this equation can be written as
\bea
&&\dpr_{\o_1}^{\xx_1} \lft[\lft(\prod_{h=2}^n e^{\l A
A_{-\o_1\o_h} \D^{-1}(\xx_1-\xx_h|\zz)}
\rgt)\lft(\prod_{h=1}^ne^{-\l A A_{-\o_1\o'_h}
\D^{-1}(\xx_1-\yy_h|z)}\rgt) G^{(2n)}_{\uo,\uo'}(\ux,\uy) \rgt]= \nn\\
&&=\sum_{k=1}^n (-1)^{k-1}\d_{\o_1,\o'_k}\d(\xx_1-\yy_k)\cdot\\
&&\cdot \lft(\prod_{h=2}^n e^{\l A A_{-\o_1\o_h}
\D^{-1}(\yy_k-\xx_h|\zz)} \rgt) \lft(\prod_{h=1\atop h\neq k}^n
e^{-\l AA_{-\o_1\o'_h} \D^{-1}(\yy_k-\yy_h|\zz)}\rgt)
G^{(2n-2)}_{\uo_1,\uo'_k}(\ux_1,\uy_{k})\nn
\eea
and hence we arrive at a formula for $G^{(2n)}$ in terms of
$G^{(2n-2)}$:
\bea
&& G^{(2n)}_{\uo,\uo'}(\ux,\uy) =\sum_{k=1}^n (-1)^{k-1}
S^{(2)}_{\o_1,\o'_k}(\xx_1-\yy_k)
G^{(2n-2)}_{\uo_1,\uo'_k}(\ux_1,\uy_{k})
\cdot\\
&&\hskip-1cm \cdot \lft(\prod_{h=2}^n e^{\l A A_{-\o_1\o_h}
\D^{-1}(\yy_k-\xx_h|\xx_1-\xx_h)} \rgt) \lft(\prod_{h=1\atop h\neq
k}^ne^{-\l AA_{-\o_1\o'_h} \D^{-1}(\yy_k-\yy_h|\xx_1-\yy_h)}\rgt)
\nn
\eea
where all $Z$ factors disappeared. Such an iterative relation is
clearly solved by \pref{sol} together with \pref{sol2}.

Let us now assume that the expression \pref{sol} is correct; we
want to check if the corresponding truncated correlation functions
satisfy the identity \pref{SD10}, for $n>1$. First of all, we
remind the connection between the two kind of functions. In order
to abridge the notation, we put $\f_j^+\defi\f^+_{\xx_j,\o_j} $
and $\f^-_j\defi\f^-_{\yy_j,\o'_j}$; furthermore, if $X_j\subset
X=\{1,\ldots,n\}$, we define $\f_{X_j}^\e\defi \prod_{k\in
X_j}\f_k^\e$, with the factors ordered with decreasing $k$, if
$\e=+$, and with increasing $k$ otherwise. Expanding $e^\WW(0,\f)$
in powers of $\WW$, we find:
\bea\lb{ge}
&&{\dpr^{2n}e^\WW\over \dpr\f^+_X \dpr\f^-_X
}(0,0) =\sum_{m=0}^n{1\over m!}\sum_{X_1,\ldots,X_m}^* \sum_{\p\in
\PP_X^{X_1,\ldots,X_m}} (-1)^\p\cdot\\
&&{\dpr^{2|X_1|}\WW\over \dpr\f^+_{X_1} \dpr\f^-_{\p(X_1)}}(0,0)
\dots {\dpr^{2|X_m|}\WW\over \dpr\f^+_{X_m}
\dpr\f^-_{\p(X_m)}}(0,0)\nn
\eea
where $\sum_{X_1,\ldots,X_m}^*$ denotes the sum over all the
possible partitions of $X$ into $m$, distinguishable and non empty
subsets $X_1,\ldots,X_m$. Furthermore, $\PP_X^{X_1,\ldots,X_m}$ is
the quotient set of the permutations of the elements of $X$,
$\PP_X$, where two elements are identified if they differ only for
a permutation in $\PP_{X_1}\otimes\cdots\otimes \PP_{X_m}$.

To find out the explicit expression of the truncated functions, we
define
\bea
&&V(\xx_s|P_t)\defi \l A A_{-\o_s\o'_t}
\D^{-1}(\xx_s-\yy_t|\xx_s-\xx_t)\nn\\
&&V_{s,t}\defi V(\xx_s|P_t)-V(\yy_s|P_t)
\eea
and, exploiting the analogy of the expression \pref{sol2} with the
partition function of a lattice gas, we perform the Mayer
expansion:
\bea
&& \GG^{(2n)}_{\uo,\uo'}(\ux,\uy) = \lft[\prod_{s,t\in X}^{s<t}
\lft(e^{V_{s,t}}-1+1\rgt)\rgt] \prod_{j=1}^n
S^{(2)}_{\o_j,\o'_j}(\xx_j,\yy_j)=\\
&&= \sum_{m=0}^n{1\over m!}\sum_{X_1,\ldots,X_m}^* \prod_{i=1}^m
\sum_{g\in\CC(X_i)} \prod_{\la s,t\ra\in g} \lft(
e^{V_{s,t}}-1\rgt) \prod_{k\in X_i}
S^{(2)}_{\o_k,\o'_k}(\xx_k,\yy_k)\nn
\eea
where the link $\la s,t\ra$ is the order pair of the elements
$s,t\in X$; $\CC(X)$ is the set of the graphs containing a path
which connects every element of $X$; when $X_j$ is made of only
one point, there is no possible graph; as usual, the product over
empty sets gives 1 by definition. Therefore  the comparison with
\pref{ge} gives the following expression for the truncated
functions:
\be \lb{SOL1}
S^{(2n)}_{\uo,\uo'}(\ux,\uy)=\sum_{\p\in P_X}(-1)^\p {\tilde
S}^{(2n)}_{\uo,\p(\uo')}(\ux,\p(\uy))
\ee
with
\bea\lb{ant}
{\tilde S}^{(2n)}_{\uo,\uo'}(\ux,\uy)= \sum_{g\in\CC(X)}
\prod_{\la s,t\ra\in g} \lft( e^{V_{s,t}}-1\rgt) \prod_{k\in X}
S^{(2)}_{\o_k,\o'_k}(\xx_k,\yy_k)
\eea
We now perform some manipulations. From the previous expression we
get
\bea
&&\lft(\prod_{s\in X}^{s\neq 1}e^{-V(\xx_1|P_s)}\rgt) e^{-{\l A
A_- \D^{-1}(\xx_1-\yy_1|\zz)}}
{\tilde S}^{(2n)}_{\uo,\uo'}(\ux,\uy) =\nn\\
&&=\d_{\o_1,\o'_1}g_{\o_1}(\xx_1-\yy_1)\lft(\prod_{k=2}^n
S^{(2)}_{\o_k,\o'_k}(\xx_k-\yy_k)\rgt)\cdot\\
&&\cdot \sum_{g\in\CC(X)} \prod_{s\in X}^{\la s,1\ra\notin g}
\lft( e^{-V(\xx_1|P_s)}\rgt) \lft(\prod_{s\in X}^{\la s,1\ra\in
g}e^{-V(\yy_1|P_s)}-e^{-V(\xx_1|P_s)}\rgt) \prod_{\la s,t\ra \in
g}^{s,t\neq 1} \lft(e^{V_{s,t}}-1\rgt)\nn
\eea
and therefore, by taking a derivative w.r.t. $\xx_1$, we find
\bea
&&\dpr_{\o_1}^{\xx_1} \lft[\lft(\prod_{s\in X}^{s\neq
1}e^{-V(\xx_1|P_s)}\rgt) e^{-\l A A_- \D^{-1}(\xx_1-\yy_1|\zz)}
{\tilde S}^{(2n)}_{\uo,\uo'}(\ux,\uy)\rgt] =\nn\\
&&=\d_{\o_1,\o'_1}g_{\o_1}(\xx_1-\yy_1)\lft(\prod_{k=2}^n
S^{(2)}_{\o_k,\o'_k}(\xx_k-\yy_k)\rgt)\cdot\\
&&\cdot \sum_{h\in X}^{h\neq 1}\sum_{g\in\CC(X)} \lft(\prod_{s\in
X}^{s\neq h\atop\la s,1\ra\notin g}e^{-V(\xx_1|P_s)}\rgt)
\prod_{s\in X}^{s\neq h\atop\la s,1\ra\in
g} \lft(e^{-V(\yy_1|P_s)}-e^{-V(\xx_1|P_s)}\rgt) \cdot\nn\\
&&\cdot\prod_{\la s,1\ra\in g}^{s,t\neq 1} \lft(e^{V_{s,t}}-1\rgt)
 (-1)^{\Th[\la h,1\ra\notin g]}e^{-V(\xx_1|P_h)}(\dpr_1
V)(\xx_1|P_h)\nn
\eea
where $\Th[\cdot]$ is equal to $1$ if the relation $[\cdot]$ is
true; $\Th[\cdot]$ is zero otherwise. If the graph $g\in \CC(X)$
does not contain the link $\la h,1\ra$, then also the graph
$g'\defi g\cup \la h,1\ra$ is in $\CC(X)$; because of the factor
$(-1)^{\Th[\la h,1\ra\notin g]}$ their contribution cancel each
other. We call  $\CC_{\la h,1\ra}(X)$ the remaining set of graphs:
it is made of the graphs in $\CC(X)$ which are no longer in
$\CC(X)$ if the link $\la h,1\ra$ is erased. Clearly they can be
also constructed by joining with the link $\la h,1\ra$ two graphs
$g_1\in \CC(X_1)$ and $g_h\in \CC(X_h)$, for any choice of
disjoint $X_1$ and $X_h$ s.t. $1\in X_1$, $h\in X_h$ and $X_1\cup
X_h=X$. Because of these considerations we arrive at the
expression
\bea
&&\dpr_{\o_1}^{\xx_1} \lft[\lft(\prod_{s\in
X}^{s\neq1}e^{-V(\xx_1|P_s)}\rgt) e^{-{\l A
A_-}\D^{-1}(\xx_1-\yy_1|z)}{\tilde S}^{(2n)}_{\uo,\uo'}(\ux,\uy)
\rgt] =\\
&& = \lft(\prod_{s\in X}^{s\neq 1}e^{-V(\xx_1|P_s)}\rgt) e^{-{\l A
A_-} \D^{-1}(\xx_1-\yy_1|z)} \lft(\prod_{k=1}^n
S^{(2)}_{\o_k,\o'_k} (\xx_k-\yy_k)\rgt)\cdot\nn\\
&&\cdot \sum_{h\in X}^{h\neq 1}\sum_{X_1,X_h}^{**}
\sum_{g_1\in\CC(X_1)\atop g_h\in\CC(X_h)}(\dpr_1 V)(\xx_1|P_h)
\lft(\prod_{(s,t)\in g_1}e^{V_{s,t}}-1\rgt) \lft(\prod_{(s,t)\in
g_h}e^{V_{s,t}}-1\rgt)\nn
\eea
where $\sum_{X_1,X_h}^{**}$ is the same of $\sum_{X_1,X_h}^{*}$,
with the further constraint that $1\in X_1$ and $h\in X_h$ (such a
notation is abusive, but quite clear): therefore $X_1, X_h$ is
ordered, and there is no factor $(1/2!)$. As consequence, for
$n>1$, $S^{(2n)}_{\uo,\uo'}(\ux,\uy)$ satisfy \pref{SD10}, that,
after suitable derivatives in the fields, reads
\bea
&&\dpr_{\o_1}^{x_1}S^{(2n)}_{\uo,\uo'}(\ux,\uy) =\l A
\sum_{h=2}^n\sum_{X_1,X_h}^{**}\sum_{\p\in P^{X_1,X_h}_X}(-1)^\p
M^{n,h}_{X_1,X_h}(\ux,\uy_{\circ\p})+\\
&&+\l A \lft[\sum_{h=2}^nA_{-\o_1\o_h} g_{-\o_1}(\xx_1-\xx_h)
-\sum_{h=1}^nA_{-\o_1\o'_h} g_{-\o_1}(\xx_1-\yy_h)\rgt]
S^{(2n)}_{\uo,\uo'}(\ux,\uy)\nn
\eea
with
\bea
&&M^{n,h}_{X_1,X_h}(\ux,\uy) \defi \lft[A_{-\o_1\o_h}
g_{-\o_1}(\xx_1-\xx_h) -A_{-\o_1\o'_h}
g_{-\o_1}(\xx_1-\yy_h)\rgt]\cdot \nn\\
&& \cdot S^{(2|X_1|)}_{\uo_{X_1},\uo'_{X_1}}(\ux_{X_1},\uy_{X_1})
S^{(2|X_h|)}_{\uo_{X_h},\uo'_{X_h}}(\ux_{X_h},\uy_{X_h})
\eea
\hfill\qed\hskip1em\null


\end{document}